\documentclass[longauth]{aa}  

\usepackage{amsmath}
\usepackage{amssymb}
\usepackage{booktabs}
\usepackage{color}
\usepackage{colortbl}
\usepackage[english]{babel}
\usepackage{enumitem}
\usepackage{epsfig}
\usepackage{gensymb}
\usepackage{graphicx}
\usepackage{hyperref}
\usepackage{ifthen}
\usepackage{latexsym}
\usepackage{lscape}
\usepackage{natbib}
\usepackage{rotating}
\usepackage{subfig}
\usepackage{txfonts}

\hypersetup{pagebackref=true, colorlinks=true, linkcolor=red, citecolor=blue, urlcolor=blue, bookmarks=false} 

\def\gsimeq{\hbox{\raise0.5ex\hbox{$>\lower1.06ex\hbox{$\kern-1.07em{\sim}$}$}}} 
\def\lsimeq{\hbox{\raise0.5ex\hbox{$<\lower1.06ex\hbox{$\kern-1.07em{\sim}$}$}}} 

\bibpunct{(}{)}{;}{a}{}{,} 

\begin{document}

\title{The VLA-COSMOS 3~GHz Large Project:\\ AGN and host-galaxy properties out to z$\lesssim$6 }

\author{I.~Delvecchio\inst{1}\thanks{email: ivand@phy.hr}
\and V.~Smol\v{c}i\'c\inst{1}
\and G.~Zamorani\inst{2}
\and C.~Del~P.~Lagos\inst{3,4}
\and S.~Berta\inst{1}\thanks{Visiting scientist}
\and J.~Delhaize\inst{1}
\and N.~Baran\inst{1}
\and D.~M.~Alexander\inst{5}
\and D.~J.~Rosario\inst{5}
\and V.~Gonzalez-Perez\inst{6}
\and O.~Ilbert\inst{7}
\and C.~G.~Lacey\inst{8}
\and O.~Le F{\`e}vre\inst{7}
\and O.~Miettinen\inst{1}
\and M.~Aravena\inst{9}
\and M.~Bondi\inst{10}
\and C.~Carilli\inst{11,12}
\and P.~Ciliegi\inst{2}
\and K.~Mooley\inst{13}
\and M.~Novak\inst{1}
\and E.~Schinnerer\inst{14}
\and P.~Capak\inst{15}
\and F.~Civano\inst{16}
\and N.~Fanidakis\inst{8}
\and N.~Herrera~Ruiz\inst{17}
\and A.~Karim\inst{18}
\and C.~Laigle\inst{19}
\and S.~Marchesi\inst{20}
\and H.~J.~McCracken\inst{19}
\and E.~Middleberg\inst{17}
\and M.~Salvato\inst{21}
\and L.~Tasca\inst{7}
}

   \institute{Department of Physics, Faculty of Science, University of Zagreb,  Bijeni\v{c}ka cesta 32, 10000  Zagreb, Croatia.
   \and  INAF - Osservatorio Astronomico di Bologna, Via Piero Gobetti 93/3, I-40129 Bologna, Italy.
   \and  Australian Research Council Centre of Excellence for All-sky Astrophysics (CAASTRO), 44 Rosehill Street Redfern, NSW 2016, Australia. 
   \and  International Centre for Radio Astronomy Research, University of Western Australia, 35 Stirling Highway, Crawley, WA 6009, Australia. 
   \and  Centre for Extragalactic Astronomy, Department of Physics, Durham University, South Road DH1 3LE, UK.
   \and  Institute of Cosmology and Gravitation, University of Portsmouth, Dennis Sciama Building, Portsmouth PO1 3FX, UK.
   \and  Aix Marseille Universit\'e, CNRS, LAM (Laboratoire d'Astrophysique de Marseille) UMR 7326, 13388, Marseille, France.
   \and  Institute for Computational Cosmology (ICC), Department of Physics, Durham University, South Road, Durham, DH1 3LE, UK. 
   \and  N\'ucleo de Astronom\'ia, Facultad de Ingenier\'ia y Ciencias, Universidad Diego Portales, Av. Ej\'ercito 441, Santiago, Chile. 
   \and  INAF - Istituto di Radioastronomia, Via P. Gobetti 101, 40129 Bologna, Italy.
   \and  National Radio Astronomy Observatory, Socorro, NM, USA. 
   \and  Cavendish Laboratory, Cambridge, UK. 
   \and  California Institute of Technology, MC 249-17, 1200 East California Boulevard, Pasadena, CA 91125, USA.
   \and  Max Planck Institut f\"ur Astronomie, K\"onigstuhl 17, 69117 Heidelberg, Germany. 
   \and  Department of Astronomy, California Institute of Technology, MC 249-17, 1200 East California Blvd, Pasadena, CA 91125, USA. 
   \and  Harvard Smithsonian Center for Astrophysics, 60 Garden st, Cambridge MA 02138, USA.
   \and  Astronomisches Institut, Ruhr-Universit\"at Bochum, Universit\"atstrasse 150, 44801 Bochum, Germany. 
   \and  Argelander-Institut f\"ur Astronomie, Universit\"at Bonn, Auf dem H\"ugel 71,D-53121 Bonn, Germany.
   \and  Institut d’Astrophysique de Paris, Sorbonne Universit\'es, UPMC Univ Paris 06 et CNRS, UMR 7095, 98 bis bd Arago, 75014 Paris, France.
   \and  Department of Physics \& Astronomy, Clemson University, Clemson, SC 29634, USA.
   \and  Max-Planck-Institut f\"ur Extraterrestrische Physik (MPE), Postfach 1312, D-85741 Garching, Germany.
}

   \date{Received}

  \abstract{
  We explore the multiwavelength properties of AGN host galaxies for different classes of radio-selected AGN out to z$\lesssim$6 via a multiwavelength analysis of about {7\,700} radio sources in the COSMOS field. The sources were selected with the Very Large Array (VLA) at 3~GHz (10 cm) within the VLA--COSMOS 3~GHz Large Project, and cross-matched with multiwavelength ancillary data. This is the largest sample of high-redshift (z$\lesssim$6) radio sources with exquisite photometric coverage and redshift measurements available. We constructed a sample of moderate-to-high radiative luminosity AGN (HLAGN) via spectral energy distribution (SED) decomposition combined with standard X-ray and mid-infrared diagnostics. Within the remainder of the sample we further identified low-to-moderate radiative luminosity AGN (MLAGN) via excess in radio emission relative to the star formation rates in their host galaxies. 
  We show that at each redshift our HLAGN have systematically higher radiative luminosities than MLAGN and that their AGN power occurs predominantly in radiative form, while MLAGN display a substantial mechanical AGN luminosity component.
  We found significant differences in the host properties of the two AGN classes, as a function of redshift. At z$<$1.5, MLAGN appear to reside in significantly more massive and less star-forming galaxies compared to HLAGN. At z$>$1.5, we observed a reversal in the behaviour of the stellar mass distributions with the HLAGN populating the higher stellar mass tail. We interpret this finding as a possible hint of the downsizing of galaxies hosting HLAGN, with the most massive galaxies triggering AGN activity earlier than less massive galaxies, and then fading to MLAGN at lower redshifts. 
  Our conclusion is that HLAGN and MLAGN samples trace two distinct galaxy and AGN populations in a wide range of redshifts, possibly resembling the radio AGN types often referred to as radiative- and jet-mode (or high- and low-excitation), respectively, whose properties might depend on the different availability of cold gas supplies.}


  \titlerunning{AGN and galaxy properties out to z$\lesssim$6}

  \maketitle


\section{Introduction} \label{intro}

Multiwavelength observations of galaxies hosting active galactic nuclei (AGN) in the past 20 years have extensively shown that supermassive black holes (SMBHs) are one of the key ingredients in shaping the evolution of galaxies through cosmic time. In particular, it is now well-established that AGN activity and star formation in their hosts are related processes, which are likely driven by a common fuelling mechanism such as accretion of cold gas supplies (e.g. \citealt{Vito+14}). In the local Universe, hints of such a connection have been suggested by the empirical correlations found between black hole mass and galaxy properties (e.g. \citealt{Magorrian+98}; \citealt{Gebhardt+00}; \citealt{Ferrarese+02}; \citealt{Gultekin+09}). In the distant Universe, this co-evolution scenario is supported by the similarity between volume-averaged cosmic star formation history and black hole accretion history, which both peak at z$\sim$2 and decline towards the local Universe (e.g. \citealt{Madau+14} for a review). Semi-
analytic models (e.g. \citealt{Bower+06}) and numerical simulations interpret these correlations as originated from a long-lasting ($\gtrsim$ Gyr) self-regulation process between the SMBH and its host (e.g. \citealt{Bower+06}; \citealt{Croton+06}; \citealt{Hopkins+08}; \citealt{Lagos+08}; \citealt{Menci+08}), which occurs in two flavours: quasar mode (QSO mode) and radio mode. 

On the one hand, the quasar mode is usually associated with a radiatively efficient phase of SMBH accretion through isotropically distributed ionising winds and molecular outflows as means to prevent the runaway growth of SMBHs (e.g. \citealt{Booth+09}; \citealt{Dubois+14}). On the other hand, a subsequent radio-mode phase is usually invoked to prevent further episodes of galaxy star formation through mechanical feedback, such as collimated and relativistic jets (\citealt{Monaco+00}; \citealt{Dubois+14}), in order to regulate the galaxy stellar mass (\citealt{Croton+06}; \citealt{Marulli+08}; \citealt{Hopkins+10}) and to reproduce the galaxy colours observed in the local Universe (\citealt{Strateva+01}).

Though it is today widely accepted that the evolution of active SMBHs is connected to the evolution of their hosts, the underlying mechanisms explaining the transition between these two key stages of the AGN and galaxy life cycles are still poorly constrained. Testing this paradigm is challenging as this is supposed to be a quick transition (on timescales of $\sim$100 Myr) from highly accreting to fading AGN. Promising studies of individual smoking-guns, mostly X-ray and optically detected AGN (e.g. \citealt{Farrah+12}; \citealt{Cicone+14}; \citealt{Perna+15}; \citealt{Brusa+15}; \citealt{Brusa+16}), have shown compelling evidence of ongoing AGN feedback, but are currently limited to a small number of candidates. 
A combined study with large AGN samples is necessary to constrain the role of AGN feedback in a more statistical sense, and to shed light on the connection between AGN and their hosts at different cosmic epochs.

A fundamental, complementary perspective in the framework of the AGN-host evolution comes from radio observations. 
Indeed, radio observations are essential to capture possible signatures of relativistic jets powered by a central SMBH (e.g. \citealt{Hogan+15}), which are detected via the synchrotron emission of the jet (e.g. \citealt{Miller+93}). In addition, radio continuum emission may arise from the diffusion of cosmic ray electrons produced in supernovae and their remnants in high-mass star-forming regions, and this emission has been calibrated on star-forming galaxies to provide an almost dust-unbiased star formation rate (SFR) indicator (\citealt{Condon92}; \citealt{Yun+01}; \citealt{Bell03}). This underlines the great potential of radio observations in unveiling a mixture of AGN and star-forming galaxies. Nevertheless, radio observations need to be supplemented by multiwavelength data to fully characterise the nature of the radio sources. 

Outstanding progress has been made through the analysis of large samples of radio-selected AGN in the local Universe (\citealt{Smolcic+09b}; \citealt{Best&Heckman12}). For instance, \citet{Smolcic+09b} identified a twofold population of radio-emitting AGN, namely high-excitation and low-excitation radio galaxies (HERGs and LERGs, respectively), on the basis of the presence of high- or low-excitation lines in their optical spectra taken from the Sloan Digital Sky Survey (SDSS; \citealt{York+00}). Interestingly, the author found that HERGs preferentially live in galaxies within the green valley (in terms of optical {\sc [NUV-$r$]} colours and stellar mass, $M_{\star}$), while LERGs usually populate the red sequence of massive and passive systems. Such a dichotomy observed in the host-galaxy properties between HERGs and LERGs may reflect physically different modes of SMBH accretion and presumably different stages of AGN-galaxy evolution (\citealt{Hardcastle+06}; see \citealt{Heckman+14} for a review). 
While SMBHs in HERGs are thought to accrete via cold gas inflows from galaxy mergers or secular processes (\citealt{Sijacki+07}; \citealt{DiMatteo+08}; \citealt{Bournaud+12}), AGN activity in LERGs is probably induced by a continuous gas inflow coming from the atmosphere of the hot halo (\citealt{Bower+06}; \citealt{Ellison+15}). In the former case, accretion is radiatively efficient and covers a wide range of the electromagnetic spectrum (up to X-ray frequencies), while for the latter scenario the feedback is predominantly mechanical and does not outshine the host\ galaxy in most bands except radio. 

In their comprehensive study, \citet{Hickox+09} have thoroughly investigated the AGN, host galaxy, and environmental properties of X-ray, mid-IR (MIR), and radio-selected AGN at 0.25$<$z$<$0.8 in the B\"ootes field (\citealt{Jannuzi+99}). In particular, \citet{Hickox+09} defined as ``radio AGN'' those sources with (rest-frame) 1.4~GHz luminosity $L_{\rm 1.4~GHz} > $ 10$^{24.8}$ W Hz$^{-1}$ to minimise the contamination from star-forming galaxies. These authors found that most radio-selected AGN have very low accretion rates (Eddington ratio $\lambda_{\rm Edd} \lesssim$ 10$^{-3}$) and populate overdense regions similarly to the most massive galaxies (e.g. \citealt{Georgakakis+07}; \citealt{Silverman+09}; \citealt{Coil+09}). In contrast, X-ray and MIR selected AGN are characterised by active star formation and less dense environments. These results, which have been corroborated through a similar analysis up to intermediate redshifts ($z\lesssim1.4$; see \citealt{Goulding+14}), strongly suggest that various AGN selection criteria might be sensitive to physically distinct classes of AGN and galaxies. In particular, the peculiarity of black hole and galaxy properties observed in radio-selected AGN stands out more than in any other AGN sample. For these reasons, it is now widely recognised that a multiwavelength investigation of radio-selected sources is essential to constrain the AGN-galaxy properties in a key stage of their cosmic evolution. 

In this paper, we exploit the largest compilation of high-redshift (z$\lesssim$6) radio-selected galaxies in the Cosmic Evolution Survey (COSMOS; \citealt{Scoville+07}) field. The Karl G. Jansky Very Large Array (VLA) observations were conducted at 3~GHz (10 cm) over the entire COSMOS field, in the framework of the VLA-COSMOS 3~GHz Large Project (PI: V. Smol\v{c}i\'c, \citealt{smolcic+17a}), reaching a 1$\sigma$ sensitivity of 2.3$~\mu$Jy~beam$^{-1}$. The rich multiwavelength (X-ray to radio) data set of photometry and redshifts available in the COSMOS field allows us to investigate the physical properties of these sources from a panchromatic perspective. 
The main goals of the present work are twofold:\ first, to provide a value-added catalogue that includes classification and physical properties for each 3~GHz VLA-selected source in the COSMOS field, and, second, to explore the multiwavelength properties of AGN hosts for different classes of radio-selected AGN out to z$\lesssim$6.

The paper is structured as follows. In Sect. 2 we describe our sample selection and the cross-match with ancillary photometry. In Sect. 3 we decompose the multiwavelength spectral energy distribution (SEDs), while the classification of our sample is discussed in Sect. 4. A brief description of the value-added catalogue is given in Sect. 5. Sect. 6 illustrates the average radio-selected AGN host-galaxy properties out to z$\lesssim$6, while the interpretation of our results are presented and discussed in detail in Sect. 7. We list our concluding remarks in Sect. 8. In Appendix A we show the results of infrared stacking, while Appendix B shows a portion of the value-added 3~GHz radio catalogue including some physical parameters used in this work. Throughout this paper, magnitudes are given in the AB system (\citealt{Oke74}). We assume a \citet{Chabrier03} initial mass function (IMF) and a flat cosmology with $\Omega_{\rm m}$ = 0.30, $\Omega_{\rm \Lambda}$ = 0.70, and H$\rm _0$ = 70 km s$^{-1}$ Mpc$^{-1}$ (\citealt{Spergel+03}).


\section{Sample selection} \label{sample}

\subsection{3~GHz radio sources} \label{radio-multi}

Radio data at 3~GHz were collected from 384 hours of observations with VLA over 2.6 deg$^2$, reaching an average rms sensitivity of 2.3 $\mu$Jy~beam$^{-1}$ and an angular resolution of about $0\farcs75$. A detailed description of the survey strategy, data reduction, and radio source catalogue is given in \citet{smolcic+17a}.
The catalogue includes 10\,830 radio sources, identified at peak surface brightness $\geq$5$\sigma$, out of which 67 are multi-components. The present catalogue represents the deepest compilations of radio sources available to date across an area of 2.6 deg$^2$. Our sample covers a wide redshift range (0$<$z$\lesssim$6, see Sect. \ref{redshift}) and is around three times larger than the 1.4~GHz sample taken from the Westerbork Synthesis Radio Telescope (WSRT, 3\,172 sources) in the NOAO Deep Wide-Field Survey (NDWFS, \citealt{deVries+02}). Moreover, our sample outnumbers  the previous 1.4~GHz VLA-COSMOS survey by a factor of about four (2\,865 sources;\ see \citealt{schinnerer+07}, \citeyear{schinnerer+10}) and by more than one order of magnitude the 1.4~GHz VLA survey in the Extended \textit{Chandra}-Deep Field South (E-CDFS; 883 sources, \citealt{Miller+13}) survey. 

\begin{figure}
\begin{center}
    \includegraphics[width=\linewidth]{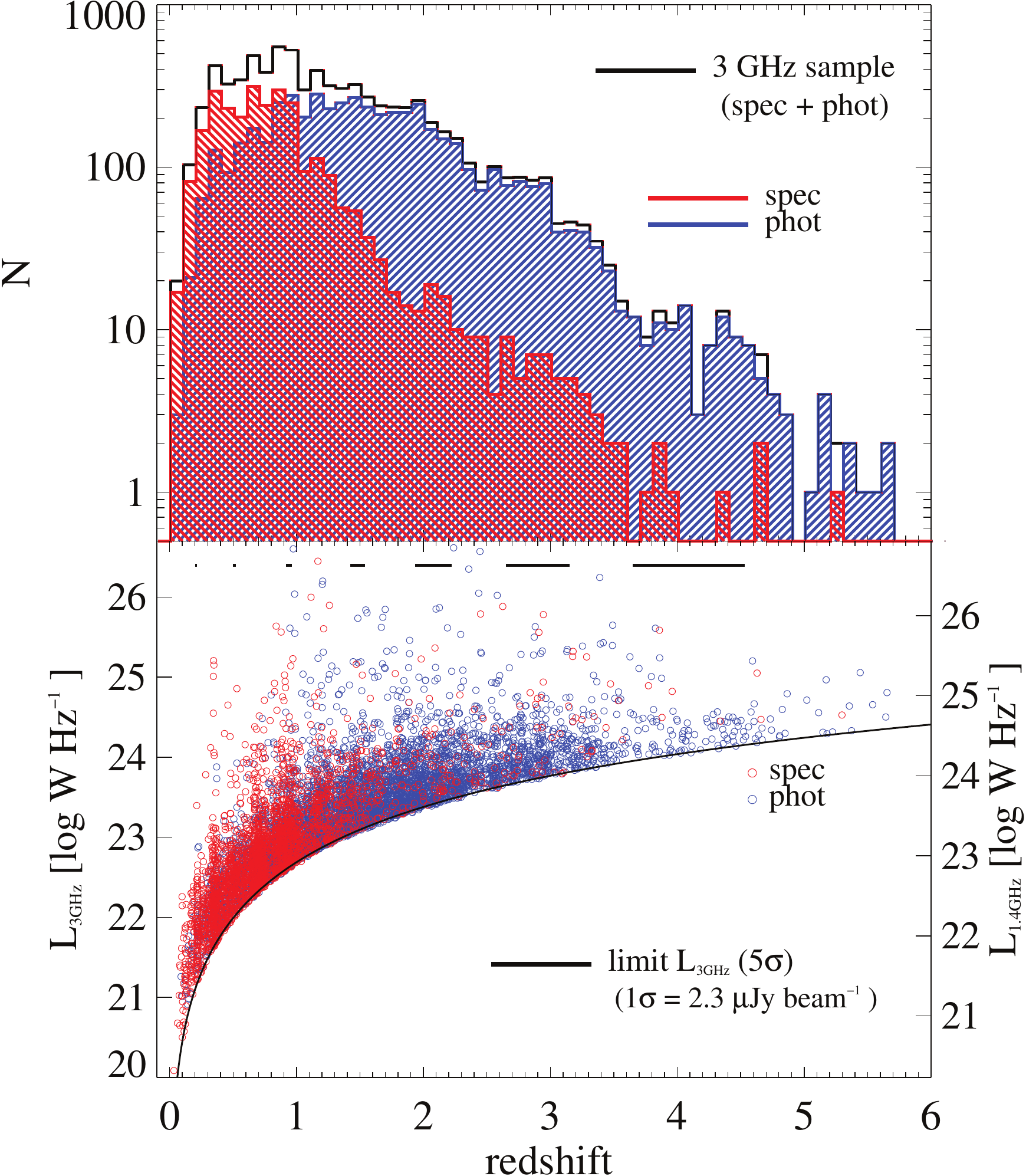}
\end{center}
 \caption{\small Top panel: redshift distribution of our {7\,729} radio sources. Spectroscopic and photometric redshifts are shown in red and blue, respectively, while the black line is the sum of the two. The scale of the $y$ axis is logarithmic. Bottom panel: circles show the rest-frame 3~GHz luminosity as a function of redshift, both spectroscopic (red) and photometric (blue). The horizontal bars indicates the average $\pm$1$\sigma$ uncertainty range of the photometric redshifts in the various redshift bins. The corresponding 1.4~GHz luminosity (scaled by using $\alpha$=--0.7) is shown for comparison on the right $y$ axis. The black solid line indicates the 5$\sigma$ luminosity limit at 3~GHz. }
   \label{fig:z_dist}
\end{figure}

We derived (rest-frame) 3~GHz radio luminosity ($L_{\rm 3\, GHz}$) for radio sources with multiwavelength counterparts and redshifts (see Sects. \ref{multi_lambda} and \ref{redshift}). Under the assumption of purely synchrotron emission, the radio spectrum behaves like a power law $S_{\nu} \propto \nu ^{\alpha}$, where the spectral index $\alpha$ is set to the observed 1.4--3~GHz slope for sources detected also at 1.4~GHz (about 30\%) in the 1.4~GHz VLA-COSMOS survey (\citealt{schinnerer+10}); the spectral index is set to --0.7, which is consistent with a non-thermal synchrotron index, (e.g. \citealt{Condon92}; see also \citealt{smolcic+17a}) if the sources are detected at 3~GHz alone. In Fig. \ref{fig:z_dist} (bottom panel) we show $L_{\rm 3\, GHz}$ as a function of redshift with respect to the 5$\sigma$ luminosity limit. For comparison, we show the corresponding 1.4~GHz luminosity $L_{\rm 1.4\, GHz}$ on the right $y$ axis. Our sample clearly spans a wide luminosity range (up to 4--5~dex), which allows us 
to investigate the multiwavelength properties of our sample in different radio luminosity regimes.
 
In particular, we are able to detect 2.5 times intrinsically fainter sources (under the assumption $\alpha$=--0.7), at a given redshift, compared to the previous 1.4~GHz VLA-COSMOS survey (\citealt{schinnerer+07}, \citeyear{schinnerer+10}).

\subsection{Optical to (sub)millimetre photometry} \label{multi_lambda}

The COSMOS field benefits from an exquisite photometric data set, covered from the X-rays to the submillimetre domain\footnote{An exhaustive overview of the COSMOS field and multiwavelength data products is available at: \url{http://cosmos.astro.caltech.edu/} }. Cross-matching our 3~GHz selected sample to existing ancillary data is essential to derive physical properties of galaxies. The multiwavelength photometry is taken from the COSMOS2015 catalogue (\citealt{Laigle+16}), which combines optical photometry\footnote{Optical photometry is taken from Subaru Hyper-Suprime Cam observations over the full 2 deg$^2$ (\citealt{Capak+07}), and also from the Canada-France-Hawaii Telescope Legacy Survey (CFHT-LS; \citealt{McCracken+01}) in the central 1 deg$^2$.}, the most recent UltraVISTA (DR2\footnote{DR2 replaces the previous DR1 by \citet{McCracken+12}. A detailed description of the survey and data products can be retrieved at: \url{http://ultravista.org/release2} }) data over the central 1.5 
deg$^2$ in the near-infrared (NIR) bands $Y$, $J$, $H$, and $K_s$\footnote{Outside the UltraVISTA coverage, NIR photometry includes CFHT $H$ and $K_s$ observations obtained with the WIRCam (\citealt{McCracken+01}).}, and MIR photometry obtained from the Infrared Array Camera (IRAC), which is recently complemented by deeper IRAC 3.6 and 4.5$~\mu$m observations with the \textit{Spitzer} Large Area Survey with Hyper-Suprime-Cam (SPLASH; \citealt{Steinhardt+14}; P.~Capak et al. in prep.). In addition, this data set has been cross-matched with 24$~\mu$m photometry (\citealt{LeFloch+09}) from the Multi-Band Imaging Photometer for \textit{Spitzer} (MIPS). \citet{Laigle+16} provides further details.

The cross-match to associate a possible optical-NIR counterpart with each radio source is fully described in \citet{smolcic+17b} (see their Sect. 3). First, they excluded stars and masked regions in the COSMOS2015 catalogue because of the less accurate optical photometry, which reduces the effective area of the COSMOS field to 1.77~deg$^2$ and our 3~GHz selected sample to 8\,696 radio sources. Secondly, they performed a nearest-neighbour matching, by selecting for each radio source only candidate counterparts within $0\farcs8$ searching radius and, at the same time, requiring a false match probability (i.e. probability of being a spurious association) lower than 20\%. This approach yields an average expected fraction of spurious associations of about 1\% (see \citealt{smolcic+17b}). The percentage of radio sources with multiple optical-NIR counterparts within $0\farcs8$ is around 1\%, for which the cut in false-match probability ensures the selection of the most probable counterpart. After this cut, our 
final sample consists of {7\,729} radio sources with optical-NIR counterparts, corresponding to about {89\%} of our radio-selected sample within the common 1.77 deg$^2$.

To enrich the spectral coverage of our analysis and derive robust star formation rates (SFRs) for as many sources as possible, we used also \textit{Herschel} photometry at far-infrared and submillimetre wavelengths provided in the COSMOS2015 catalogue. \textit{Herschel} imaging covers the entire COSMOS field with the Photoconductor Array Camera and Spectrometer (PACS; 100 and 160$~\mu$m, \citealt{Poglitsch+10}) and Spectral and Photometric Imaging Receiver (SPIRE; 250, 350, and 500$~\mu$m, \citealt{Griffin+10}) data, as part of the PACS Evolutionary Probe (PEP; \citealt{Lutz+11}) and the \textit{Herschel} Multi-tiered Extragalactic Survey (HerMES; \citealt{Oliver+12}). \textit{Herschel} fluxes were extracted and de-blended by using 24$~\mu$m positional priors and unambiguously associated with the corresponding optical-NIR counterpart via 24$~\mu$m sources listed in both catalogues. In total, the number of radio sources with ($\geq$3$\sigma$) \textit{Herschel} detection in at least one PACS or SPIRE band are {
4\,836/7\,729} ({63\%}). This percentage decreases with redshift, being {87\%} at z$<$0.3 and {45\%} at z$>$3.5.

To obtain reliable dust-based SFRs also in potential high-redshift candidates (z$>$3), where \textit{Herschel} observations are incomplete even towards ultraluminous infrared galaxies (ULIRGs, i.e. having rest-frame 8-1000$~\mu$m infrared luminosity $L_{\rm IR} \geq$ 10$^{12}$ L$_\odot$, e.g. \citealt{Sanders+96}), photometry at longer wavelengths is essential. For around 115 radio sources, we retrieved additional (sub)millimetre photometry from at least one of the following data sets: JCMT/SCUBA-2 at 450 and 850$~\mu$m (\citealt{Casey+13}), LABOCA at 870$~\mu$m (F.~Navarrete et al. priv. comm.), Bolocam (PI: J.~Aguirre), JCMT/AzTEC (\citealt{Scott+08}) and ASTE/AzTEC (\citealt{Aretxaga+11}) at 1.1 mm, MAMBO at 1.2 mm (\citealt{Bertoldi+07}), and interferometric observations at 1.3~mm with ALMA (PI: M. Aravena, M.~Aravena et al. in prep.) and PdBI (\citealt{Smolcic+12}; \citealt{Miettinen+15}). The (sub)mm positions were cross-matched to the COSMOS2015 positions via a nearest 
neighbour matching, using $1\farcs$ searching radius (the smallest beam width of the (sub)mm data we collected). A thorough visual inspection of the counterpart associations has been performed for the 1.3~mm detected ALMA sources (68\% of the (sub)mm photometry we collected), which is detailed in \citet{brisbin+17} and \citet{miettinen+17}.

We also collected X-ray data from the \textit{Chandra}-COSMOS (\citealt{Elvis+09}; \citealt{Civano+12}) and COSMOS-Legacy catalogues (\citealt{Civano+16}). The optical-NIR counterparts of X-ray sources were matched via a maximum likelihood algorithm and are presented in \citet{Marchesi+16}. We matched their catalogue to our 3~GHz selected sample of {7\,729} optical-NIR counterparts via COSMOS2015 IDs. This match yields {903} X-ray sources, corresponding to {12\% (903/7\,729)} of our radio sample, and to {32\% (903/2\,804)} of the X-ray sources with optical-NIR association in unmasked areas.

\subsection{Spectroscopic and photometric redshifts} \label{redshift}

We collected photometric redshifts for the {7\,729} radio sources with a counterpart in the COSMOS2015 catalogue. Photometric redshift estimates are included in the catalogue and were derived using the {\sc Le Phare} SED-fitting code (\citealt{Arnouts+99}; \citealt{Ilbert+06}) following the procedure detailed in \citeauthor{Ilbert+09} (\citeyear{Ilbert+09}, \citeyear{Ilbert+13}). Based on the comparison with the spectroscopic redshifts available in the COSMOS field, \citet{Laigle+16} report an average photometric redshift accuracy of $\left \langle |\Delta z/(1 + z)| \right \rangle =$ 0.021 for $K_s>$22, which becomes less than 0.010 for brighter sources. 

For X-ray sources, we used a different set of photometric redshifts from M.~Salvato et al. (in prep.), which are more suitable for AGN-dominated sources as they account for AGN variability and adopt additional AGN templates (\citealt{Salvato+09}, \citeyear{Salvato+11}).

An exhaustive list of spectroscopic redshifts was compiled (April 2015, Salvato et al. in prep.) and made internally accessible to the COSMOS team. Most of the spectroscopic redshifts used in this paper were taken from the $z$COSMOS survey (\citealt{Lilly+07}, \citeyear{Lilly+09}), either the public $z$COSMOS-bright or the proprietary $z$COSMOS-deep database, the DEep Imaging Multi-Object Spectrograph (DEIMOS, Capak et al. in prep.), and the FOcal Reducer and low dispersion Spectrograph (FORS2, \citealt{Comparat+15}). M.~Salvato et al. (in prep.) provide for a full reference list. 

For each radio source with a multiwavelength counterpart in this spectroscopic compilation, we replaced the photometric redshifts with new spectroscopic values only in case of secure or very secure measurements\footnote{The reliability of each spectroscopic redshift is determined by its quality flag. In case of spectroscopic redshift from the zCOSMOS survey (\citealt{Lilly+07}, \citeyear{Lilly+09}), we followed the prescription recommended on the zCOSMOS IRSA webpage: \url{https://irsa.ipac.caltech.edu/data/COSMOS/spectra/z-cosmos/Z-COSMOS_INFO.html} For the other surveys we selected quality flag $Qf \geq$ 3 and discarded less reliable spectroscopic redshifts from our analysis.}. In addition, we included the latest spectroscopic redshifts from the VIMOS Ultra Deep Survey (VUDS; \citealt{lefevre+15}; \citealt{Tasca+16}), from which we found 25 associations to our radio sources.

After these checks, the number of radio sources with spectroscopic redshift is {2\,734}/{7\,729} (around {35\%}). Every radio source with multiwavelength counterpart has its own redshift estimate. Fig. \ref{fig:z_dist} (top panel) shows the final redshift distribution for our {7\,729} radio sources. The number of spectroscopic and photometric redshifts are comparable out to z$\sim$1, while photometric redshifts become more numerous at higher redshift.
We tested the accuracy of the photometric redshifts in our sample based on the spectroscopic measurements available for {2\,734} sources. We found a median $\left \langle |\Delta z/(1 + z)| \right \rangle =$ 0.010, which becomes as high as 0.035 at z$>$3. Therefore, the proved accuracy of the photometric redshifts allows us to push our analysis out to z$\lesssim$6, even if the number of sources at z$>$4 is relatively small ({84} sources).


\section{SED-fitting decomposition of 3~GHz sources} \label{sed_fitting}

In this section, we fit the multiwavelength SEDs of our radio sources to disentangle the AGN emission from that related to the host-galaxy.
It is well known that radio-selected samples contain distinct galaxy populations (e.g. \citealt{Condon84}; \citealt{Windhorst+85}; \citealt{Gruppioni+99}) in terms of star formation and AGN properties. Therefore, fitting the multiwavelength SEDs may provide meaningful results only if AGN and galaxy templates are both taken into account. 

We used both the SED-fitting code {\sc magphys}\footnote{The original {\sc magphys} code is publicly available at this link: \url{http://www.iap.fr/magphys/magphys/MAGPHYS.html}} (\citealt{daCunha+08}), and the three-component SED-fitting code {\sc sed3fit} by \citet{Berta+13}, which accounts for an additional AGN component\footnote{The three-component SED-fitting code {\sc sed3fit} can be retrieved from \url{http://cosmos.astro.caltech.edu/page/other-tools}}. The aforementioned references provide for a full description of these SED-fitting codes. Here we briefly outline the main prescriptions that are relevant for our analysis.

The {\sc magphys} code is designed to reproduce a variety of galaxy SEDs, from weakly star-forming to starbursting galaxies, over a wide redshift range\footnote{For extensive application of the {\sc magphys} code in deriving physical properties of galaxies, see also \citet{Smith+12}; \citet{Rowlands+14}; \citet{Michalowski+14}; \citet{Hayward+15}.}. This code relies on the energy balance between the dust-absorbed stellar continuum and the reprocessed dust emission at infrared wavelengths. This recipe ensures that optical and infrared emission originating from star formation are linked in a self-consistent manner, but does not account for a possible AGN emission component. The three-component SED-fitting code presented by \citet{Berta+13} combines the emission from stars, dust heated by star formation, and a possible AGN-torus component from the library of \citeauthor{Feltre+12} (\citeyear{Feltre+12}, see also \citealt{Fritz+06}). This approach results in an effectively simultaneous three-component fit. For each best-fit parameter, the code provides a corresponding probability distribution function (PDF), which enables the user to obtain reliable confidence ranges for parameter estimates (see e.g. Calistro~Rivera et al., in prep., for a similar SED-fitting technique).

We decomposed each observed SED by using the best available redshift (either spectroscopic or photometric, see Sect. \ref{redshift}) as input, and we derived integrated galaxy properties, such as SFR and $M_{\star}$, for each individual source. The SFR was derived from the total IR (rest 8-1000$~\mu$m) luminosity taken from the best-fit galaxy SED (i.e. corrected for a possible AGN emission), assuming a \citet{Kennicutt98} conversion factor scaled to a \citet{Chabrier03} IMF. We note that about {37\%} of our sample are not $\geq$3$\sigma$ detected in any \textit{Herschel} bands. To obtain better constrained IR luminosities, we performed SED-fitting using the nominal PACS and SPIRE 3$\sigma$ upper limits, which are equal to 5.0 (100$~\mu$m), 10.2 (160$~\mu$m), 8.1 (250$~\mu$m), 10.7 (350$~\mu$m), and 15.4 (500$~\mu$m) mJy, including confusion noise (\citealt{Lutz+11}; \citealt{Oliver+12}). We modified the $\chi^2$ calculation to correctly account for those \textit{Herschel} bands that have only upper 
limits, similar to the approach adopted by \citet{dacunha+15}.

As a sanity check, we verified that the IR luminosities based on our three-component fit are in good agreement with those calculated independently using a different set of IR templates (from \citealt{Chary+01}; \citealt{Dale+02}; \citealt{Siebenmorgen+07}; \citealt{Polletta+07}; \citealt{Wuyts+08}; \citealt{Elbaz+11}; \citealt{Nordon+12}, see \citealt{Berta+13} for a comprehensive discussion). We briefly discuss the comparison with the \textit{Herschel} fluxes derived through stacking in Sect. \ref{ir_stacking}. The $M_{\star}$ is derived from the SED decomposition itself, which allows us to obtain robust estimates if the optical-NIR SED is dominated by the host-galaxy light (e.g. \citealt{Bongiorno+12}).

In order to quantify the relative incidence of a possible AGN component, we fitted each individual SED, both with the three-component approach and the {\sc magphys} code. The fit obtained with the AGN is preferred if the reduced $\chi^2$ value of the best fit is significantly (at $\geq99$\% confidence level, on the basis of a Fisher test) smaller than that obtained from the fit without the AGN; see \citet{Delvecchio+14} for details. From our analysis, we found that {1\,169} out of {7\,729} radio sources (about 15\%) show a $\geq99$\% significant AGN component in their best fit.

We extensively tested this technique against independent AGN indicators in the COSMOS field, such as MIR colours and X-rays (see \citealt{Delvecchio+14}). For instance, \citet{Lanzuisi+15} showed that the AGN radiative luminosities derived from SED decomposition were consistent (1$\sigma$=0.4~dex) with those calculated from X-ray spectra and assuming a set of bolometric corrections (e.g. \citealt{Lusso+12}). Moreover, the unprecedented accuracy of photometric redshifts and the photometric coverage exploited in this work, strengthened by our sizeable sample, further increase the reliability of our method. However, if the galaxy light outshines the AGN in the full optical-to-mm SED, this statistical technique becomes progressively less effective in identifying AGN.

\begin{figure}
\begin{center}
         \includegraphics[width=3.0in]{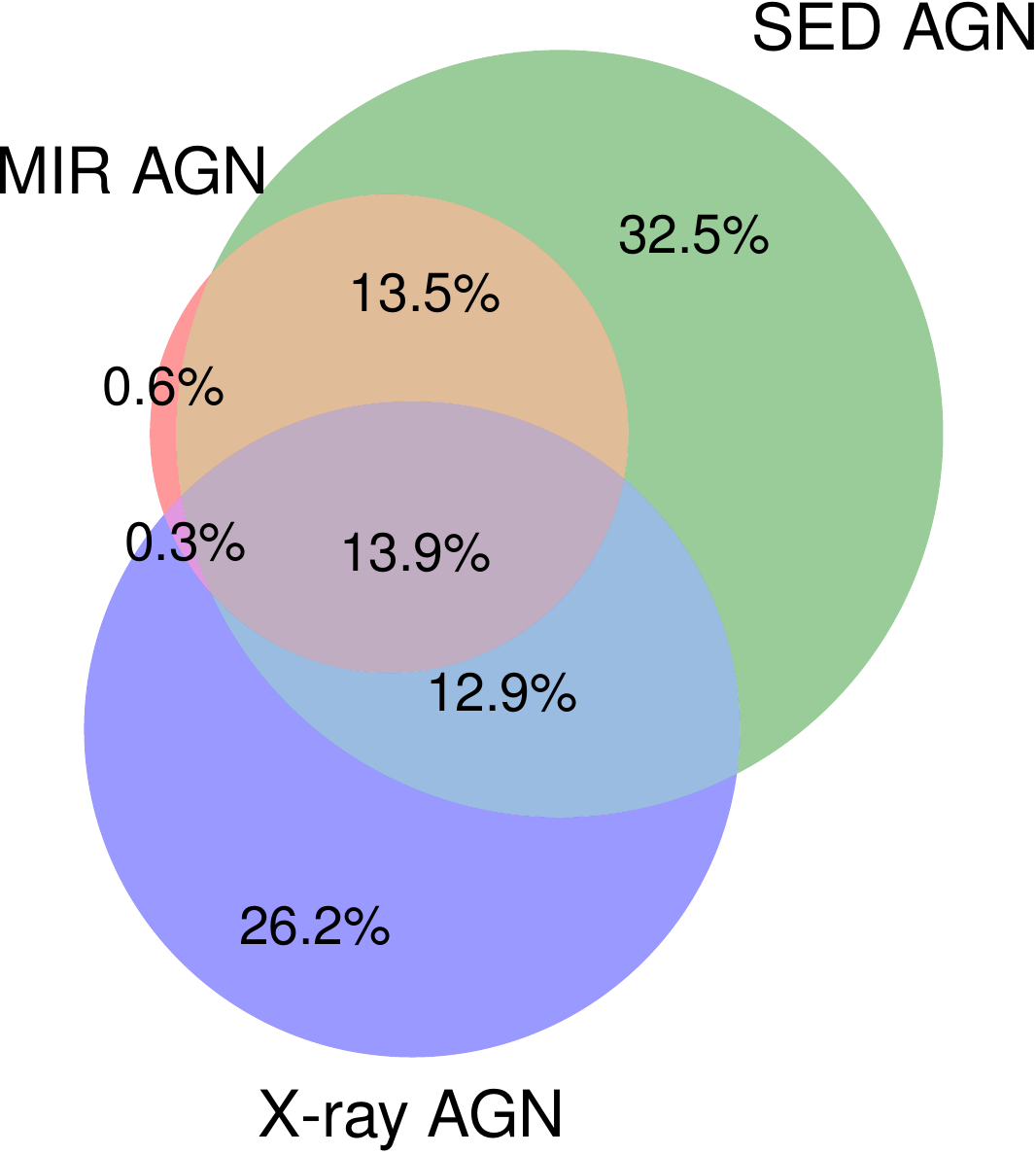}
\end{center}
 \caption{\small Venn diagram illustrating the percentages of the 1\,604 HLAGN in our sample identified from different AGN diagnostics: X-rays (blue), MIR (red) and SED decomposition (green). Areas roughly scale with percentages. }
   \label{fig:venn}
\end{figure}


\section{Classification of 3~GHz radio sources} \label{classification}

We combine SED-fitting decomposition (Sect. \ref{sed_fitting}) with other multiwavelength AGN diagnostics to reach a more complete census of AGN in our sample. These additional indicators are taken from X-ray, MIR and radio data, which allow us to identify two main populations of radio-selected AGN in our sample: moderate-to-high radiative luminosity AGN (HLAGN) and low-to-moderate radiative luminosity AGN (MLAGN). Hereafter, we refer to these populations as HLAGN (X-ray, MIR, and SED-selected AGN) and MLAGN (radio-excess sources that are not HLAGN), as {explained} in the next sections. 

This naming convention comes from the idea that the selection criteria based on SED-fitting, X-ray, and MIR data preferentially select higher luminosity AGN, where the term ``luminosity'' here refers to the AGN radiative luminosity (L$\rm_{rad, AGN}$), which is a proxy of the SMBH accretion rate (BHAR; e.g. \citealt{Alexander+12}). This classification does not translate into a sharp threshold in the accretion efficiency (or Eddington ratio) between HLAGN and MLAGN, but rather reflects the reliability of the adopted diagnostics in identifying such AGN populations\ combined with the sensitivity of our survey at various wavelengths. Moreover, the tags ``low to moderate'' and ``moderate to high'' intentionally imply a potential overlap in L$\rm_{rad, AGN}$ between the two classes at various redshifts. However, at a given redshift, these differently selected AGN display significantly distinct distributions of AGN luminosity, as detailed in Sect. \ref{naming_convention}. Therefore, the present 
classification should be considered as observationally based, and aimed at dissecting our radio sources on the basis of AGN diagnostics that are known to be luminosity dependent. A detailed investigation on the distribution of radio-selected AGN as a function of their intrinsic Eddington ratio will be presented in a forthcoming paper (Delvecchio et al., in prep.).

In Sects. \ref{hlagn} and \ref{llagn} we describe in more detail the multiwavelength diagnostics used to identify MLAGN and HLAGN, respectively, while in Sect. \ref{further_tests} we justify this naming convention by studying their L$\rm_{rad, AGN}$ distributions.

\subsection{Moderate-to-high radiative luminosity AGN} \label{hlagn}

As previously mentioned, the so-called HERG population identified in the local Universe consists of highly accreting SMBHs on the basis of the presence of high-excitation lines in their optical spectra (e.g. \citealt{Smolcic+08}), which implies radiatively efficient accretion. In order to detect potential HERG analogues in our sample, we combine SED decomposition (Sect. \ref{sed_fitting}) with X-ray and MIR indicators. All these selection criteria are sensitive to an excess of emission likely arising from accretion onto the central SMBH rather than from star formation. As a consequence, despite the different biases intrinsic to each criterion, all of these criteria preferentially select higher radiative luminosity AGN. The selection criteria used to identify this AGN population are briefly summarised below. \citet{smolcic+17b} provide a detailed description of the X-ray and MIR-based AGN indicators.

\begin{figure*}
\begin{center}
         \includegraphics[width=180mm,keepaspectratio]{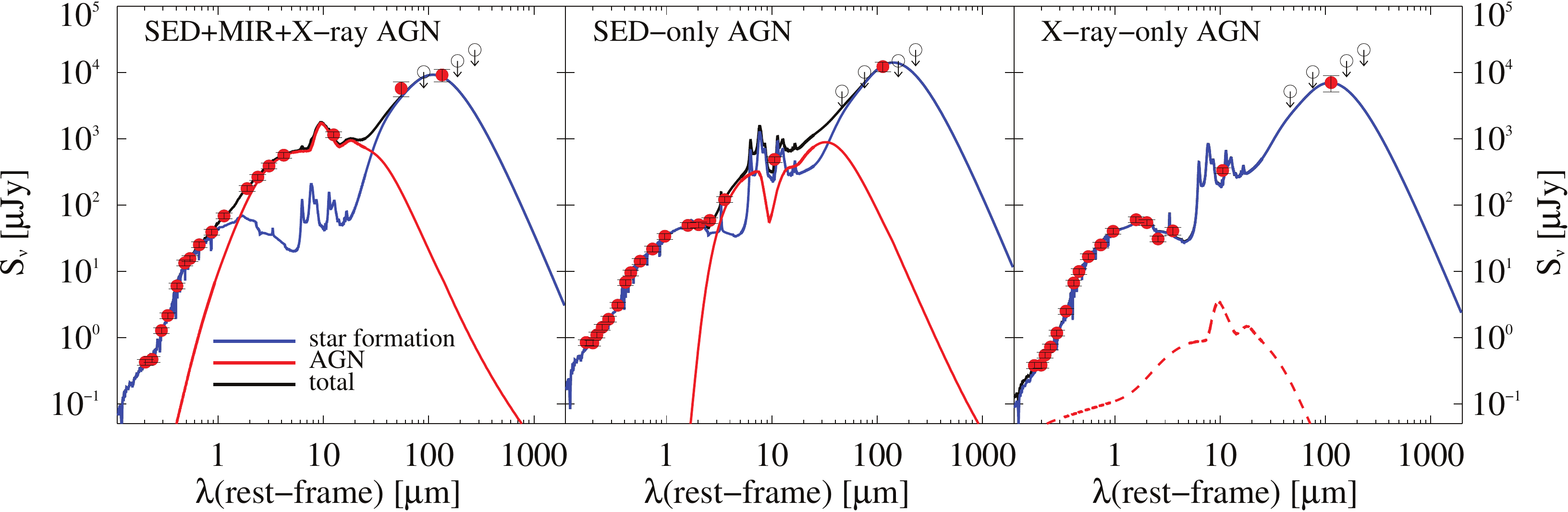}
\end{center}
 \caption{\small Three examples of best-fit SEDs of HLAGN selected from different criteria. Coloured lines represent the corresponding best-fit templates of AGN (red), galaxy star formation (blue), and the sum of the two (black). (Left panel) AGN identified from X-rays, MIR colours, and SED fitting. (Central panel) AGN identified only from SED-fitting. (Right panel) AGN identified only from X-rays. The red dashed line indicates that the AGN component is $<$99\% significant on the basis of the Fisher test (see text for more details). Red circles indicate the optical to far-IR (FIR) photometry (rest-frame), while downward pointing arrows represent 3$\sigma$ upper limits in the \textit{Herschel} bands. }
   \label{fig:sed_comparison}
\end{figure*}

First, SED-fitting decomposition identifies {1\,169} sources with $\geq$99\% significant AGN component in their global SED, SED AGN hereafter (see Sect. \ref{sed_fitting}).

Second, we used X-ray luminosities ($L_{\rm x}$) in the rest-frame [0.5--8] keV. The $L_{\rm x}$ estimates were calculated for the {903} X-ray detected sources, by assuming a fixed X-ray spectral slope $\Gamma=1.8$, and correcting for nuclear obscuration on the basis of the measured hardness ratio (e.g.~\citet{Xue+10}). We identified {855} sources with X-ray luminosity $L_{\rm x} \geq $10$^{42}$~erg~s$^{-1}$ as X-ray AGN (e.g. \citealt{Szokoly+04}). We verified that the X-ray emission expected from recent $L_{\rm x}$--SFR relations (taken from \citealt{Symeonidis+14}) is always negligible for our X-ray AGN (about a few percent). On the basis of the aforementioned relation, about 30 X-ray detected sources with $L_{\rm x} < $10$^{42}$ erg s$^{-1}$ show an X-ray excess and, therefore, could be considered low-luminosity X-ray AGN. However, we prefer to apply the same cut at $L_{\rm x} \geq $10$^{42}$ erg s$^{-1}$ for all our sources to avoid potential contamination from outliers with respect to the $L_{\rm x}$--
SFR relation.

Third, MIR colours can be very useful in identifying AGN, both unobscured and heavily obscured. \citet{Donley+12} proposed a conservative criterion to select AGN, on the basis of the MIR colour-colour diagram drawn from a combination of the four \textit{Spitzer}-IRAC (3.6, 4.5, 5.8, and 8.0$~\mu$m) bands. We followed Eqs. (1) and (2) of their paper to identify AGN at z$<$2.7, while at higher redshift we applied the additional conditions stated in their Eqs. (3) and (4) to minimise the contamination from high-redshift starbursts without AGN. This method is highly reliable for bright AGN, but becomes incomplete at $L_{\rm x}<$10$^{44}$ erg s$^{-1}$. In total, {455} out of {7\,729} radio sources (about 6\%) satisfy the \citet{Donley+12} criterion, and therefore are classified as MIR AGN.

Hereafter, we will use the term ``moderate-to-high radiative luminosity AGN'' (HLAGN) to collectively refer to the union of X-ray, MIR, and SED-selected AGN identified in our sample, for a total of {1\,604} objects (21\% of the radio sample). 

Figure \ref{fig:venn} shows the percentages of HLAGN classified from each criterion: the percentage of AGN that fulfills all the criteria simultaneously is only about {14\%} of the full HLAGN population. This small overlap further suggests that different AGN diagnostics are sensitive to distinct AGN populations. This overlap increases with increasing X-ray luminosity, which is 7\% for $10^{42}<L_{\rm x}<$10$^{43}$~erg s$^{-1}$, 25\% for $10^{43}<L_{\rm x}<$10$^{44}$~erg s$^{-1}$, and 49\% for $L_{\rm x}>$10$^{44}$~erg s$^{-1}$. These relatively small percentages are mainly driven by the incompleteness of the MIR classification, as the \citet{Donley+12} criterion is very conservative. We checked that the agreement between X-ray and SED-fitting diagnostics is as high as {21\%} for $10^{42}<L_{\rm x}<$10$^{43}$~erg s$^{-1}$, {52\%} for $10^{43}<L_{\rm x}<$10$^{44}$~erg s$^{-1}$, and 79\% for $L_{\rm x}>$10$^{44}$~erg s$^{-1}$.

In Fig. \ref{fig:sed_comparison} we illustrate some examples of best-fit SEDs, showing different levels of agreement between the AGN diagnostics described above. In all the panels, red circles indicate the (rest-frame) multiwavelength photometry, while downward pointing arrows set the 3$\sigma$ upper limits in the \textit{Herschel} bands. Solid lines represent the best-fit templates of AGN (red), galaxy star formation (blue), and the sum of the two (black). 

The left panel shows the SED of an unambiguous AGN, successfully identified from X-rays, MIR-colours, and SED-fitting decomposition. The central panel shows an AGN identified only from SED decomposition. Indeed, galaxies hosting heavily obscured AGN might be undetected in the X-rays, but also misclassified from MIR colours since the \citet{Donley+12} criterion is highly incomplete at $L_{\rm x} < 10^{44}$ erg s$^{-1}$. However, neither SED-fitting decomposition nor MIR colours can identify an AGN when the optical-IR SED is outshined by the host-galaxy light (right panel), although the X-ray luminosity suggests the presence of a moderately luminous X-ray AGN ($L_{\rm x} \sim$10$^{43}$ erg s$^{-1}$). We looked at the observed distribution of the X-ray to optical-UV index, defined as $\alpha_{\rm ox}$~=~$-$Log[$L_{\rm 2~keV}$~/~$L_{\rm 2500~\AA}$]/2.605, where $L_{\rm 2500~\AA}$ and $L_{\rm 2~keV}$ are the rest-frame monochromatic luminosities at 2500\AA~and 2 keV, respectively (e.g. \citealt{Zamorani+81}). We 
verified that the observed distribution of $\alpha_{\rm ox}$ for HLAGN identified solely from X-rays peaks at $\alpha_{\rm ox} \sim$1, unlike the average value $\alpha_{\rm ox} \sim$1.37 found for X-ray selected AGN in the COSMOS field (\citealt{Lusso+10}). The lower $\alpha_{\rm ox}$ suggests that HLAGN identified only from X-rays are optically fainter than the rest of X-ray AGN in the COSMOS field, as expected from their galaxy-dominated SEDs. This is also confirmed by the fact that in more than 80\% of them, the optical-NIR photometry has been fitted without AGN templates when calculating the photometric redshifts (see \citealt{Marchesi+16}; M.~Salvato et al. in prep.).

By using different and complementary tracers of highly accreting AGN, we can build a more representative (though not 100\% complete) sample of HLAGN. Our analysis would certainly benefit from optical-NIR spectroscopy to identify AGN at lower intrinsic luminosities. Unfortunately, the spectral lines used to calculate the emission line ratios [\ion{O}{III}]/H$\beta$ and [\ion{N}{II}]/H$\alpha$) in the BPT diagram (from \citealt{Baldwin+81}) are detected only in a low percentage (about 5\%) of our radio sample, mostly at z$<$0.5, as some optical lines (e.g. H$\alpha$) would be redshifted outside the observed spectral window at higher redshifts. For consistency, in this work we preferred to make use of AGN selection criteria that are applicable to the entire sample. 

For our 1\,604 HLAGN, we took the SFR and $M_{\star}$ estimates from the best-fit solution obtained with the three-component SED-fitting code by \citet{Berta+13}. This approach allows us to account for a possible AGN contamination in galaxy parameter estimates and to study uncertainties and degeneracies, which is important when comparing galaxy properties between AGN and non-AGN hosts. However, we checked that the AGN contribution to the total (8-1000$~\mu$m) IR luminosity is very small (a few percent) for most of the HLAGN sample, as argued by previous studies of X-ray and IR-selected AGN (e.g. \citealt{Mullaney+11}; \citealt{Santini+12}; \citealt{Rosario+12}).

\begin{figure}
\begin{center}
         \includegraphics[width=3.5in]{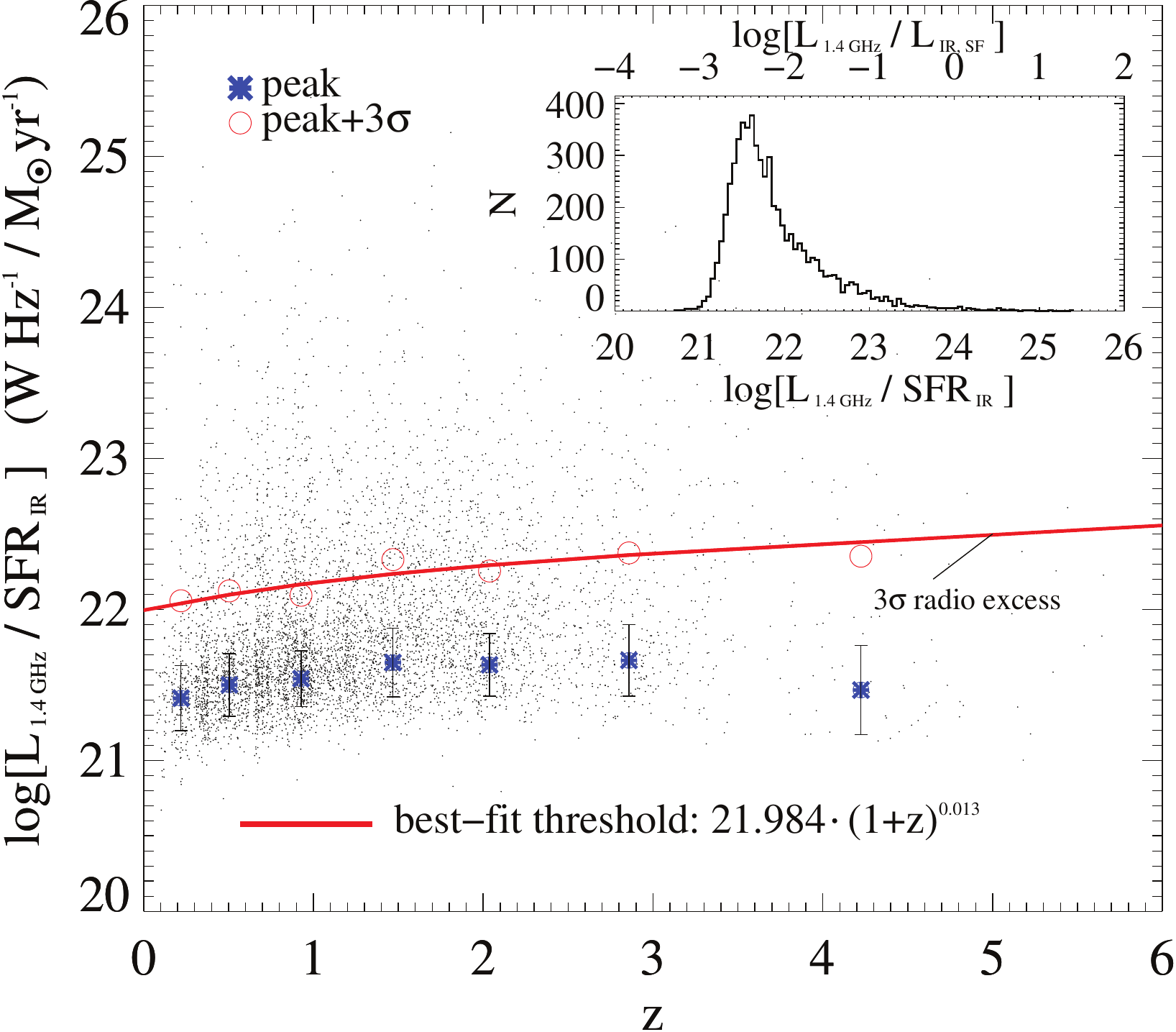}
\end{center}
 \caption{\small Distribution of the ratio between $L_{\rm 1.4~GHz}$ and SFR$_{\rm IR}$ as a function of redshift (black points) for sources not classified as HLAGN (79\% of our sample). The blue filled circles (and errors) indicate the peak (and dispersion) of the Gaussian distribution identified in a given redshift bin, while the corresponding 3$\sigma$ deviation is set by the red open circles. The red solid line indicates the redshift-dependent threshold derived by fitting the open circles at each redshift bin. Sources above the red line are identified as ``low to moderate radiative luminosity AGN'' (hereafter MLAGN) via radio excess. The full histogram of $L_{\rm 1.4~GHz}$/SFR$_{\rm IR}$ is shown in the top right corner. }
   \label{fig:radio_excess}
\end{figure}

\subsection{Low-to-moderate radiative luminosity AGN via radio excess} \label{llagn}

Radio sources that are not classified as HLAGN in Sect. \ref{hlagn} do not show evidence of AGN activity according to X-ray, MIR, or SED decomposition. However, this does not necessarily mean that SMBH accretion is not occuring at all, but rather that the AGN diagnostics described above might fail in detecting signatures of less efficient accretion episodes. As mentioned in the previous sections, radio observations are crucial for chasing such an elusive AGN population.

To identify lower radiative luminosity AGN in our sample, we first considered the 3~GHz selected sources that are not classified as HLAGN (i.e. 79\%). For each of them, the $L_{\rm IR}$ obtained via SED fitting without AGN (\citealt{daCunha+08}) has been converted to IR-based SFR (SFR$_{\rm IR}$ hereafter) by assuming a \citet{Kennicutt98} scaling factor and a \citet{Chabrier03} IMF. To identify the possible AGN contribution in radio emission, we analysed the ratio between the 1.4~GHz radio luminosity $L_{\rm 1.4\, GHz}$ and the SFR$_{\rm IR}$ for each source. Fig. \ref{fig:radio_excess} shows the distribution of their ratio (i.e. $L_{\rm 1.4\,GHz}$/SFR$_{\rm IR}$, in logarithmic scale) as a function of redshift (black points). Typical 1$\sigma$ uncertainties of the observed ratio are of the order of 0.15~dex. The histogram in the top right corner shows the distribution of our sources as a function of $L_{\rm 1.4\, GHz}$/SFR$_{\rm IR}$. The distribution is skewed towards high values of the ratio, suggesting 
that AGN 
activity might be contributing to the integrated radio emission. However, the average $L_{\rm 1.4\, GHz}$/SFR$_{\rm IR}$ also increases with redshift, which partly explains the skewness of the observed distribution. To quantify these factors, we split our sample into seven redshift bins (0.01$<$z$<$0.3, 0.3$<$z$<$0.7, 0.7$<$z$<$1.2, 1.2$<$z$<$1.8, 1.8$<$z$<$2.5, 2.5$<$z$<$3.5, and 3.5$<$z$<$5.7) and fit each single distribution with a log-normal function that reproduces the peak and the negative part of the observed distribution well. The values of the peak (and dispersion) of the Gaussian function identified at each redshift bin are represented in Fig. \ref{fig:radio_excess} with blue filled circles (and relative errors). The position of the peak generally shifts to higher $L_{\rm 1.4\, GHz}$/SFR$_{\rm IR}$ ratios with increasing redshift, which justifies our choice of a redshift-dependent approach. Moreover, some recent studies (e.g. \citealt{Magnelli+15}; \citealt{Delhaize+17}) have found a slight, but 
significant decrease of q$_{\rm IR}$ (proportional to SFR$_{\rm IR}$/$L_{\rm 1.4\, GHz}$) with redshift through a careful treatment of non-detections via stacking or double-censored survival analysis.

We calculated the 3$\sigma$ deviation from the peak of the log-normal function at each redshift bin (see red open circles in Fig. \ref{fig:radio_excess}). By fitting the open circles with a power-law function (red solid line in Fig. \ref{fig:radio_excess}), we derived the analytical expression that describes a redshift-dependent threshold in radio excess as follows:
\begin{equation}
   \log{\left(\frac{L_{\rm 1.4\, GHz}}{\rm SFR_{IR}}\right)}_{\rm excess} = 21.984 \times (1+z)^{0.013} .
   \label{eq:excess}
\end{equation}
From this expression, we set a threshold above which the radio emission shows a $>$3$\sigma$ excess compared to that expected from star formation. The aforementioned threshold identifies {1\,333} low-to-moderate radiative luminosity AGN (hereafter MLAGN) via radio excess, corresponding to 17\% of our radio sample. The percentage of MLAGN, which would have been identified through a single threshold at all redshifts, would be around 18\% instead of the 17\% found from Eq. \ref{eq:excess}; this would not affect our results.

The choice of a 3$\sigma$ radio excess imposed in each redshift bin ensures a negligible contamination from star-forming galaxies without radio excess (about 0.15\%). On the other hand, our selection might miss a significant number of potential MLAGN in our sample, which is estimated to be around {1\,000} sources ({75\%} of the identified sample of MLAGN) based on the difference between the distribution below the threshold and the best-fitting Gaussian profile, in all redshift bins. A comparison with other definitions of radio excess found in the literature is presented in Sect. \ref{literature}.

The \citet{Kennicutt98} $L_{\rm IR}$--SFR$_{\rm IR}$ conversion assumes that the total IR luminosity arises entirely from optically thick, dust-obscured regions. While this assumption is reasonable in highly star-forming galaxies, such as those detected by \textit{Herschel} (\citealt{Wuyts+11}; \citealt{Magnelli+13}), this is not true for weakly star-forming (or passive) systems, where a significant portion of the IR luminosity may originate from ($>$ few Gyr) old stellar populations (e.g. \citealt{Groves+12}). 
We verified that the unobscured SFR derived from the UV galaxy emission (e.g. \citealt{Papovich+07}) is around 5\% of the obscured SFR$_{\rm IR}$, on average, therefore its contribution would not significantly affect our definition of radio excess presented above. 

Moreover, our threshold is calibrated on a radio-selected sample, which is expected to bias the observed distribution towards higher values of $L_{\rm 1.4\, GHz}$/SFR$_{\rm IR}$ compared to the true (i.e. unbiased) distribution. These arguments suggest that our definition of radio excess is likely to be fairly conservative and the number of MLAGN selected in this way should more properly be considered as a lower limit.

\bigskip

In summary, by combining multiwavelength AGN diagnostics, we managed to isolate two populations of AGN in our 3~GHz selected sample. We identified {1\,604} HLAGN (21\%) and {1\,333} MLAGN (17\%), which collectively make our sample of radio-selected AGN. The remainder of the sample (62\%) is characterised in detail in \citet{smolcic+17b}.
Moreover, we found that about 30\% of HLAGN shows also a $\geq$3$\sigma$ radio excess. We checked that the relative numbers of HLAGN identified from each criterion and shown in Fig. \ref{fig:venn} do not change between sources with and without significant radio excess. A more detailed analysis of the average galaxy and AGN properties for the aforementioned AGN classes (MLAGN, HLAGN, including the subsample with radio excess) is presented in Sect. \ref{results}.

\subsection{Comparison with radio-based classifications} \label{literature}

In this section we compare our source classification with other independent methods from the literature.

\subsubsection{Comparison with VLBI sources} \label{VLBI}

We cross-matched our 3~GHz selected sample of {7\,729} sources with the Very Large Baseline Interferometry (VLBI) 1.4~GHz source catalogue from N.~Herrera Ruiz et al. (in prep.). The authors targeted the radio sources selected by the 1.4~GHz VLA-COSMOS survey (\citealt{schinnerer+10}) with VLBI at $\lesssim0\farcs01$ angular resolution, reaching a 1$\sigma$ sensitivity of 10$~\mu$Jy~beam$^{-1}$ in the central part of the field. They detected 468 radio sources at signal-to-noise ratio higher than 5.5. A total of {354} matches have been found within $0\farcs4$ (half-beam size of 3~GHz VLA observations).
Given the high angular resolution, VLBI is sensitive to the radio emission on circum-nuclear scales (from $d\sim$20 pc at $z$=0.1 to $d\sim$80 pc at $z$=2), likely arising from an AGN. 

Interestingly, we found that {91\%} of VLBI sources are classified as AGN by our method, where {55\%} are MLAGN and 36\% are HLAGN. About {88\%} of the HLAGN also show a $>$3$\sigma$ radio excess compared to the SFR$_{\rm IR}$ (Sect. \ref{llagn}). These notably high percentages of VLBI sources classified as AGN (both HLAGN and MLAGN) in our sample demonstrate the high reliability of our classification method.

\begin{table}
   \caption{\small Comparison between our classification method and that presented in \citet{Bonzini+13}. For this check, we cut our sample at total 1.4~GHz flux $S_{\rm 1.4} > $ 37$~\mu$Jy to match the radio selection adopted by the authors. }
\begin{tabular}{l cccc }
\hline
\hline
 Classifications                &             &               &   (B13)       &            \\
 \hline
 (this work,                    &             &     RQ AGN    &    RL AGN     &    SFGs    \\
 $S_{\rm 1.4} > $ 37$~\mu$Jy)   &   total     &     (609)     &    (865)      &   (3099)  \\
\hline
HLAGN                           &  (1044)     &       609     &      169      &     266     \\
MLAGN                           &  (1032)     &        0      &      569      &     463     \\
Rest of the sample              &  (2497)     &        0      &      127      &     2370      \\
\hline

\end{tabular}

\label{table:b13}
\end{table}

\begin{figure}
\begin{center}
    \includegraphics[width=3.4in]{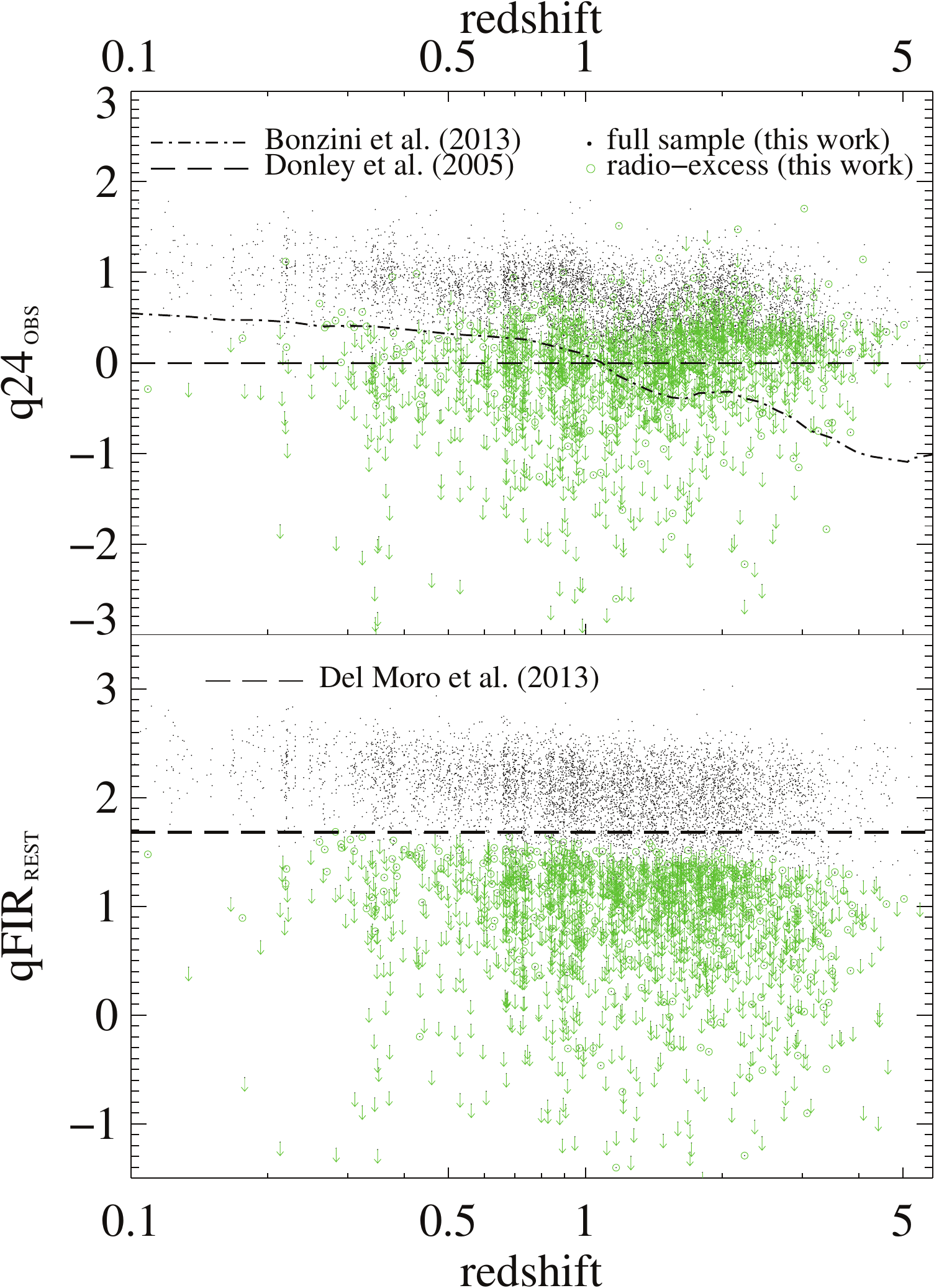}
\end{center}
 \caption{\small Redshift distribution of q$_{\rm 24, obs}$ (top panel) and q$_{\rm FIR}$ (bottom panel) for our 3~GHz sample (black dots). The subsample with radio excess is highlighted with green circles. The dash-dotted line (top panel) indicates the radio-excess threshold by \citeauthor{Bonzini+13} (\citeyear{Bonzini+13}, B13), while the horizontal dashed line indicates the threshold in q$_{\rm 24, obs}$ defined by \citet{Donley+05}. The dashed line of the bottom panel indicates the threshold in the rest-frame q$_{\rm FIR}$ identified by \citet{DelMoro+13}. }
   \label{fig:q24}
\end{figure}

\subsubsection{Comparison with \citet{Bonzini+13}} \label{bonzini}

In a similar study, \citeauthor{Bonzini+13} (\citeyear{Bonzini+13}, B13 hereafter) carried out a panchromatic analysis of about 800 high-redshift (z$\leq$4) radio sources in the E-CDFS, selected with VLA at 1.4~GHz. They separated radio sources into radio-loud AGN (RL-AGN), radio-quiet AGN (RQ-AGN) and star-forming galaxies (SFGs) on the basis of the observed 24$~\mu$m-to-1.4~GHz flux ratio (also called q$_{24,\rm obs}$, see \citealt{Donley+05}). Briefly, B13 selected RL-AGN that are below the 2$\sigma$ deviation from the average q$_{24,\rm obs}$ at a given redshift. {In case q$_{24,\rm obs}$ was an upper limit, the authors selected RL-AGN that are below the 1$\sigma$ deviation from the average value.} Within and above the 2$\sigma$ deviation, they selected RQ-AGN {that are not RL-AGN and at the same time fulfilling either X-ray or MIR diagnostics (similar to those discussed in Sect. \ref{hlagn})}; the rest of the sample was classified as SFGs. On top of these criteria, B13 applied further checks (see their 
Sect. 3.5.1) to improve the sample characterisation, which led the authors to reclassify 11 sources from SFGs to RQ or RL AGN in their sample.

Despite the larger area, our 3~GHz data in COSMOS are deeper than the E-CDFS data at 1.4~GHz, which allows us to compare our classification with B13 in the same flux density range. First, we scaled our 3~GHz flux density to the observed 1.4~GHz for each source (as discussed in Sect. \ref{radio-multi}). Secondly, we cut our sample at total 1.4~GHz flux $S_{\rm 1.4~GHz} > $ 37$~\mu$Jy as in B13, yielding {4\,573} sources (around 59\% of the full sample) and we computed q$_{24,\rm obs}$ for all of them. Thirdly, we applied the same criteria of B13 to identify RQ-AGN, RL-AGN and SFGs in our sample (including their additional AGN diagnostics for consistency, see Sect. \ref{add_diagnostics}), and show the numbers in Table \ref{table:b13}. 

This comparison suggests that the HLAGN and MLAGN identified in this work fairly overlap with the RQ-AGN and RL-AGN classes, respectively, even if our classification tends to classify more objects as AGN ({45\%}) than the B13 classification ({32\%}). The RQ-AGN percentage found by B13 (24\%) is higher than that listed in Table \ref{table:b13} ({13\%}), and is likely driven by the higher incompleteness of X-ray observations in the COSMOS field towards moderately luminous X-ray AGN ($10^{42}<$ $L_{\rm x}< 10^{44}$ erg s$^{-1}$), especially at high redshifts (z$>$2). The aim of this comparison is not to invalidate either of the methods, but simply to clarify how different nomenclatures compare to each other. The main difference in the source classification lies in the different definitions of radio excess. In their work, B13 used a redshift-dependent threshold in q$_{\rm 24, obs}$, which was calibrated on the M82 template. The dash-dotted line in Fig. \ref{fig:q24} (top panel) shows the 
redshift-dependent threshold defined by B13. Black symbols indicate the distribution of our 3~GHz radio sources ({7\,729}) as a function of q$_{\rm 24, obs}$ and redshift with our radio-excess sources ({1\,814} in total, being {1\,333} in MLAGN and {481} in HLAGN) highlighted in green. Downward pointing arrows indicate (5$\sigma$) upper limits due to non-detection at 24$~\mu$m. The black dashed line sets the threshold q$_{\rm 24, obs}<$ 0 adopted by \citet{Donley+05} to identify radio-excess sources. \citet{DelMoro+13} already pointed out that q$_{\rm 24, obs}$ is a reliable tracer of radio excess, although it is not complete. We confirm this statement, as most of the sources below the q$_{\rm 24, obs}$ threshold (by B13) also satisfy our radio-excess definition. However, the B13 criterion at z$>$1 becomes even more stringent than that proposed by \citet{Donley+05}. This decreasing trend of q$_{\rm 24, obs}$ with redshift is driven by the shape of the M82 template SED, which is rather peculiar compared to 
the average SED of star-forming galaxies at 0$<$z$<$3, implying a steeper decline with redshift compared to what is observed in our sample. However, the percentage of z$>$2 sources in B13 is relatively small compared to our sample, which implies that the M82 template SED shape should not have a large effect on the source classification in the E-CDFS sample.

\subsubsection{Comparison with \citet{DelMoro+13}} \label{del_moro}

An alternative method to search for radio excess is by using the (rest-frame) FIR-to-radio flux ratio $q_{\rm FIR}$ (e.g. \citealt{Sargent+10}), where the FIR flux refers to the rest-frame range 42.5--122.5$~\mu$m, and the radio flux refers to the rest-frame 1.4~GHz. Recently, \citet{DelMoro+13} used $q_{\rm FIR}<$ 1.68 in star-forming galaxies within the GOODS-North field to identify sources with $>3\sigma$ radio excess. 

As proposed by \citeauthor{DelMoro+13} (\citeyear{DelMoro+13}, see their Figure 5), we calculate the q$_{\rm FIR}$ for our 3~GHz sources and show in Fig. \ref{fig:q24} (bottom panel) their distribution with redshift with respect to the threshold set by \citet{DelMoro+13}. \citet{DelMoro+13} calibrated the q$_{\rm FIR}$ threshold on a sample of sources detected at both 1.4~GHz and 24$~\mu$m, which are, therefore, on average more star-forming than our radio-selected galaxies in COSMOS. If limiting our sample to 24$~\mu$m detected sources, we estimate the percentage of radio-excess sources to be around 13\%, which is in agreement with the percentage found by \citet{DelMoro+13}. We checked that the average q$_{\rm FIR}$ values of our 3~GHz sources also detected at 24$~\mu$m closely resemble those of \citet{DelMoro+13} in the common redshift range. This is expected, given that the radio-to-24$~\mu$m sensitivity limits are similar between the GOODS-North and COSMOS fields. For comparison with \citet{DelMoro+13}, 
we calculated the $q_{\rm FIR}$ for our radio-excess sources and verified that 100\% of these sources satisfy the condition $q_{\rm FIR}<$ 1.68, while around 72\% of sources with $q_{\rm FIR}<$ 1.68 satisfy our definition of radio excess. This check further supports the reliability of our definition of radio excess.

\subsection{Further tests of the source classification} \label{further_tests}

The classification scheme presented in Sects. \ref{hlagn} and \ref{llagn} was based on a few assumptions that we test in this section. In particular, we detail the motivation for our naming convention (Sect. \ref{naming_convention}), show how our classification would change if considering additional AGN diagnostics (Sect. \ref{add_diagnostics}), and compare the IR luminosities derived for \textit{Herschel}-undetected sources against IR stacking (Sect. \ref{ir_stacking}).

\begin{figure*}
\begin{center}
         \includegraphics[width=180mm,keepaspectratio]{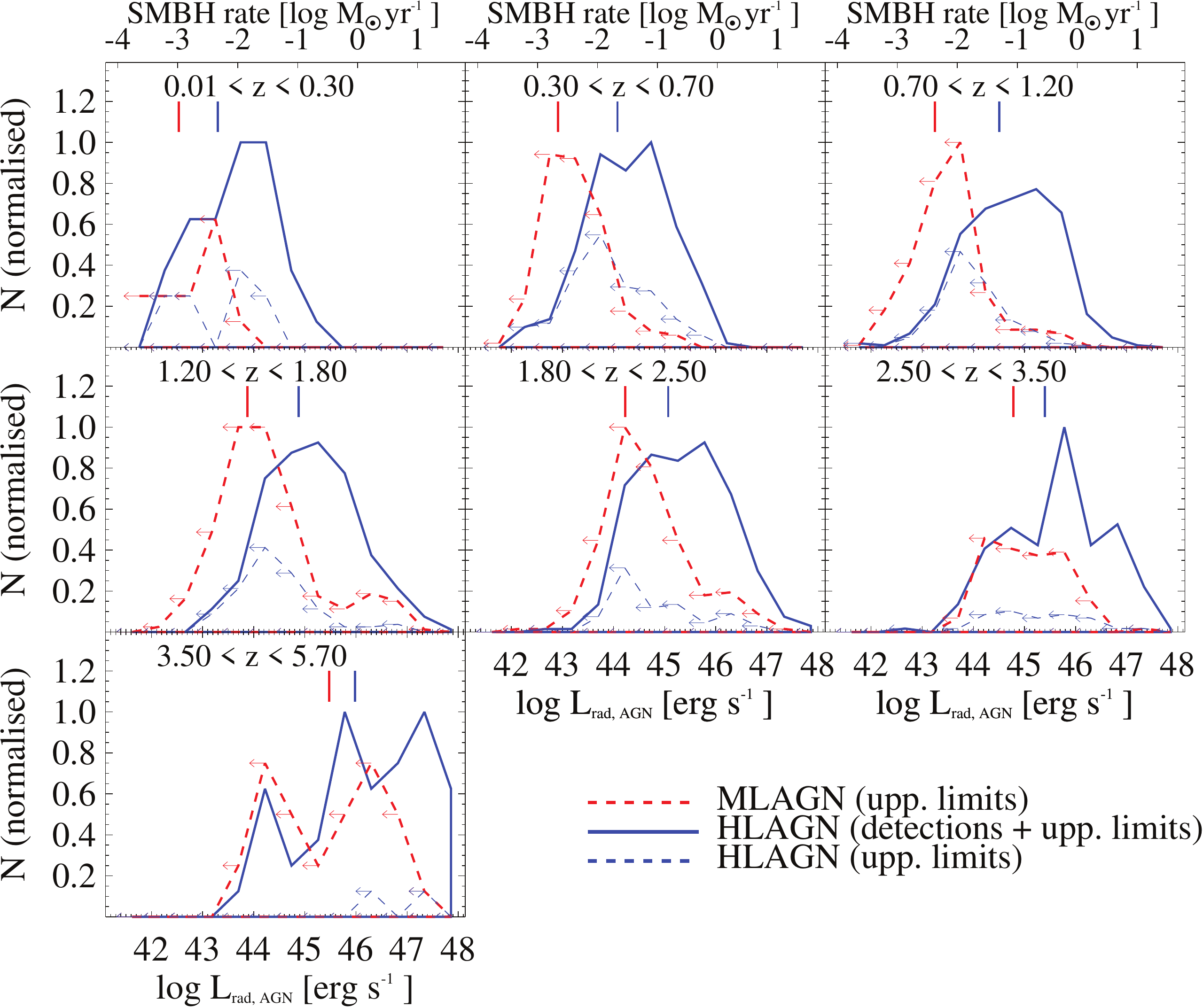}
\end{center}
 \caption{\small Normalised distribution of L$\rm_{rad, AGN}$ (or BHAR), as a function of redshift, separately for MLAGN (red dashed) and HLAGN (blue solid). The subsample of HLAGN not identified as ``SED-AGN'' ({27\%} of all HLAGN) is represented by the blue dashed distribution. The left-pointing arrows indicate upper limits at 90\% confidence level in L$\rm_{rad, AGN}$ for the corresponding distribution. }   
\label{fig:dist_lradagn}
\end{figure*}

\subsubsection{The choice of the naming convention } \label{naming_convention}
 
Our samples of HLAGN and MLAGN include AGN identified through different diagnostics. As shown in Fig. \ref{fig:venn}, only {14\%} of HLAGN meet simultaneously the three diagnostics (X-ray, MIR, and SED decomposition), suggesting that various AGN selection criteria might be sensitive to a broad range of AGN luminosities. We study the distribution of MLAGN and HLAGN as a function of AGN radiative luminosity (L$\rm_{rad, AGN}$), as derived from SED-fitting decomposition, converted to BHAR following Eq. 1 from \citet{Alexander+12} and assuming a canonical mass-energy efficiency conversion of 10\% (e.g. \citealt{Marconi+04}). Each estimate of L$\rm_{rad, AGN}$ is calculated from the corresponding best-fit AGN template (Sect. \ref{sed_fitting}) obtained from SED-fitting decomposition. This parameter should be considered an upper limit if the AGN component is not significant from SED-decomposition, which is the case for all MLAGN (by definition) and for about {27\%} of HLAGN identified solely from X-ray or MIR 
criteria. In the latter case, the L$\rm_{rad, AGN}$ has been taken from the 95$^{\rm th}$ percentile of the corresponding PDF obtained from the {\sc sed3fit} code, which is equivalent to an upper limit at 90\% confidence level.

Fig. \ref{fig:dist_lradagn} shows the L$\rm_{rad, AGN}$ distribution separately for MLAGN (red dashed line), HLAGN (blue solid line), and the subsample of HLAGN not identified from SED fitting (blue dashed line) in seven redshift bins. The vertical lines represent the median of the corresponding distribution at each redshift bin. The median values show a significant difference (around 1~dex) in L$\rm_{rad, AGN}$ between MLAGN and HLAGN at all redshifts. Despite the non-negligible overlap between the two distributions, we stress that the L$\rm_{rad, AGN}$ estimates for MLAGN consist of upper limits, which implies that the difference between the true distributions, at a given redshift, is even more significant. 

This test suggests that, at each redshift, HLAGN are significantly more powerful than MLAGN and justifies the naming convention proposed in this work in a statistical sense.

\subsubsection{Additional AGN diagnostics}  \label{add_diagnostics}

As carried out in B13, we applied additional diagnostics to verify the robustness of our AGN identification method. 
\begin{itemize}
  \item Optical spectra: we searched for sources flagged as broad line AGN in the optical spectra taken from the zCOSMOS-Bright survey (\citealt{Lilly+07}, \citeyear{Lilly+09}); we found {15} in total, which were all pre-classified as HLAGN on the basis of X-ray and mid-IR criteria. 
 \item VLBI\ sample:\ as mentioned in Sect. \ref{VLBI}, {354} radio sources selected at 3~GHz were also identified in the VLBI sample (N.~Herrera Ruiz et al. in prep.) available in the COSMOS field. Of these, {30 (8\%)} were not identified as either MLAGN or HLAGN, although they are likely to be AGN.
 \item Inverted radio spectra: we found a total of {11} radio sources not classified as radio-selected AGN, but that have inverted radio spectra ($\alpha>$0), where the spectral index $\alpha$ is set to the observed 1.4--3~GHz slope. This feature may indicate the presence of a compact radio core (e.g. \citealt{kellermann+69}).
 \item Hardness ratio: we found a total of {seven} X-ray sources that were not classified as AGN in our sample with a positive hardness ratio ($HR>$0), indicating the likely presence of obscured AGN (e.g. \citealt{Brusa+10}).
 {\item Polycyclic aromatic hydrocarbon (PAH) features: sources lying in the IRAC colour-colour wedge discussed in B13 (see their Sect. 3.5.5) likely display PAH-dominated SEDs, which are typical of star-forming galaxies. We found 444 sources that fulfil this criterion with only five (26) that are classified as MLAGN (HLAGN) in our sample.}
 \end{itemize}

The above-mentioned criteria suggest that some of our sources might be misclassified. We found that {46} radio sources ({30} from VLBI detection, {9} from inverted radio spectra, {and seven} from the hardness ratio) should be reclassified from non-AGN to AGN in our sample. However, these criteria are applicable to a very low percentage of our sample, meaning that by incorporating them we would likely introduce a bias against sources with no available diagnostics.  Reclassifying these {46} sources (0.6\% of our sample) would have no impact on our main results and conclusions. {We also found 31 radio-selected AGN displaying PAH-dominated SEDs in the MIR; even though this criterion does not rule out a potential AGN contribution in other bands, the percentage of possibly misclassified AGN is minimal.} For these reasons, we limited our AGN selection criteria to those presented in Sects. \ref{hlagn} and \ref{llagn}.

\subsubsection{Testing radio excess with infrared stacking}  \label{ir_stacking}

As discussed in Sect. \ref{sed_fitting}, the $L_{\rm IR}$ and SFR estimates are less robust for sources without \textit{Herschel} detection, which constitute about one-third of our 3~GHz selected sample. For these sources, we fitted their SEDs using the nominal upper limits in the PACS and SPIRE bands (as described in Sect. \ref{sed_fitting}). As a sanity check, we compared the \textit{Herschel} fluxes used for SED fitting with those obtained via \textit{Herschel} stacking. The difference between the two approaches enables us to test the robustness of our radio-excess definition.

We looked at the 79\% of sources \textit{not} classified as HLAGN, shown in Fig. \ref{fig:radio_excess}, and considered only those without $\geq$3$\sigma$ \textit{Herschel} detection in any PACS or SPIRE band, which amounts to  {2\,203} in total. As carried out in previous papers (e.g. \citealt{Santini+12}; \citealt{Rosario+12}; \citealt{Bonzini+13}), we split this sample in different redshift bins (see Table \ref{table:stack}) and we stacked the PACS and SPIRE images at the optical-NIR position of each source in the same bin using a stacking tool from \citet{Bethermin+10}\footnote{This IDL routine is described in detail in \citet{Bethermin+10} and can be retrieved at \url{https://www.ias.u-psud.fr/irgalaxies/downloads.php}.}.

For PACS images, we performed a mean stacking on the residual maps at 100 and 160$~\mu$m, from which all detections were removed to avoid contamination by nearby brighter sources. Point spread function (PSF) photometry was performed on the final stacked images, using the uncertainty maps (\citealt{Lutz+11}) as weights. The stacked fluxes were corrected for aperture and correlated noise\footnote{The full PACS documentation is available at \url{www.mpe.mpg.de/resources/PEP/DR1\_tarballs/}.}. The SPIRE images\footnote{The SPIRE images for the COSMOS field were taken from the HeDAM database: \url{http://hedam.lam.fr/HerMES/index/download}.} at 250, 350, and 500$~\mu$m were already given in units of surface brightness (Jy~beam$^{-1}$), hence we inferred the stacked flux directly from the value of the central pixel in the stacked image. To evaluate the uncertainty on the stacked fluxes, we performed a bootstrapping analysis (e.g. \citealt{Shao+10}; \citealt{Santini+12}; \citealt{Rosario+12}). Briefly, a set of $N$ 
sources, where $N$ is equal to the number of stacked sources per redshift bin, is randomly chosen 1\,000 times, allowing repetition of the same source. To mitigate the possible contamination from a few brighter outliers, we set the stacked flux to the median of the distribution obtained from bootstrapping, while the error on the flux was drawn by the 16$^{\rm th}$ and 84$^{\rm th}$ percentiles of the same distribution. For PACS images, we also corrected the errors for high-pass filtering effects. We did not correct the stacked fluxes for possible blending in the optical-NIR images. In every redshift bin we obtained $>$2$\sigma$ detection in two to five \textit{Herschel} bands. The final stacked fluxes and corresponding uncertainties are summarised in Table \ref{table:stack}.

We re-fitted the {2\,203} SEDs again with {\sc magphys}, using the corresponding stacked fluxes as detections for
each source. The resulting $M_{\star}$ estimates agree very well (no offset, 1$\sigma$ dispersion is about 0.15~dex) with those obtained in Sect. \ref{sed_fitting}. The newly derived $L_{\rm IR}$ estimates are generally lower (by 50\%) than those obtained in Sect. \ref{sed_fitting}, but for 40\% of the stacked sample we found slightly higher $L_{\rm IR}$. This is mainly because while in most cases the fluxes obtained from \textit{Herschel} stacking (see Table \ref{table:stack}) are lower than the flux values used for SED fitting, in a few bins they are instead slightly higher, especially at z$\gtrsim$1.5. However, we verified that using the $L_{\rm IR}$ estimates derived from stacking would only minimally affect the overall distribution of $L_{\rm 1.4\, GHz}$/SFR$_{\rm IR}$ (Fig. \ref{fig:radio_excess}). In particular, we checked that our classification would remain 
unchanged for 90\% of the sources, while the remainder of the sample would move either from SFGs to MLAGN (5\%) or vice versa (5\%). 

The purpose of this test was to quantify the impact of using different sets of \textit{Herschel} fluxes on the source classification. The general agreement obtained between the two methods ensures the robustness of our classification. As a consequence, we decided to keep using the \textit{Herschel} upper limits introduced in Sect. \ref{sed_fitting} through the rest of this work.


\section{Catalogue description} \label{catalog}

The value-added catalogue presented in this section includes classification and selected physical properties used in this work for our 3~GHz radio sample with optical-NIR counterparts ({7\,729} sources in total). We also list the individual criteria used in this work to classify our radio sources (columns 14 to 17). This way, any user can easily retrieve our classification or adjust it to a different set of selection criteria. 
The catalogue will be made available through the COSMOS IPAC/IRSA database\footnote{\url{http://irsa.ipac.caltech.edu/Missions/cosmos.html}}. Here we describe its structure, following the same format of Table \ref{table:catalogue}.

\begin{itemize}
 \item (1) Identification number of the radio source (ID).
 \item (2) Right ascension (J2000) of the radio source.
 \item (3) Declination (J2000) of the radio source. \item (4) Best redshift available for the source.
 \item (5) Origin of the redshift: spectroscopic (``spec'') if available, photometric (``phot'') otherwise.
 \item (6) 3~GHz integrated radio flux density [$\mu$Jy].
 \item (7) 3~GHz (rest-frame) radio luminosity [log W~Hz$^{-1}$].
 \item (8) 1.4~GHz (rest-frame) radio luminosity [log W~Hz$^{-1}$], obtained as described in Sect. \ref{radio-multi}.
 \item (9) Star formation infrared (8-1000$~\mu$m rest-frame) luminosity derived from SED fitting [log $L_{\odot}$]. If the source is classified as HLAGN, this value represents the portion of the total infrared luminosity arising from star formation, while it corresponds to the total IR luminosity otherwise (see Sect. \ref{sed_fitting}).
 \item (10) Flag for \textit{Herschel} detection at $\geq$3$\sigma$, in at least one band: ``true'' if detected, ``false'' if only upper limits are available.
 \item (11) Stellar mass derived from SED-fitting decomposition [log $M_{\odot}$]. The value is drawn from the fit with AGN if the source is classified as HLAGN and otherwise from the fit without AGN. Calculated with a \citet{Chabrier03} IMF.
 \item (12) Star formation rate [$M_{\odot}$yr$^{-1}$] obtained from the total infrared luminosity listed in column (9), assuming the \citet{Kennicutt98} conversion factor, and scaled to a \citet{Chabrier03} IMF.
 \item (13) Rest-frame {\sc [NUV-$r$]} colour obtained from the best-fitting galaxy template and corrected for dust attenuation (AB magnitude).
 \item (14) X-ray-AGN: ``1'' if true, ``0'' otherwise.
 \item (15) MIR AGN: ``1'' if true, ``0'' otherwise.
 \item (16) SED-AGN: ``1'' if true, ``0'' otherwise.
 \item (17) Radio-excess: ``1'' if true, ``0'' otherwise.
 \item (18) Class: moderate-to-high radiative luminosity AGN (HLAGN), low-to-moderate radiative luminosity AGN (MLAGN), or neither of the two (empty space). A source is classified as HLAGN if (14)=1 $\vee$ (15)=1 $\vee$ (16)=1, while it is classified as MLAGN if (14,15,16)=(0,0,0) $\wedge$ (17)=1. 
 \end{itemize}

\begin{table}
   \centering
   \caption{\small Number of 3~GHz radio sources studied in this work as a function of redshift and AGN class: MLAGN, HLAGN, and HLAGN with radio excess (in brackets). For each redshift bin we report the mean redshift $\langle z \rangle$ of the corresponding population. }
\begin{tabular}{l ccc }
\hline
\hline
 redshift bin         &  $\langle z \rangle $  &     MLAGN          &       HLAGN         \\
                      &                        &                     &    (radio-excess)    \\
\hline
0.01 $\leq z <$ 0.30  &          0.21          &        22           &        36 (9)       \\ 
0.30 $\leq z <$ 0.70  &          0.51          &        221          &       232 (66)      \\ 
0.70 $\leq z <$ 1.20  &          0.94          &        375          &       416 (135)     \\ 
1.20 $\leq z <$ 1.80  &          1.48          &        350          &       350 (98)     \\ 
1.80 $\leq z <$ 2.50  &          2.08          &        225          &       307 (84)      \\
2.50 $\leq z <$ 3.50  &          2.89          &        111          &       217 (74)      \\ 
3.50 $\leq z <$ 5.70  &          4.21          &         29          &        46 (15)       \\ 
\hline

\end{tabular}

\label{table:class}
\end{table}

\begin{figure*}
\begin{center}
         \includegraphics[width=180mm,keepaspectratio]{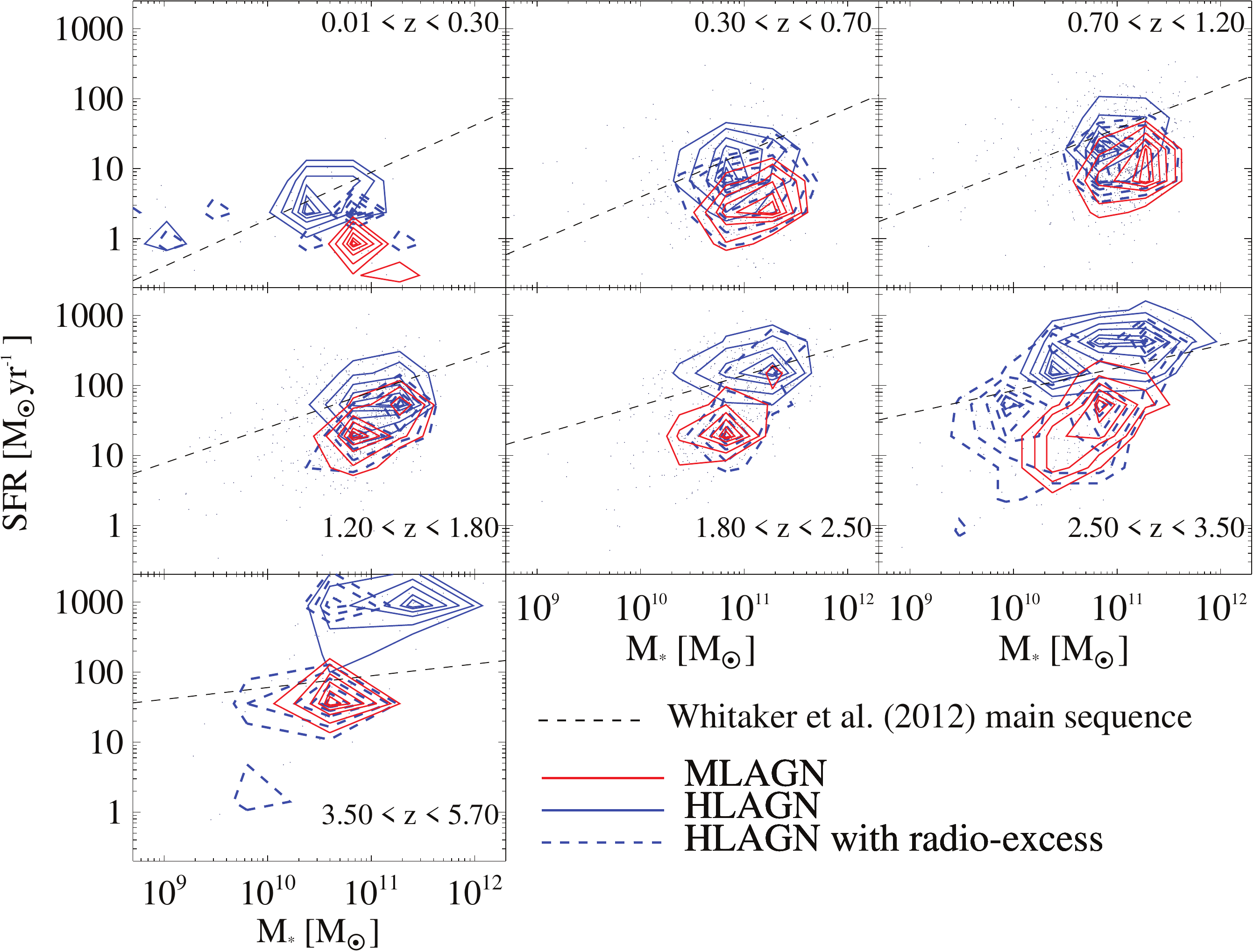}
\end{center}
 \caption{\small Distribution of HLAGN and MLAGN in the SFR--$M_{\star}$ plane as a function of redshift (black dots). The 2D density contours highlight the distribution of MLAGN (red), HLAGN (blue), and the subsample of HLAGN with radio excess (blue dashed contours), respectively. The 2D density contours enclose the sources (from outer to inner contours) with density levels $>$35, $>$50, $>$68, $>$80, $>$90, and $>$95\% of the maximum 2D density. The black dotted lines indicate the MS relation (\citealt{Whitaker+12}) at different redshifts. The highest redshift bin contains only a few tens of sources, which explains the noise seen in the 2D density contours. }   
\label{fig:sed_mass_plane_classes}
\end{figure*}


\section{Results} \label{results}
 
In this section we present the average AGN and galaxy properties for our 3~GHz radio sources classified as AGN host galaxies. In particular, we focus our analysis on MLAGN (17\%), HLAGN (21\%), and also the subsample of HLAGN with radio excess (6\%). 
We show, for these classes, the location in the SFR--$M_{\star}$ plane (Sect. \ref{sfr_mass_plane}) and the distributions of AGN and galaxy properties of their hosts (Sect. \ref{histograms}) at different cosmic epochs.

\begin{table*}
   \centering
   \caption{\small Results from the two-sample Kolmogorov-Smirnov (K-S) test. The table shows, in different redshift bins, the probability {\sc P(K-S)} that the distributions of a given parameter ($M_{\star}$, SFR or {\sc [NUV-$r$]}) for HLAGN and MLAGN (shown in Figs. \ref{fig:histogal1}, \ref{fig:histogal2} and \ref{fig:histogal3}) are drawn from the same parent distribution. A lower probability indicates a more significant difference. Values in brackets report the significance of the K-S test in units of $\sigma$. }
\begin{tabular}{l ccc }
\hline
\hline
                      &                       &       {\sc P(K-S)}         &             \\
Redshift bin          &         $M_{\star}$         &         SFR         &      {\sc [NUV-$r$]}  \\
                      &   \%    ($\sigma$)    &    \%  ($\sigma$)   &      \%   ($\sigma$)        \\
\hline
0.01 $\leq z <$ 0.30   &   0.76 (2.67)                &        2.97 (2.17)             &    0.11 (3.25)    \\
0.30 $\leq z <$ 0.70   &   2.07$\times10^{-3}$ (4.26) &      $<10^{-20}$ ($>$10)       &    $<10^{-20}$ ($>$10)    \\    
0.70 $\leq z <$ 1.20   &   3.84$\times10^{-4}$ (4.62) &      $<10^{-20}$ ($>$10)       &    $<10^{-20}$ ($>$10)    \\
1.20 $\leq z <$ 1.80   &   34.3 (0.95)                &      $<10^{-20}$ ($>$10)       &    3.50$\times10^{-14}$ (8.16)    \\
1.80 $\leq z <$ 2.50   &   4.51$\times10^{-7}$ (5.86) &      $<10^{-20}$ ($>$10)       &    1.06$\times10^{-4}$ (4.88)     \\
2.50 $\leq z <$ 3.50   &   1.17 (2.52)                &    1.52$\times10^{-13}$ (7.97) &    1.99$\times10^{-8}$ (6.36)      \\
3.50 $\leq z <$ 5.70   &   13.5 (1.49)                &   4.44$\times10^{-3}$ (4.08)   &    10.7 (1.61)   \\
\hline

\end{tabular}

\label{table:ks}
\end{table*}

\subsection{The SFR--$M_{\star}$ plane of radio-selected AGN} \label{sfr_mass_plane}

Fig. \ref{fig:sed_mass_plane_classes} shows the 2D density contours in SFR--$M_{\star}$ plane for our samples of MLAGN (red), HLAGN (blue), and for the subsample of HLAGN with radio excess (blue dashed contours). Black dots represent our joint sample of aforementioned AGN at different redshifts. The 2D density contours enclose the sources (from outer to inner contours) with density levels $>$35, $>$50, $>$68, $>$80, $>$90, and $>$95\% of the maximum 2D density for a given class and redshift bin. 
The black dashed line marks the so-called main sequence (MS) of star-forming galaxies (taken from \citealt{Whitaker+12}), which is known to evolve positively with redshift (e.g. \citealt{Noeske+07}; \citealt{Elbaz+11}; \citealt{Speagle+14}; \citealt{Schreiber+15}). 

The $M_{\star}$ and SFR estimates for each source were computed directly by the three-component fit, hence already correcting for a possible AGN contamination (Sect. \ref{sed_fitting}). Typical 1$\sigma$ uncertainties on $M_{\star}$ and SFR are of the order of 0.1~dex, but for \textit{Herschel} undetected sources the uncertainty in SFR is around 0.2~dex.

Table \ref{table:class} summarises the number of sources shown in Fig. \ref{fig:sed_mass_plane_classes} for each class and redshift bin. These numbers show that the redshift distribution for both HLAGN and MLAGN peaks around z$\sim$1. The two AGN populations become comparable around z$\sim$1.5, while at higher and lower redshift the HLAGN generally outnumber MLAGN. The percentage of HLAGN with a $>$3$\sigma$ radio excess is roughly constant with redshift (around 25--35\% of the HLAGN sample).

Our HLAGN and MLAGN appear to lie in different regions of the SFR--$M_{\star}$ plane, at various redshifts. At low redshift (z$<$0.3), the two AGN classes show rather distinct $M_{\star}$ distributions, where MLAGN are more than two times more massive than HLAGN (10$^{11}$ $M_{\odot}$ versus 10$^{10-10.5}~M_{\odot}$). This difference is consistent with that found by previous studies in the local Universe (e.g. \citealt{Smolcic+09a}; \citealt{Best&Heckman12}). However, we notice that HLAGN with radio excess predominantly lie in the high-$M_{\star}$ tail of the HLAGN population and closely resemble the distribution of MLAGN at this redshift (z$<$0.3). Therefore, this subsample seems to show intermediate SFR and $M_{\star}$ distributions between the two AGN categories. At higher redshift (0.3$<$z$<$1.8), the two AGN populations show a larger overlap in the SFR--$M_{\star}$ plane than that observed at lower redshift. However, the bulk of HLAGN is generally located around the MS relation, while MLAGN 
preferentially 
lie in the lower part of the MS with typically lower SFRs (by a factor of 2--3) compared to HLAGN. The subsample of HLAGN with radio excess is mostly concentrated between the locations of the two main AGN classes. At even higher redshift (z$>$1.8) the overlap between the two distributions decreases, with HLAGN, largely located above the MS relation, having on average both higher M$_{\star}$ and SFR than the MLAGN. A detailed description of the distributions of host-galaxy properties is given in Sect. \ref{histograms}. The highest redshift bin contains only a few tens of sources, which explains the noise seen in the 2D density contours. 

This difference in SFR is consistent with the different percentages of \textit{Herschel}-detected sources between the two AGN classes: 57\% of HLAGN has a $>$3$\sigma$ detection in at least one \textit{Herschel} band, while this percentage decreases to only 17\% for MLAGN. These numbers slightly decrease with redshift for either classes, from {81\% (27\%)} at z$<$0.3 to {50\% (17\%)} in the highest redshift bin for HLAGN (MLAGN). These numbers further suggest that the host galaxies of HLAGN are typically star forming at all redshifts. 
Although this work does not aim to investigate the nature of the sources not classified as AGN, we checked that the subsample of sources that are neither HLAGN nor MLAGN (i.e. 62\%) are mostly located on the MS relation at all redshifts, thus resembling the distribution of HLAGN in the SFR--$M_{\star}$ plane. This suggests that the remainder of our sample might consist of mostly star-forming galaxies.

A comprehensive study of the radio-AGN population in the SFR--M$_{\star}$ plane has been presented by \citet{Bonzini+15}. They found that most of RQ AGN show significant star formation in their hosts, and typically (75\%) lie along the MS relation, likewise SFGs, at various redshifts. Moreover, \citet{Bonzini+15} found that the majority of RL AGN reside in less star-forming galaxies, which are predominantly located below the MS. Despite the different nomenclature and sample selection used by the authors (see Table \ref{table:b13} for a comparison), the qualitative agreement with their results is reassuring.

\begin{figure*}[!t]
\begin{center}
         \includegraphics[width=7in]{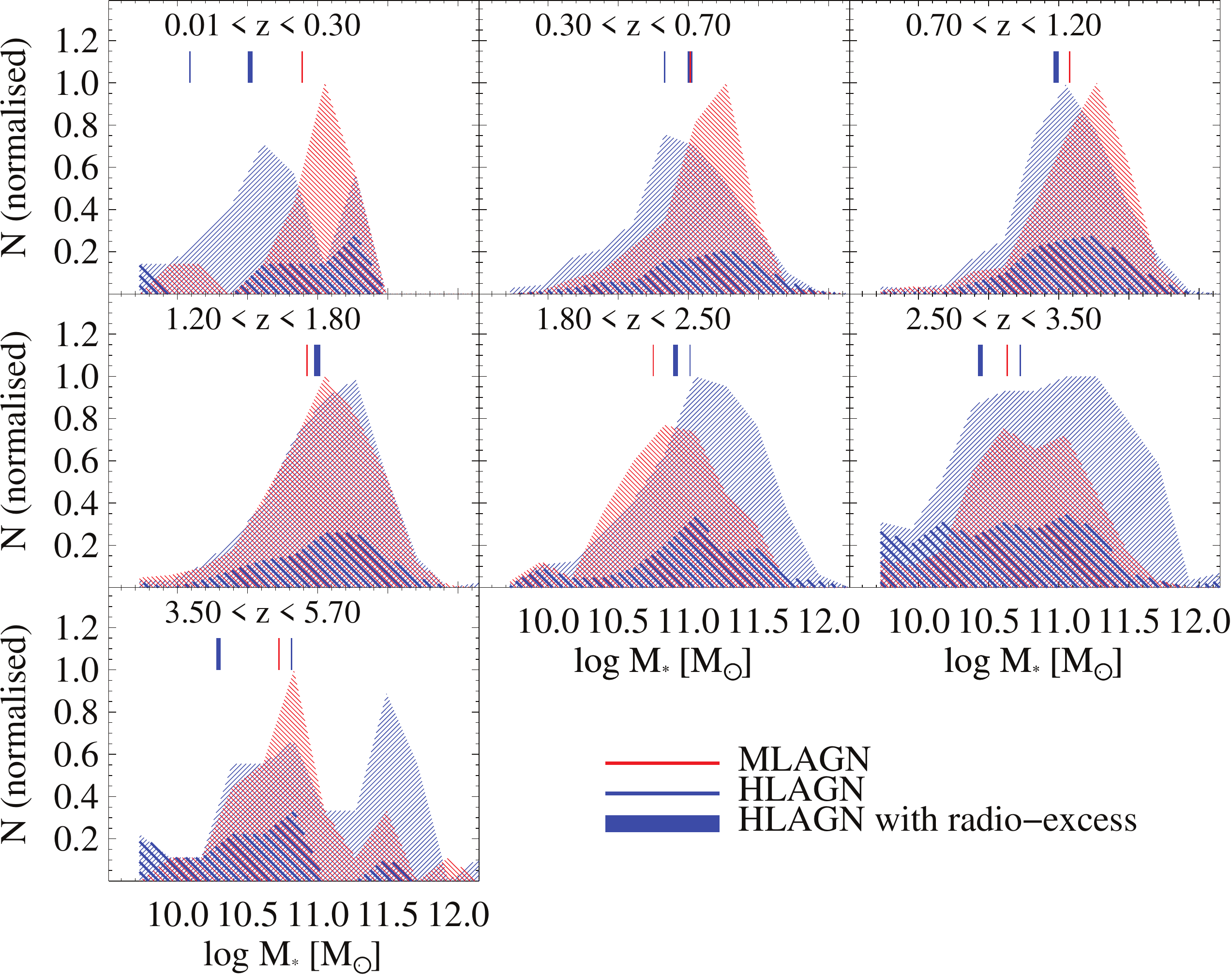}
\end{center}
 \caption{\small Normalised distributions of $M_{\star}$, as a function of redshift. Radio classes are highlighted as follows: HLAGN (blue), HLAGN subsample with radio excess (blue thicker distribution), and MLAGN (red). Vertical lines show the median value for MLAGN (red), HLAGN (blue) and their subsample with radio excess (blue thicker). }   
\label{fig:histogal1}
\end{figure*}

\begin{figure*}[!t]
\begin{center}
         \includegraphics[width=7in]{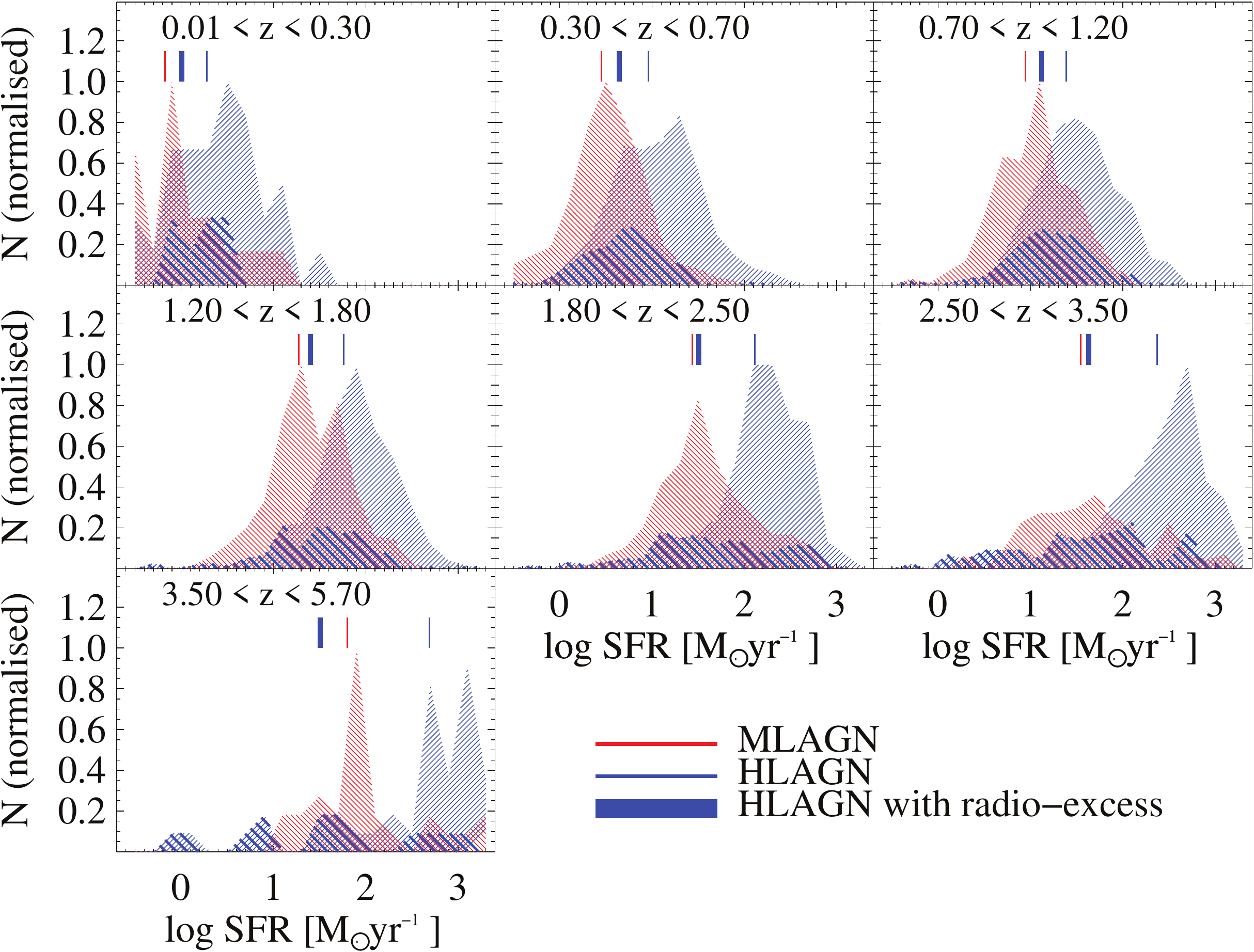}
\end{center}
 \caption{\small Normalised distributions of the SFR as a function of redshift. Radio classes are highlighted as follows: HLAGN (blue), HLAGN subsample with radio excess (blue thicker distribution), and MLAGN (red). Vertical lines show the median value for MLAGN (red), HLAGN (blue), and their subsample with radio excess (blue thicker).}   
\label{fig:histogal2}
\end{figure*}

\begin{figure*}[!t]
\begin{center}
         \includegraphics[width=7in]{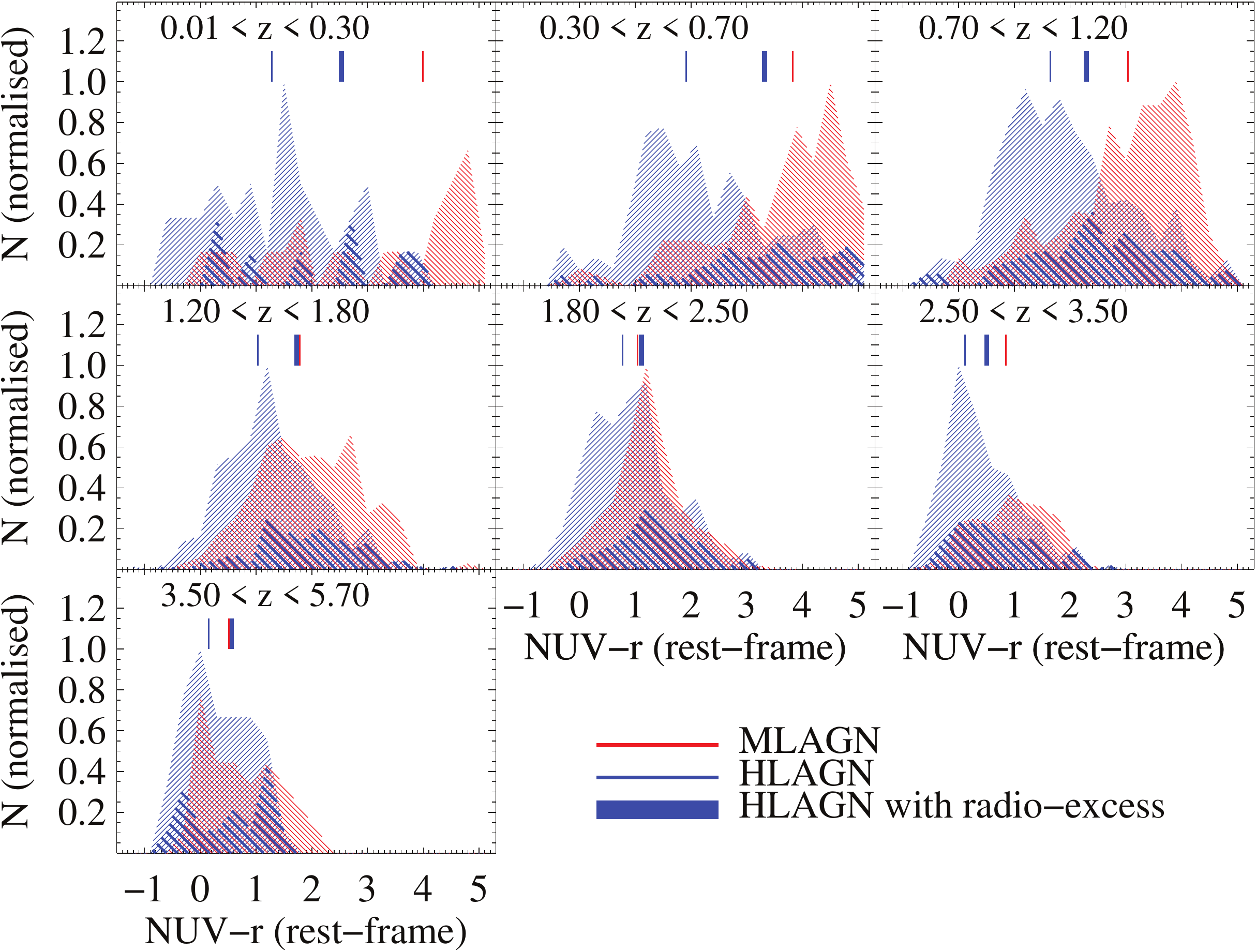}
\end{center}
 \caption{\small Normalised distributions of the rest-frame {\sc [NUV--$r$]} colours, corrected for dust attenuation, as a function of redshift. Radio classes are highlighted as follows: HLAGN (blue), HLAGN subsample with radio excess (blue thicker distribution), and MLAGN (red). Vertical lines show the median value for MLAGN (red), HLAGN (blue), and their subsample with radio excess (blue thicker).}   
\label{fig:histogal3}
\end{figure*}

\subsection{Physical properties of AGN hosts} \label{histograms}

In this section we investigated the distributions of galaxy and AGN properties for the populations of HLAGN and MLAGN. We applied a two-sample Kolmogorov-Smirnov (K-S) test to quantify the difference or similarity
between two distributions. This statistical test allows {us} to determine if two input data sets could be drawn from a common parent distribution without any assumption about its shape. We run this test to evaluate the difference between HLAGN and MLAGN in terms of various galaxy properties. The probabilities {\sc P(K-S)} obtained for each parameter and redshift bin are listed in Table \ref{table:ks}, along with the corresponding $\sigma$ level. Smaller values of {\sc P(K-S)} indicate a more significant difference between the two data sets.

\subsubsection{Distribution of galaxy properties}  \label{results_gal}

Figs. \ref{fig:histogal1}, \ref{fig:histogal2}, and \ref{fig:histogal3} show the distributions of $M_{\star}$, SFR, and rest-frame {\sc [NUV--$r$]} colours, respectively, for the following classes: MLAGN (red), HLAGN (blue), and the subsample of HLAGN with radio excess (blue thicker distribution). Vertical lines show the median value of the corresponding distribution. The distributions are shown in seven redshift bins out to z$\lesssim$6 and are normalised to the highest maximum value of the two distributions.

As mentioned in Sect. \ref{sfr_mass_plane}, the $M_{\star}$ distributions of MLAGN at low redshift are skewed towards higher $M_{\star}$ compared to HLAGN, and the difference remains significant up to z$\sim$1 at $\gtrsim$99\% level (see Table \ref{table:ks}). At z$\sim$1.5 (4$^{\rm th}$ redshift bin) the two distributions appear more similar. 
At higher redshifts (z$\sim$2), we observe a possible reversal of the $M_{\star}$ distributions with the HLAGN populating the high-$M_{\star}$ tail. At this redshift, the two-sample K-S test finds an almost {6}$\sigma$ difference between the two distributions. However, we are not able to confirm or disclaim this trend at z$>$2.5, given the tentative significance (about 2$\sigma$) of the results obtained from the K-S test. A more detailed discussion and interpretation of these trends is presented in Sect. \ref{discussion}. As seen in the SFR--$M_{\star}$ plane (Fig. \ref{fig:sed_mass_plane_classes}), the subsample of HLAGN with radio excess overlaps significantly with the distribution of MLAGN, showing intermediate $M_{\star}$ between the two AGN classes (except in the highest redshift bin). 

In Fig. \ref{fig:histogal2} we show the same plots for the SFR, obtained by integrating the best-fit galaxy template over the range (rest-frame) 8--1000$~\mu$m, and by assuming a \citet{Kennicutt98} scaling factor and a \citet{Chabrier03} IMF. As already seen in Fig. \ref{fig:sed_mass_plane_classes}, we confirm that the HLAGN with radio excess populate the lower tail of the SFR distribution, overlapping significantly with MLAGN. The difference between HLAGN and MLAGN in SFR remains visible and $\gtrsim99$\% significant  in the higher redshift bins, up to z$\sim$3, as well. 

Rest-frame {\sc [NUV-$r$]} colours were calculated from the best-fit galaxy template of each source and also corrected for dust attenuation. The distribution of {\sc [NUV-$r$]} colours shown in Fig, \ref{fig:histogal3} confirms that most of HLAGN have blue or green rest-frame optical colours ({\sc [NUV-$r$]}$<$3.5; \citealt{Ilbert+10}) at all redshifts. On the other hand, MLAGN are more pronounced towards quiescent systems ({\sc [NUV-$r$]}$>$3.5; \citealt{Ilbert+10}), at least up to z$\sim$1. This bimodality in the colour distributions becomes progressively less pronounced at higher redshift, also showing that the host-galaxies of MLAGN become, on average, more star forming with increasing redshift (see Sect. \ref{ir_stacking}). It is recognised that the number density of quiescent galaxies (selected via optical colours) at $M_{\star}>$10$^{10}$~$M_{\odot}$ decreases with increasing redshift (e.g \citealt{Brammer+11}, \citealt{Ilbert+13}). However, the difference in {\sc [NUV-$r$]} 
between HLAGN and MLAGN remains highly significant up to z$\sim$3.5, while it disappears at 3.5$<$z$<$5.7. The subsample of HLAGN with radio excess shows intermediate colours between the rest of HLAGN and the population of MLAGN.


\begin{figure*}[!t]
\begin{center}
\includegraphics[width=7in]{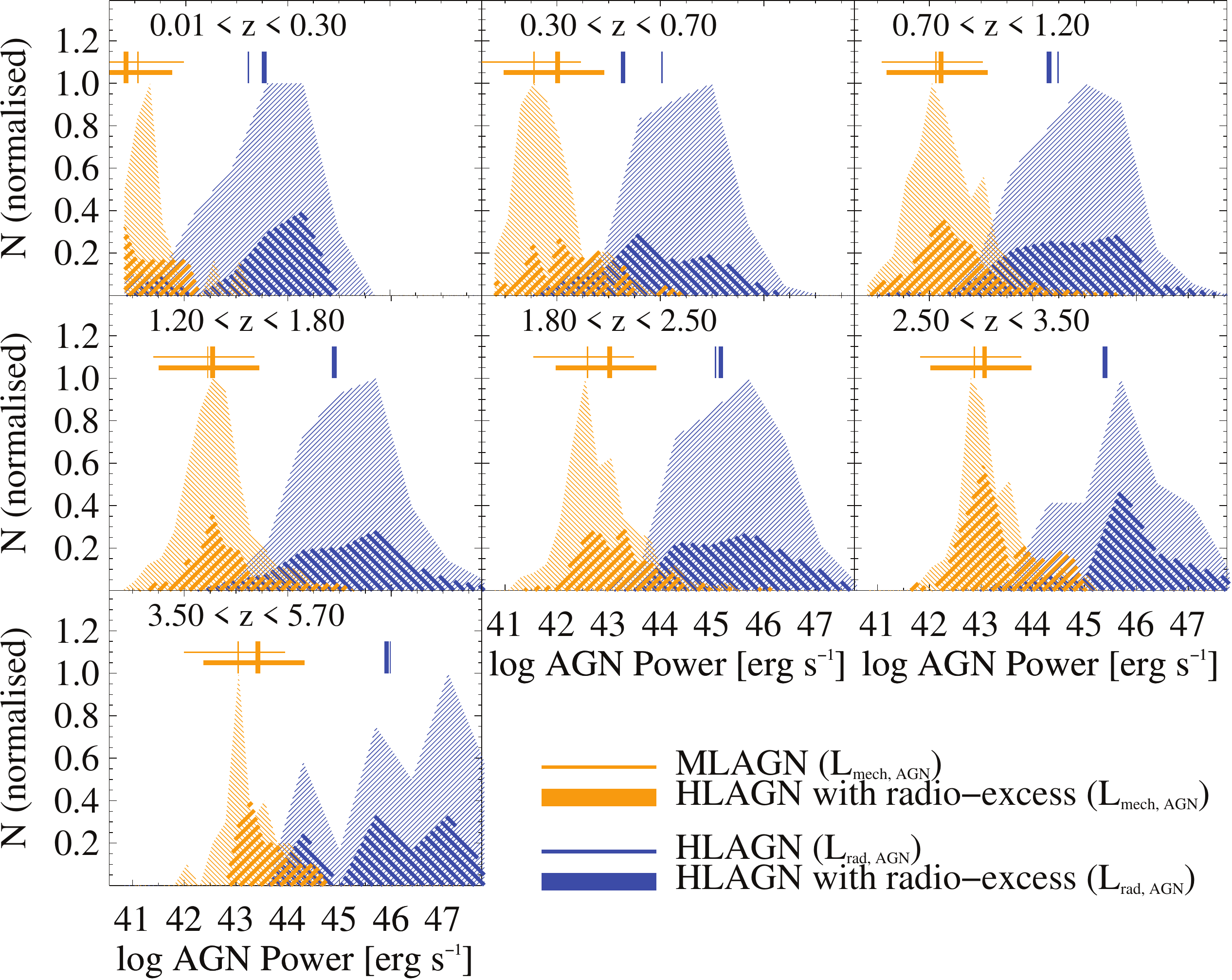}
\end{center}
 \caption{\small Normalised distributions of AGN power, both radiative (L$\rm_{rad, AGN}$) and mechanical (L$\rm_{mech, AGN}$) as a function of redshift. The distributions of L$\rm_{rad, AGN}$ are shown for HLAGN (blue) and HLAGN with radio excess (blue thicker), while the distributions of L$\rm_{mech, AGN}$ are shown only for MLAGN (orange) and HLAGN with radio excess (orange thicker). The median value of each distribution is indicated with a vertical line with same colour and thickness of the corresponding histogram. The orange horizontal lines around the median value of the AGN mechanical power show the range within which the median could shift if accounting for all of the uncertainties on the $L_{\rm 1.4\, GHz}$--L$\rm_{mech, AGN}$ relation (see text for details). The normalisations are set separately for the two L$\rm_{mech, AGN}$ (red) and the two L$\rm_{rad, AGN}$ (blue) distributions. See the text for details.}
\label{fig:histogramagn}
\end{figure*}

\subsubsection{Distribution of AGN properties} \label{results_agn}

Fig. \ref{fig:histogramagn} shows the distribution of both AGN radiative and mechanical power for our classes of AGN. 

We calculated the AGN radiative power (L$\rm_{rad, AGN}$) of HLAGN from the best-fit AGN template obtained with the three-component SED-fitting code {\sc sed3fit} for sources both with and without radio excess. The typical uncertainties on L$\rm_{rad, AGN}$ are around 0.4~dex for sources with ($\geq$99\%) significant AGN component (i.e. SED-AGN, see Sect. \ref{classification}), while  we took the L$\rm_{rad, AGN}$ from the 95$^{\rm th}$ percentile of the corresponding PDF obtained from the {\sc sed3fit} code for HLAGN not identified as such from SED decomposition (30\% of HLAGN); this is equivalent to an upper limit at 90\% confidence level on the AGN radiative luminosity. Fig. \ref{fig:histogramagn} shows the normalised distributions of L$\rm_{rad, AGN}$, separately for HLAGN (blue) and for the subsample with radio excess (blue thicker). The 
distributions of L$\rm_{rad, AGN}$ cover a broad range ($>$3~dex) in each redshift bin, which is around 10$^{42-45}$ erg~s$^{-1}$ at z$<$0.3, and 10$^{44-47}$ erg~s$^{-1}$ at z$\sim$3. As mentioned in Sect. \ref{naming_convention}, the percentage of HLAGN identified from SED fitting does not depend on the presence of a radio excess. Therefore, applying the upper limits at 90\% confidence level on L$\rm_{rad, AGN}$ does not affect the ratio of the L$\rm_{rad, AGN}$ distributions between HLAGN and their subsample with radio excess. 

We calculated the rest-frame 1.4~GHz radio luminosity $L_{\rm 1.4\, GHz}$ for each source by scaling its radio flux from 3~GHz to 1.4~GHz and taking the observed 1.4--3~GHz spectral index $\alpha$, as explained in Sect. \ref{radio-multi}. The presence of a $>$3$\sigma$ radio excess suggests that a notable portion of the radio emission is not arising from star formation processes in the host, but possibly from the central SMBH. For this reason, each $L_{\rm 1.4\, GHz}$ measurement was scaled to the portion associated with AGN activity, based on the deviation of the observed $L_{\rm 1.4\, GHz}$ - to - SFR$_{\rm IR}$ ratio from the peak of the Gaussian function (associated with star formation) at the corresponding redshift bin (blue points in Fig. \ref{fig:radio_excess}). 

We converted the AGN-related radio emission to AGN mechanical power (L$\rm_{mech, AGN}$) of the radio jet, by assuming the redshift-independent relation by \citet{Willott+99}, which is based on theoretical grounds and adopted in other studies (e.g. \citealt{Merloni+08}; \citealt{LaFranca+10}, see \citealt{Best&Heckman12} for a review). We used this relation expressed in terms of $L_{\rm 1.4\, GHz}$ (see Eq. 1 from \citealt{Heckman+14}). \citet{Willott+99} combined all of the uncertainties on this relation into a single factor, $f_{\rm W}$, which can range between 1 and 20. This scaling factor is still a matter of debate in the literature (\citealt{Godfrey+16}). Nonetheless, following the approach of numerous studies (e.g. \citealt{Merloni+07}; \citealt{Smolcic+09a}; \citealt{LaFranca+10}; \citealt{Best&Heckman12}; \citealt{Pracy+16}), we make the simplistic assumption that this relation holds at all radio luminosities $L_{\rm 1.4\, GHz}$ that are probed by our sample.

The normalised distributions of L$\rm_{mech, AGN}$ are shown in Fig. \ref{fig:histogramagn} for both MLAGN (orange) and HLAGN with radio excess (orange thicker distribution). The typical range in L$\rm_{mech, AGN}$ probed by the distributions is about 2~dex wide in each redshift bin, which is around 10$^{41-43}$ erg~s$^{-1}$ at z$<$0.3 and 10$^{43-45}$ erg~s$^{-1}$ at z$\sim$3. Fig. \ref{fig:histogramagn} shows the distributions of L$\rm_{mech, AGN}$ by taking $f_{\rm W}$=5, which is consistent with the relation derived by \citet{Daly+12}. The vertical lines indicate the median value of the corresponding distribution. However, we calculated the range within which the median value could shift, by changing f$_{\rm W}$ between f$_{\rm W}$=1 and f$_{\rm W}$=20, which is shown by the orange horizontal lines around the median. 

Our analysis seems to suggest that the overall AGN properties observed for HLAGN with radio excess are similar to the rest of HLAGN in terms of radiative power and are also consistent with MLAGN in terms of mechanical power. The subsample of HLAGN with radio excess ({6}\% of our parent 3~GHz radio sample) is particularly interesting because it enables a direct comparison between radiative and mechanical AGN power for the same sources. Despite the uncertainties on the relation proposed by \citet{Willott+99}, we show that the AGN mechanical power in HLAGN with radio excess is typically lower than (or at most marginally comparable to) the AGN radiative power, depending on f$_{\rm W}$, although in some cases L$\rm_{mech, AGN}$ can exceed L$\rm_{rad, AGN}$ (see \citealt{Heckman+14}). While the AGN power of HLAGN occurs predominantly in radiative form, MLAGN display a substantial mechanical AGN luminosity component. These properties may suggest that HLAGN and MLAGN samples qualitatively resemble radio AGN 
types often referred to as radiative mode (or HERG) and jet mode (or LERG), respectively. In addition, we note that MLAGN have significantly lower L$\rm_{rad, AGN}$ than HLAGN with radio excess, despite both classes {showing} a relatively high radio loudness. As a consequence, a simple RL--RQ separation would not allow such direct insight into the fundamental properties of AGN.


\section{Discussion} \label{discussion}

A radio-based selection allows us to study a mixture of galaxy populations that are powered by either star formation, AGN activity, or both. It is also crucial to exploit multiwavelength ancillary data to reach a more comprehensive perspective of the nature of our sources. In this work, we made use of this approach to derive integrated AGN and galaxy properties and compare them between different AGN classes, and over a wide range of radio luminosity and redshift. In this section, we discuss and interpret our findings in the context of current AGN and galaxy evolutionary scenarios.

\subsection{Radio emission in HLAGN and MLAGN }

The origin of radio emission in the sub-mJy radio population is still a matter of debate. Recent studies based on interferometric radio observations of sub-mJy radio sources have the potential to shed light on this issue. For example, \citet{HerreraRuiz+16} analysed in detail the interferometric images of three RQ-AGN obtained with VLBI in the COSMOS field, which are part of the sample described in Sect. \ref{VLBI}. The comparison between VLBI and VLA fluxes suggested that 50--75\% of the radio emission in these sources is arising from non-thermal AGN activity. We note that these sources would have been classified as HLAGN with radio excess according to our method. Similar conclusions were reached by \citet{Chi+13} and \citet{Maini+16} on different samples of RQ-AGN, supporting the idea that some radio sources could be predominantly powered by AGN activity.

At low redshift (z$<$0.3), independent hints on the origin of radio emission in the sub-mJy radio population were provided by \citet{Kimball+11}, who constructed the 6$~$GHz radio luminosity function from a sample of QSO host galaxies at 0.2$<$z$<$0.3. They concluded that radio emission in sources with 6$~$GHz luminosity $L_{\rm 6~GHz} >$10$^{22.5}$ W~Hz$^{-1}$ was AGN related, while in fainter sources it was mainly driven by star formation. However, the parent sample analysed by the authors consists of optically identified QSOs from the SDSS, which are systematically more powerful, relative to our 3~GHz sample of AGN in COSMOS, due to our smaller comoving volume covered at z$<$0.3. Support for radio emission that is powered by AGN activity was also provided by \citet{White+15}, who studied a sample of RQ-QSOs at 1.4~GHz flux $S_{\rm 1.4~GHz}<$1~mJy. The authors stress, however, that their analysis may be biased towards the brightest optically identified QSOs. 

A complementary view on this topic benefits from deeper radio surveys, which can push this analysis to higher redshifts and to intrinsically fainter radio sources. For example, \citet{Bonzini+15} and \citet{Padovani+15} investigated the origin of radio emission in RQ and RL AGN in the E-CDFS down to 37$~\mu$Jy (5$\sigma$). They found a mixture of AGN and SFGs contributing to the sub-mJy radio population, where RQ AGN is predominantly powered by star formation. We checked this by exploiting a larger sample of 3~GHz selected sources with optical-NIR counterparts, counting in total {1\,604} HLAGN and {1\,333} MLAGN (Sect. \ref{classification}) out to z$\lesssim$6. The analysis presented in Sects. \ref{hlagn} and \ref{llagn} suggests that roughly 70\% of the HLAGN does not show a $\geq$3$\sigma$ radio excess, which might suggest that radio and infrared emission in HLAGN are commonly (to a certain amount) powered by star formation in their hosts, as proposed by previous studies (e.g.~\citealt{Moric+10}; \citealt{Baldi+13}; \citealt{Padovani+15}). 
However, the radio excess detected for the remaining 30\% is a potential signature of radio-selected AGN activity, possibly linked to jet-mode (or radio-mode) feedback, often referred to in the literature (e.g. \citealt{Hardcastle+07}; \citealt{Best&Heckman12}; \citealt{Heckman+14}). On the other hand, radio emission in our sample of MLAGN is predominantly arising from non-thermal radiation likely ascribed to AGN activity, rather than star formation in their hosts. These results agree with the conclusions presented by \citet{Padovani+15} and \citet{Bonzini+15} for a sample of RQ-AGN and RL-AGN, supporting the composite nature of the sub-mJy radio source population (e.g. \citealt{Smolcic+08}; \citealt{Padovani+11}; \citealt{Baldi+14}).

\subsection{Radio AGN in the context of galaxy evolution} \label{interpretation}

We attempt to interpret the nature of our HLAGN and MLAGN populations in the framework of AGN-galaxy evolution. As suggested by previous authors (e.g. \citealt{Hardcastle+07}; \citealt{Smolcic+09b}; \citealt{Best&Heckman12}; \citealt{Padovani+15}), in the local Universe the HERG and LERG classes show a clear dichotomy in terms of AGN and host-galaxy properties. These findings have been interpreted within a self-consistent evolutionary scenario, where HERG and LERG trace earlier and later stages, respectively, of galaxies' life cycle (see \citealt{Heckman+14} for a comprehensive review). 

At higher redshift, \citet{Merloni+08} proposed a model to reproduce the kinetic and radiative luminosity function of AGN in which the highly efficient accretion onto the SMBH can produce both kinetic and radiative feedback (e.g. \citealt{Veilleux+13}), which are consistent with the AGN properties observed for our HLAGN with and without radio excess, respectively. Nevertheless, the power from weakly accreting SMBHs ($\lambda_{\rm Edd} \leq$ 10$^{-2}$, also named ``advection-dominated accretion flow'', ADAF; e.g. \citealt{Blandford+77}) is mainly in the form of kinetic feedback (\citealt{Bower+06}; \citealt{Fanidakis+11}), linking to the properties of our MLAGN population. 

Semi-analytic models predict different accretion modes between highly and weakly accreting AGN. On the one hand, highly accreting AGN have been usually connected to a fast gas accretion mode in galaxy halos in which the free fall times are usually longer than the cooling times. On the other hand, weakly accreting AGN are in the regime of slow gas accretion, where cooling time is much larger than the free fall time (e.g. \citealt{Fanidakis+11}, \citeyear{Fanidakis+12}).

From an observational point of view, we found that galaxies hosting MLAGN are more massive, redder, and less star forming compared to HLAGN, at least up to z$\sim$1. In particular, the most massive galaxies ($M_{\star}\sim$10$^{11}$~$M_{\odot}$) at these redshifts typically host MLAGN, while the $M_{\star}$ distributions of HLAGN and MLAGN become comparable at z$\sim$1.5 and display a reversal at z$\sim$2. This trend is unlikely to be driven by the incompleteness in M$_{\star}$, as the optical-NIR selected sample in the COSMOS field is $>$80\% complete at M$_{\star}>$10$^{9.7}$M$_{\sun}$ out to z$\sim$4 (see \citealt{Davidzon+17}). We stress that this $M_{\star}$ behaviour is observed in our radio-selected sample, while it might not be the same for differently selected samples of AGN. A more comprehensive analysis combining panchromatic samples of X-ray, MIR, and radio-selected AGN would be crucial to test the widespread validity of this finding. This possible hint of ``downsizing'' (e.g.~\citealt{Cowie+96}) 
links back to the known anti-hierarchical growth of galaxies over cosmic time with the most massive systems evolving earlier and faster than their lower mass counterparts (see also \citealt{Bundy+06}; \citealt{Fontanot+09}). In particular, this M$_{\star}$ behaviour is expected if the most massive galaxies trigger higher radiative luminosity AGN activity earlier than less massive galaxies and then fade to lower radiative luminosity AGN at lower redshifts. The same qualitative argument is proposed in the evolution of AGN with the number density of powerful AGN ($L_{\rm x}>$10$^{44}$ erg~s$^{-1}$) peaking earlier in cosmic time compared to lower luminosity AGN (e.g. \citealt{Barger+05}; \citealt{Hasinger+05}; \citealt{Silverman+08}; \citealt{Ueda+14}). 

Different studies of AGN host galaxies have argued that AGN accretion preferentially occurs in gas-rich galaxies (\citealt{Vito+14}), and that the percentage of galaxies hosting X-ray AGN increases with infrared luminosity $L_{\rm IR}$ (e.g. \citealt{Bongiorno+12}; \citealt{Santini+12}). This is consistent with the increasing gas fraction observed in MS galaxies from low to high redshift (e.g. \citealt{Daddi+10}; \citealt{Saintonge+12}; \citealt{Tacconi+13}), and explained via the Schmidt--Kennicutt relation (\citealt{Schmidt59}; \citealt{Kennicutt98}). The increasing gas fraction from low to high redshift might explain the higher occurence of higher radiative luminosity AGN in massive galaxies ($M_{\star}\sim$10$^{11}$~$M_{\odot}$) at higher redshift (z$\sim$2). Indeed, we showed that galaxies hosting HLAGN are mostly on the MS relation (see also \citealt{Rosario+12}), which implies a large availability of cold gas supplies, and possibly a more efficient fuelling mechanism of the central SMBH (i.e. with 
higher accretion rates), compared to the physical processes taking place in MLAGN. 

According to this scenario, less star-forming galaxies are less likely to host an active SMBH. Interestingly, we found that most of MLAGN reside in weakly star-forming galaxies, which are typically located a factor of 2--3 below the MS. A plausible interpretation is that the difference in cold gas reservoirs leads HLAGN and MLAGN to be mainly powered by different accretion mechanisms. This raises the question of what triggers AGN activity in these two AGN populations. Shedding light on this issue requires a thorough investigation of the Eddington ratio distributions between these two AGN classes, which will be presented in a future work (I.~Delvecchio et al., in prep.). 

It is worth noting that HLAGN with radio excess show intermediate $M_{\star}$, {\sc [NUV-$r$]} and SFR distributions between MLAGN and the rest of HLAGN, especially at z$<$1. Under the assumption that, in a stable phase, the AGN feedback occurs predominantly in either radiative or mechanical form, the population of HLAGN with radio excess might coincide with a transitional phase of AGN feedback.

According to semi-analytic models (e.g. \citealt{Croton+06}; \citealt{Marulli+08}, \citealt{Hopkins+08}), AGN feedback is one of the possible means to track the AGN host galaxies from the blue cloud of star-forming systems to the red sequence of passive galaxies, passing through a transition (often referred to as ``green valley''), where the star formation is weaker but not yet stopped. According to this possible scenario, MLAGN and HLAGN with radio excess might represent intrinsically the same galaxies but that are observed at different stages of their AGN duty cycle, in which the energy produced via accretion onto the SMBH is emitted in either radiative or mechanical forms. The lower level of star formation in MLAGN might be a consequence of AGN-driven feedback, where the radio emission powered by the AGN could limit or hamper the galaxy star formation. This scenario is supported by studies of radio-selected AGN, where jet-induced feedback can strongly impact the molecular gas supplies of the host-
galaxy (e.
g. \citealt{Feruglio+10}; \citealt{Morganti+13}; \citealt{Combes+13}). In this context, it is possible that the population of HLAGN with radio excess probes a particular stage of the radio-mode feedback phase, where the molecular gas in the host galaxy is not yet depleted, and no evident impact in the integrated properties of the galaxy should be detectable during this transition (see Figs. \ref{fig:histogal1}, \ref{fig:histogal2}, and \ref{fig:histogal3}). This scenario is also supported by recent spectroscopic observations of powerful outflows detected in X-ray-MIR selected AGN, some of which show a significant radio excess (e.g. \citealt{Perna+15}; \citealt{Lonsdale+15}; \citealt{Brusa+16}). For these reasons, we stress that our sample of HLAGN with radio excess could be ideal to investigate the impact of AGN feedback, both radiative and mechanical, in a statistical sense and in a wide redshift range.


\section{Conclusions} \label{conclusions}

This work presents a multiwavelength analysis of radio-selected AGN host-galaxy properties out to z$\lesssim$6. Our sample consists of about {7\,700} radio sources selected at 3~GHz in the COSMOS field, and cross-matched with optical-NIR counterparts. The exquisite photometry and redshifts available enabled us to use multiwavelength diagnostics to identify two main AGN populations in our sample: HLAGN (21\%, out of which 30\% also shows a $>$3$\sigma$ radio-excess) and MLAGN (17\%). We analysed the average properties of their host galaxies at different cosmic epochs and summarise our main conclusions as follows:

\begin{enumerate}
 
 \item We tested our source classification method against independent criteria used in recent radio-based studies (e.g. \citealt{DelMoro+13}; \citealt{Bonzini+13}; \citealt{Padovani+15}; N.~Herrera Ruiz et al., in prep.), finding a good agreement and demonstrating the robustness of our method.
 
 \item We provided a value-added catalogue containing the classification and the main physical properties discussed in this work for each radio source ($M_{\star}$, SFR, {\sc [NUV-$r$]} colours, $L_{\rm 3~GHz}$ and $L_{\rm IR, SF}$).
 
 \item Our HLAGN and MLAGN lie in different regions of the SFR--$M_{\star}$ plane, where the former are, on average, less massive and more star forming than the latter at various redshifts. We analysed in detail the observed distributions of various galaxy properties, finding significantly higher SFR and bluer {\sc [NUV-$r$]} colours in HLAGN compared to MLAGN at all redshifts. Nevertheless, the $M_{\star}$ distribution is mainly populated by MLAGN at the highest $M_{\star}$ values ($M_{\star}$>10$^{11}$~$M_{\odot}$) at z$<$1, while the two AGN classes equally contribute to the highest $M_{\star}$ at z$\sim$1.5, and display a {6}$\sigma$ reversal in the $M_{\star}$ behaviour at z$\sim$2.
 
 \item Our results are consistent with radio emission predominantly arising from star formation in around 70\% of HLAGN, while the remaining 30\% shows a $\geq$3$\sigma$ radio excess that is likely attributable to AGN activity. The fractional AGN contribution to the radio emission in MLAGN is expected to be around 80--90\%.
 
\end{enumerate}
 
 Overall, the differences in galaxy properties seen between these two AGN classes suggest that HLAGN and MLAGN samples trace two distinct galaxy populations in a wide range of redshifts. This might reflect the presence of two different driving mechanisms of AGN activity, which is possibly linked to the different availability of cold gas supplies in their hosts. In this scenario, the subsample of HLAGN with radio excess might coincide with a transitional phase during the AGN duty cycle, in which AGN activity occurs in both radiative and mechanical forms.


\begin{acknowledgements}
The authors are grateful to the anonymous referee for his/her careful reading and useful comments, which improved the content of this manuscript. ID, VS, JD, MN, and OM acknowledge the European Union's Seventh Framework programme under grant agreement 337595 (ERC Starting Grant, ``CoSMass''). NB acknowledges the European Union’s Seventh Framework programme under grant agreement 333654 (CIG, ‘AGN feedback’). CL is funded by a Discovery Early Career Researcher Award (DE150100618). Parts of this research were conducted by the Australian Research Council Centre of Excellence for All-sky Astrophysics (CAASTRO), through project number CE110001020. MB and PC acknowledge supports from the PRIN-INAF 2014. MA acknowledges partial support from FONDECYT through grant 1140099. AK acknowledges support by the Collaborative Research Council 956, sub-project A1, funded by the Deutsche Forschungsgemeinschaft (DFG). This work was partially supported by NASA Chandra grant number GO3-14150C and GO3-14150B (FC, SM).

ID is grateful to B. Magnelli for useful suggestions and to Takamitsu Miyaji for his help with CSTACK.

\end{acknowledgements}

 \bibliographystyle{aa} 
 \bibliography{biblio} 

\begin{thebibliography}{183}
\expandafter\ifx\csname natexlab\endcsname\relax\def\natexlab#1{#1}\fi

\bibitem[{{Alexander} \& {Hickox}(2012)}]{Alexander+12}
{Alexander}, D.~M. \& {Hickox}, R.~C. 2012, \nar, 56, 93

\bibitem[{{Aretxaga} {et~al.}(2011){Aretxaga}, {Wilson}, {Aguilar}, {Alberts},
  {Scott}, {Scoville}, {Yun}, {Austermann}, {Downes}, {Ezawa}, {Hatsukade},
  {Hughes}, {Kawabe}, {Kohno}, {Oshima}, {Perera}, {Tamura}, \&
  {Zeballos}}]{Aretxaga+11}
{Aretxaga}, I., {Wilson}, G.~W., {Aguilar}, E., {et~al.} 2011, \mnras, 415,
  3831

\bibitem[{{Arnouts} {et~al.}(1999){Arnouts}, {Cristiani}, {Moscardini},
  {Matarrese}, {Lucchin}, {Fontana}, \& {Giallongo}}]{Arnouts+99}
{Arnouts}, S., {Cristiani}, S., {Moscardini}, L., {et~al.} 1999, \mnras, 310,
  540

\bibitem[{{Baldi} {et~al.}(2014){Baldi}, {Capetti}, {Chiaberge}, \&
  {Celotti}}]{Baldi+14}
{Baldi}, R.~D., {Capetti}, A., {Chiaberge}, M., \& {Celotti}, A. 2014, \aap,
  567, A76

\bibitem[{{Baldi} {et~al.}(2013){Baldi}, {Chiaberge}, {Capetti},
  {Rodriguez-Zaurin}, {Deustua}, \& {Sparks}}]{Baldi+13}
{Baldi}, R.~D., {Chiaberge}, M., {Capetti}, A., {et~al.} 2013, \apj, 762, 30

\bibitem[{{Baldwin} {et~al.}(1981){Baldwin}, {Phillips}, \&
  {Terlevich}}]{Baldwin+81}
{Baldwin}, J.~A., {Phillips}, M.~M., \& {Terlevich}, R. 1981, \pasp, 93, 5

\bibitem[{{Barger} \& {Cowie}(2005)}]{Barger+05}
{Barger}, A.~J. \& {Cowie}, L.~L. 2005, \apj, 635, 115

\bibitem[{{Bell}(2003)}]{Bell03}
{Bell}, E.~F. 2003, \apj, 586, 794

\bibitem[{{Berta} {et~al.}(2013){Berta}, {Lutz}, {Santini}, {Wuyts}, {Rosario},
  {Brisbin}, {Cooray}, {Franceschini}, {Gruppioni}, {Hatziminaoglou}, {Hwang},
  {Le Floc'h}, {Magnelli}, {Nordon}, {Oliver}, {Page}, {Popesso}, {Pozzetti},
  {Pozzi}, {Riguccini}, {Rodighiero}, {Roseboom}, {Scott}, {Symeonidis},
  {Valtchanov}, {Viero}, \& {Wang}}]{Berta+13}
{Berta}, S., {Lutz}, D., {Santini}, P., {et~al.} 2013, \aap, 551, A100

\bibitem[{{Bertoldi} {et~al.}(2007){Bertoldi}, {Carilli}, {Aravena},
  {Schinnerer}, {Voss}, {Smolcic}, {Jahnke}, {Scoville}, {Blain}, {Menten},
  {Lutz}, {Brusa}, {Taniguchi}, {Capak}, {Mobasher}, {Lilly}, {Thompson},
  {Aussel}, {Kreysa}, {Hasinger}, {Aguirre}, {Schlaerth}, \&
  {Koekemoer}}]{Bertoldi+07}
{Bertoldi}, F., {Carilli}, C., {Aravena}, M., {et~al.} 2007, \apjs, 172, 132

\bibitem[{{Best} \& {Heckman}(2012)}]{Best&Heckman12}
{Best}, P.~N. \& {Heckman}, T.~M. 2012, \mnras, 421, 1569

\bibitem[{{B{\'e}thermin} {et~al.}(2010){B{\'e}thermin}, {Dole}, {Beelen}, \&
  {Aussel}}]{Bethermin+10}
{B{\'e}thermin}, M., {Dole}, H., {Beelen}, A., \& {Aussel}, H. 2010, \aap, 512,
  A78

\bibitem[{{Blandford} \& {Znajek}(1977)}]{Blandford+77}
{Blandford}, R.~D. \& {Znajek}, R.~L. 1977, \mnras, 179, 433

\bibitem[{{Bongiorno} {et~al.}(2012){Bongiorno}, {Merloni}, {Brusa},
  {Magnelli}, {Salvato}, {Mignoli}, {Zamorani}, {Fiore}, {Rosario}, {Mainieri},
  {Hao}, {Comastri}, {Vignali}, {Balestra}, {Bardelli}, {Berta}, {Civano},
  {Kampczyk}, {Le Floc'h}, {Lusso}, {Lutz}, {Pozzetti}, {Pozzi}, {Riguccini},
  {Shankar}, \& {Silverman}}]{Bongiorno+12}
{Bongiorno}, A., {Merloni}, A., {Brusa}, M., {et~al.} 2012, \mnras, 427, 3103

\bibitem[{{Bonzini} {et~al.}(2015){Bonzini}, {Mainieri}, {Padovani},
  {Andreani}, {Berta}, {Bethermin}, {Lutz}, {Rodighiero}, {Rosario}, {Tozzi},
  \& {Vattakunnel}}]{Bonzini+15}
{Bonzini}, M., {Mainieri}, V., {Padovani}, P., {et~al.} 2015, \mnras, 453, 1079

\bibitem[{{Bonzini} {et~al.}(2013){Bonzini}, {Padovani}, {Mainieri},
  {Kellermann}, {Miller}, {Rosati}, {Tozzi}, \& {Vattakunnel}}]{Bonzini+13}
{Bonzini}, M., {Padovani}, P., {Mainieri}, V., {et~al.} 2013, \mnras, 436, 3759

\bibitem[{{Booth} \& {Schaye}(2009)}]{Booth+09}
{Booth}, C.~M. \& {Schaye}, J. 2009, \mnras, 398, 53

\bibitem[{{Bournaud} {et~al.}(2012){Bournaud}, {Juneau}, {Le Floc'h},
  {Mullaney}, {Daddi}, {Dekel}, {Duc}, {Elbaz}, {Salmi}, \&
  {Dickinson}}]{Bournaud+12}
{Bournaud}, F., {Juneau}, S., {Le Floc'h}, E., {et~al.} 2012, \apj, 757, 81

\bibitem[{{Bower} {et~al.}(2006){Bower}, {Benson}, {Malbon}, {Helly}, {Frenk},
  {Baugh}, {Cole}, \& {Lacey}}]{Bower+06}
{Bower}, R.~G., {Benson}, A.~J., {Malbon}, R., {et~al.} 2006, \mnras, 370, 645

\bibitem[{{Brammer} {et~al.}(2011){Brammer}, {Whitaker}, {van Dokkum},
  {Marchesini}, {Franx}, {Kriek}, {Labb{\'e}}, {Lee}, {Muzzin}, {Quadri},
  {Rudnick}, \& {Williams}}]{Brammer+11}
{Brammer}, G.~B., {Whitaker}, K.~E., {van Dokkum}, P.~G., {et~al.} 2011, \apj,
  739, 24

\bibitem[{{Brisbin et al.}(2017)}]{brisbin+17}
{Brisbin et al.} 2017, \aap [\eprint{submitted}]

\bibitem[{{Brusa} {et~al.}(2010){Brusa}, {Civano}, {Comastri}, {Miyaji},
  {Salvato}, {Zamorani}, {Cappelluti}, {Fiore}, {Hasinger}, {Mainieri},
  {Merloni}, {Bongiorno}, {Capak}, {Elvis}, {Gilli}, {Hao}, {Jahnke},
  {Koekemoer}, {Ilbert}, {Le Floc'h}, {Lusso}, {Mignoli}, {Schinnerer},
  {Silverman}, {Treister}, {Trump}, {Vignali}, {Zamojski}, {Aldcroft},
  {Aussel}, {Bardelli}, {Bolzonella}, {Cappi}, {Caputi}, {Contini},
  {Finoguenov}, {Fruscione}, {Garilli}, {Impey}, {Iovino}, {Iwasawa},
  {Kampczyk}, {Kartaltepe}, {Kneib}, {Knobel}, {Kovac}, {Lamareille},
  {Leborgne}, {Le Brun}, {Le Fevre}, {Lilly}, {Maier}, {McCracken}, {Pello},
  {Peng}, {Perez-Montero}, {de Ravel}, {Sanders}, {Scodeggio}, {Scoville},
  {Tanaka}, {Taniguchi}, {Tasca}, {de la Torre}, {Tresse}, {Vergani}, \&
  {Zucca}}]{Brusa+10}
{Brusa}, M., {Civano}, F., {Comastri}, A., {et~al.} 2010, \apj, 716, 348

\bibitem[{{Brusa} {et~al.}(2015){Brusa}, {Feruglio}, {Cresci}, {Mainieri},
  {Sargent}, {Perna}, {Santini}, {Vito}, {Marconi}, {Merloni}, {Lutz},
  {Piconcelli}, {Lanzuisi}, {Maiolino}, {Rosario}, {Daddi}, {Bongiorno},
  {Fiore}, \& {Lusso}}]{Brusa+15}
{Brusa}, M., {Feruglio}, C., {Cresci}, G., {et~al.} 2015, \aap, 578, A11

\bibitem[{{Brusa} {et~al.}(2016){Brusa}, {Perna}, {Cresci}, {Schramm},
  {Delvecchio}, {Lanzuisi}, {Mainieri}, {Mignoli}, {Zamorani}, {Berta},
  {Bongiorno}, {Comastri}, {Fiore}, {Kakkad}, {Marconi}, {Rosario}, {Contini},
  \& {Lamareille}}]{Brusa+16}
{Brusa}, M., {Perna}, M., {Cresci}, G., {et~al.} 2016, \aap, 588, A58

\bibitem[{{Bundy} {et~al.}(2006){Bundy}, {Ellis}, {Conselice}, {Taylor},
  {Cooper}, {Willmer}, {Weiner}, {Coil}, {Noeske}, \& {Eisenhardt}}]{Bundy+06}
{Bundy}, K., {Ellis}, R.~S., {Conselice}, C.~J., {et~al.} 2006, \apj, 651, 120

\bibitem[{{Capak} {et~al.}(2007){Capak}, {Aussel}, {Ajiki}, {McCracken},
  {Mobasher}, {Scoville}, {Shopbell}, {Taniguchi}, {Thompson}, {Tribiano},
  {Sasaki}, {Blain}, {Brusa}, {Carilli}, {Comastri}, {Carollo}, {Cassata},
  {Colbert}, {Ellis}, {Elvis}, {Giavalisco}, {Green}, {Guzzo}, {Hasinger},
  {Ilbert}, {Impey}, {Jahnke}, {Kartaltepe}, {Kneib}, {Koda}, {Koekemoer},
  {Komiyama}, {Leauthaud}, {Le Fevre}, {Lilly}, {Liu}, {Massey}, {Miyazaki},
  {Murayama}, {Nagao}, {Peacock}, {Pickles}, {Porciani}, {Renzini}, {Rhodes},
  {Rich}, {Salvato}, {Sanders}, {Scarlata}, {Schiminovich}, {Schinnerer},
  {Scodeggio}, {Sheth}, {Shioya}, {Tasca}, {Taylor}, {Yan}, \&
  {Zamorani}}]{Capak+07}
{Capak}, P., {Aussel}, H., {Ajiki}, M., {et~al.} 2007, \apjs, 172, 99

\bibitem[{{Casey} {et~al.}(2013){Casey}, {Chen}, {Cowie}, {Barger}, {Capak},
  {Ilbert}, {Koss}, {Lee}, {Le Floc'h}, {Sanders}, \& {Williams}}]{Casey+13}
{Casey}, C.~M., {Chen}, C.-C., {Cowie}, L.~L., {et~al.} 2013, \mnras, 436, 1919

\bibitem[{{Chabrier}(2003)}]{Chabrier03}
{Chabrier}, G. 2003, \pasp, 115, 763

\bibitem[{{Chary} \& {Elbaz}(2001)}]{Chary+01}
{Chary}, R. \& {Elbaz}, D. 2001, \apj, 556, 562

\bibitem[{{Chi} {et~al.}(2013){Chi}, {Barthel}, \& {Garrett}}]{Chi+13}
{Chi}, S., {Barthel}, P.~D., \& {Garrett}, M.~A. 2013, \aap, 550, A68

\bibitem[{{Cicone} {et~al.}(2014){Cicone}, {Maiolino}, {Sturm},
  {Graci{\'a}-Carpio}, {Feruglio}, {Neri}, {Aalto}, {Davies}, {Fiore},
  {Fischer}, {Garc{\'{\i}}a-Burillo}, {Gonz{\'a}lez-Alfonso},
  {Hailey-Dunsheath}, {Piconcelli}, \& {Veilleux}}]{Cicone+14}
{Cicone}, C., {Maiolino}, R., {Sturm}, E., {et~al.} 2014, \aap, 562, A21

\bibitem[{{Civano} {et~al.}(2012){Civano}, {Elvis}, {Brusa}, {Comastri},
  {Salvato}, {Zamorani}, {Aldcroft}, {Bongiorno}, {Capak}, {Cappelluti},
  {Cisternas}, {Fiore}, {Fruscione}, {Hao}, {Kartaltepe}, {Koekemoer}, {Gilli},
  {Impey}, {Lanzuisi}, {Lusso}, {Mainieri}, {Miyaji}, {Lilly}, {Masters},
  {Puccetti}, {Schawinski}, {Scoville}, {Silverman}, {Trump}, {Urry},
  {Vignali}, \& {Wright}}]{Civano+12}
{Civano}, F., {Elvis}, M., {Brusa}, M., {et~al.} 2012, \apjs, 201, 30

\bibitem[{{Civano} {et~al.}(2016){Civano}, {Marchesi}, {Comastri}, {Urry},
  {Elvis}, {Cappelluti}, {Puccetti}, {Brusa}, {Zamorani}, {Hasinger},
  {Aldcroft}, {Alexander}, {Allevato}, {Brunner}, {Capak}, {Finoguenov},
  {Fiore}, {Fruscione}, {Gilli}, {Glotfelty}, {Griffiths}, {Hao}, {Harrison},
  {Jahnke}, {Kartaltepe}, {Karim}, {LaMassa}, {Lanzuisi}, {Miyaji}, {Ranalli},
  {Salvato}, {Sargent}, {Scoville}, {Schawinski}, {Schinnerer}, {Silverman},
  {Smolcic}, {Stern}, {Toft}, {Trakhtenbrot}, {Treister}, \&
  {Vignali}}]{Civano+16}
{Civano}, F., {Marchesi}, S., {Comastri}, A., {et~al.} 2016, \apj, 819, 62

\bibitem[{{Coil} {et~al.}(2009){Coil}, {Georgakakis}, {Newman}, {Cooper},
  {Croton}, {Davis}, {Koo}, {Laird}, {Nandra}, {Weiner}, {Willmer}, \&
  {Yan}}]{Coil+09}
{Coil}, A.~L., {Georgakakis}, A., {Newman}, J.~A., {et~al.} 2009, \apj, 701,
  1484

\bibitem[{{Combes} {et~al.}(2013){Combes}, {Garc{\'{\i}}a-Burillo}, {Casasola},
  {Hunt}, {Krips}, {Baker}, {Boone}, {Eckart}, {Marquez}, {Neri}, {Schinnerer},
  \& {Tacconi}}]{Combes+13}
{Combes}, F., {Garc{\'{\i}}a-Burillo}, S., {Casasola}, V., {et~al.} 2013, \aap,
  558, A124

\bibitem[{{Comparat} {et~al.}(2015){Comparat}, {Richard}, {Kneib}, {Ilbert},
  {Gonzalez-Perez}, {Tresse}, {Zoubian}, {Arnouts}, {Brownstein}, {Baugh},
  {Delubac}, {Ealet}, {Escoffier}, {Ge}, {Jullo}, {Lacey}, {Ross}, {Schlegel},
  {Schneider}, {Steele}, {Tasca}, {Yeche}, {Lesser}, {Jiang}, {Jing}, {Fan},
  {Fan}, {Ma}, {Nie}, {Wang}, {Wu}, {Zhang}, {Zhou}, {Zhou}, \&
  {Zou}}]{Comparat+15}
{Comparat}, J., {Richard}, J., {Kneib}, J.-P., {et~al.} 2015, \aap, 575, A40

\bibitem[{{Condon}(1984)}]{Condon84}
{Condon}, J.~J. 1984, \apj, 287, 461

\bibitem[{{Condon}(1992)}]{Condon92}
{Condon}, J.~J. 1992, \araa, 30, 575

\bibitem[{{Cowie} {et~al.}(1996){Cowie}, {Songaila}, {Hu}, \&
  {Cohen}}]{Cowie+96}
{Cowie}, L.~L., {Songaila}, A., {Hu}, E.~M., \& {Cohen}, J.~G. 1996, \aj, 112,
  839

\bibitem[{{Croton} {et~al.}(2006){Croton}, {Springel}, {White}, {De Lucia},
  {Frenk}, {Gao}, {Jenkins}, {Kauffmann}, {Navarro}, \& {Yoshida}}]{Croton+06}
{Croton}, D.~J., {Springel}, V., {White}, S.~D.~M., {et~al.} 2006, \mnras, 365,
  11

\bibitem[{{da Cunha} {et~al.}(2008){da Cunha}, {Charlot}, \&
  {Elbaz}}]{daCunha+08}
{da Cunha}, E., {Charlot}, S., \& {Elbaz}, D. 2008, \mnras, 388, 1595

\bibitem[{{da Cunha} {et~al.}(2015){da Cunha}, {Walter}, {Smail}, {Swinbank},
  {Simpson}, {Decarli}, {Hodge}, {Weiss}, {van der Werf}, {Bertoldi},
  {Chapman}, {Cox}, {Danielson}, {Dannerbauer}, {Greve}, {Ivison}, {Karim}, \&
  {Thomson}}]{dacunha+15}
{da Cunha}, E., {Walter}, F., {Smail}, I.~R., {et~al.} 2015, \apj, 806, 110

\bibitem[{{Daddi} {et~al.}(2010){Daddi}, {Bournaud}, {Walter}, {Dannerbauer},
  {Carilli}, {Dickinson}, {Elbaz}, {Morrison}, {Riechers}, {Onodera}, {Salmi},
  {Krips}, \& {Stern}}]{Daddi+10}
{Daddi}, E., {Bournaud}, F., {Walter}, F., {et~al.} 2010, \apj, 713, 686

\bibitem[{{Dale} \& {Helou}(2002)}]{Dale+02}
{Dale}, D.~A. \& {Helou}, G. 2002, \apj, 576, 159

\bibitem[{{Daly} {et~al.}(2012){Daly}, {Sprinkle}, {O'Dea}, {Kharb}, \&
  {Baum}}]{Daly+12}
{Daly}, R.~A., {Sprinkle}, T.~B., {O'Dea}, C.~P., {Kharb}, P., \& {Baum}, S.~A.
  2012, \mnras, 423, 2498

\bibitem[{{Davidzon} {et~al.}(2017){Davidzon}, {Ilbert}, {Laigle}, {Coupon},
  {McCracken}, {Delvecchio}, {Masters}, {Capak}, {Hsieh}, {Tresse}, {Le Fevre},
  {Bethermin}, {Chang}, {Faisst}, {Le Floc'h}, {Steinhardt}, {Toft}, {Aussel},
  {Dubois}, {Hasinger}, {Salvato}, {Sanders}, {Scoville}, \&
  {Silverman}}]{Davidzon+17}
{Davidzon}, I., {Ilbert}, O., {Laigle}, C., {et~al.} 2017, ArXiv e-prints
  [\eprint[arXiv]{1701.02734}]

\bibitem[{{de Vries} {et~al.}(2002){de Vries}, {Morganti}, {R{\"o}ttgering},
  {Vermeulen}, {van Breugel}, {Rengelink}, \& {Jarvis}}]{deVries+02}
{de Vries}, W.~H., {Morganti}, R., {R{\"o}ttgering}, H.~J.~A., {et~al.} 2002,
  \aj, 123, 1784

\bibitem[{{Del Moro} {et~al.}(2013){Del Moro}, {Alexander}, {Mullaney},
  {Daddi}, {Pannella}, {Bauer}, {Pope}, {Dickinson}, {Elbaz}, {Barthel},
  {Garrett}, {Brandt}, {Charmandaris}, {Chary}, {Dasyra}, {Gilli}, {Hickox},
  {Hwang}, {Ivison}, {Juneau}, {Le Floc'h}, {Luo}, {Morrison}, {Rovilos},
  {Sargent}, \& {Xue}}]{DelMoro+13}
{Del Moro}, A., {Alexander}, D.~M., {Mullaney}, J.~R., {et~al.} 2013, \aap,
  549, A59

\bibitem[{{Delhaize} {et~al.}(2017){Delhaize}, {Smolcic}, {Delvecchio},
  {Novak}, {Sargent}, {Baran}, {Magnelli}, {Zamorani}, {Schinnerer}, {Murphy},
  {Aravena}, {Berta}, {Bondi}, {Capak}, {Ciliegi}, {Civano}, {Ilbert}, {Karim},
  {Laigle}, {Le Fevre}, {Marchesi}, {McCracken}, {Salvato}, {Seymour}, \&
  {Tasca}}]{Delhaize+17}
{Delhaize}, J., {Smolcic}, V., {Delvecchio}, I., {et~al.} 2017, ArXiv e-prints
  [\eprint[arXiv]{1703.09723}]

\bibitem[{{Delvecchio} {et~al.}(2014){Delvecchio}, {Gruppioni}, {Pozzi},
  {Berta}, {Zamorani}, {Cimatti}, {Lutz}, {Scott}, {Vignali}, {Cresci},
  {Feltre}, {Cooray}, {Vaccari}, {Fritz}, {Le Floc'h}, {Magnelli}, {Popesso},
  {Oliver}, {Bock}, {Carollo}, {Contini}, {Le F{\'e}vre}, {Lilly}, {Mainieri},
  {Renzini}, \& {Scodeggio}}]{Delvecchio+14}
{Delvecchio}, I., {Gruppioni}, C., {Pozzi}, F., {et~al.} 2014, \mnras, 439,
  2736

\bibitem[{{Di Matteo} {et~al.}(2008){Di Matteo}, {Colberg}, {Springel},
  {Hernquist}, \& {Sijacki}}]{DiMatteo+08}
{Di Matteo}, T., {Colberg}, J., {Springel}, V., {Hernquist}, L., \& {Sijacki},
  D. 2008, \apj, 676, 33

\bibitem[{{Donley} {et~al.}(2012){Donley}, {Koekemoer}, {Brusa}, {Capak},
  {Cardamone}, {Civano}, {Ilbert}, {Impey}, {Kartaltepe}, {Miyaji}, {Salvato},
  {Sanders}, {Trump}, \& {Zamorani}}]{Donley+12}
{Donley}, J.~L., {Koekemoer}, A.~M., {Brusa}, M., {et~al.} 2012, \apj, 748, 142

\bibitem[{{Donley} {et~al.}(2005){Donley}, {Rieke}, {Rigby}, \&
  {P{\'e}rez-Gonz{\'a}lez}}]{Donley+05}
{Donley}, J.~L., {Rieke}, G.~H., {Rigby}, J.~R., \& {P{\'e}rez-Gonz{\'a}lez},
  P.~G. 2005, \apj, 634, 169

\bibitem[{{Dubois} {et~al.}(2014){Dubois}, {Volonteri}, \& {Silk}}]{Dubois+14}
{Dubois}, Y., {Volonteri}, M., \& {Silk}, J. 2014, \mnras, 440, 1590

\bibitem[{{Elbaz} {et~al.}(2011){Elbaz}, {Dickinson}, {Hwang},
  {D{\'{\i}}az-Santos}, {Magdis}, {Magnelli}, {Le Borgne}, {Galliano},
  {Pannella}, {Chanial}, {Armus}, {Charmandaris}, {Daddi}, {Aussel}, {Popesso},
  {Kartaltepe}, {Altieri}, {Valtchanov}, {Coia}, {Dannerbauer}, {Dasyra},
  {Leiton}, {Mazzarella}, {Alexander}, {Buat}, {Burgarella}, {Chary}, {Gilli},
  {Ivison}, {Juneau}, {Le Floc'h}, {Lutz}, {Morrison}, {Mullaney}, {Murphy},
  {Pope}, {Scott}, {Brodwin}, {Calzetti}, {Cesarsky}, {Charlot}, {Dole},
  {Eisenhardt}, {Ferguson}, {F{\"o}rster Schreiber}, {Frayer}, {Giavalisco},
  {Huynh}, {Koekemoer}, {Papovich}, {Reddy}, {Surace}, {Teplitz}, {Yun}, \&
  {Wilson}}]{Elbaz+11}
{Elbaz}, D., {Dickinson}, M., {Hwang}, H.~S., {et~al.} 2011, \aap, 533, A119

\bibitem[{{Ellison} {et~al.}(2015){Ellison}, {Patton}, \&
  {Hickox}}]{Ellison+15}
{Ellison}, S.~L., {Patton}, D.~R., \& {Hickox}, R.~C. 2015, \mnras, 451, L35

\bibitem[{{Elvis} {et~al.}(2009){Elvis}, {Civano}, {Vignali}, {Puccetti},
  {Fiore}, {Cappelluti}, {Aldcroft}, {Fruscione}, {Zamorani}, {Comastri},
  {Brusa}, {Gilli}, {Miyaji}, {Damiani}, {Koekemoer}, {Finoguenov}, {Brunner},
  {Urry}, {Silverman}, {Mainieri}, {Hasinger}, {Griffiths}, {Carollo}, {Hao},
  {Guzzo}, {Blain}, {Calzetti}, {Carilli}, {Capak}, {Ettori}, {Fabbiano},
  {Impey}, {Lilly}, {Mobasher}, {Rich}, {Salvato}, {Sanders}, {Schinnerer},
  {Scoville}, {Shopbell}, {Taylor}, {Taniguchi}, \& {Volonteri}}]{Elvis+09}
{Elvis}, M., {Civano}, F., {Vignali}, C., {et~al.} 2009, \apjs, 184, 158

\bibitem[{{Fanidakis} {et~al.}(2011){Fanidakis}, {Baugh}, {Benson}, {Bower},
  {Cole}, {Done}, \& {Frenk}}]{Fanidakis+11}
{Fanidakis}, N., {Baugh}, C.~M., {Benson}, A.~J., {et~al.} 2011, \mnras, 410,
  53

\bibitem[{{Fanidakis} {et~al.}(2012){Fanidakis}, {Baugh}, {Benson}, {Bower},
  {Cole}, {Done}, {Frenk}, {Hickox}, {Lacey}, \& {Del P.~Lagos}}]{Fanidakis+12}
{Fanidakis}, N., {Baugh}, C.~M., {Benson}, A.~J., {et~al.} 2012, \mnras, 419,
  2797

\bibitem[{{Farrah} {et~al.}(2012){Farrah}, {Urrutia}, {Lacy}, {Efstathiou},
  {Afonso}, {Coppin}, {Hall}, {Lonsdale}, {Jarrett}, {Bridge}, {Borys}, \&
  {Petty}}]{Farrah+12}
{Farrah}, D., {Urrutia}, T., {Lacy}, M., {et~al.} 2012, \apj, 745, 178

\bibitem[{{Feltre} {et~al.}(2012){Feltre}, {Hatziminaoglou}, {Fritz}, \&
  {Franceschini}}]{Feltre+12}
{Feltre}, A., {Hatziminaoglou}, E., {Fritz}, J., \& {Franceschini}, A. 2012,
  \mnras, 426, 120

\bibitem[{{Ferrarese}(2002)}]{Ferrarese+02}
{Ferrarese}, L. 2002, \apj, 578, 90

\bibitem[{{Feruglio} {et~al.}(2010){Feruglio}, {Maiolino}, {Piconcelli},
  {Menci}, {Aussel}, {Lamastra}, \& {Fiore}}]{Feruglio+10}
{Feruglio}, C., {Maiolino}, R., {Piconcelli}, E., {et~al.} 2010, \aap, 518,
  L155

\bibitem[{{Fontanot} {et~al.}(2009){Fontanot}, {De Lucia}, {Monaco},
  {Somerville}, \& {Santini}}]{Fontanot+09}
{Fontanot}, F., {De Lucia}, G., {Monaco}, P., {Somerville}, R.~S., \&
  {Santini}, P. 2009, \mnras, 397, 1776

\bibitem[{{Fritz} {et~al.}(2006){Fritz}, {Franceschini}, \&
  {Hatziminaoglou}}]{Fritz+06}
{Fritz}, J., {Franceschini}, A., \& {Hatziminaoglou}, E. 2006, \mnras, 366, 767

\bibitem[{{Gebhardt} {et~al.}(2000){Gebhardt}, {Bender}, {Bower}, {Dressler},
  {Faber}, {Filippenko}, {Green}, {Grillmair}, {Ho}, {Kormendy}, {Lauer},
  {Magorrian}, {Pinkney}, {Richstone}, \& {Tremaine}}]{Gebhardt+00}
{Gebhardt}, K., {Bender}, R., {Bower}, G., {et~al.} 2000, \apjl, 539, L13

\bibitem[{{Georgakakis} {et~al.}(2007){Georgakakis}, {Nandra}, {Laird},
  {Cooper}, {Gerke}, {Newman}, {Croton}, {Davis}, {Faber}, \&
  {Coil}}]{Georgakakis+07}
{Georgakakis}, A., {Nandra}, K., {Laird}, E.~S., {et~al.} 2007, \apjl, 660, L15

\bibitem[{{Godfrey} \& {Shabala}(2016)}]{Godfrey+16}
{Godfrey}, L.~E.~H. \& {Shabala}, S.~S. 2016, \mnras, 456, 1172

\bibitem[{{Goulding} {et~al.}(2014){Goulding}, {Forman}, {Hickox}, {Jones},
  {Murray}, {Paggi}, {Ashby}, {Coil}, {Cooper}, {Huang}, {Kraft}, {Newman},
  {Weiner}, \& {Willner}}]{Goulding+14}
{Goulding}, A.~D., {Forman}, W.~R., {Hickox}, R.~C., {et~al.} 2014, \apj, 783,
  40

\bibitem[{{Griffin} {et~al.}(2010){Griffin}, {Abergel}, {Abreu}, {Ade},
  {Andr{\'e}}, {Augueres}, {Babbedge}, {Bae}, {Baillie}, {Baluteau}, {Barlow},
  {Bendo}, {Benielli}, {Bock}, {Bonhomme}, {Brisbin}, {Brockley-Blatt},
  {Caldwell}, {Cara}, {Castro-Rodriguez}, {Cerulli}, {Chanial}, {Chen},
  {Clark}, {Clements}, {Clerc}, {Coker}, {Communal}, {Conversi}, {Cox},
  {Crumb}, {Cunningham}, {Daly}, {Davis}, {de Antoni}, {Delderfield}, {Devin},
  {di Giorgio}, {Didschuns}, {Dohlen}, {Donati}, {Dowell}, {Dowell}, {Duband},
  {Dumaye}, {Emery}, {Ferlet}, {Ferrand}, {Fontignie}, {Fox}, {Franceschini},
  {Frerking}, {Fulton}, {Garcia}, {Gastaud}, {Gear}, {Glenn}, {Goizel},
  {Griffin}, {Grundy}, {Guest}, {Guillemet}, {Hargrave}, {Harwit}, {Hastings},
  {Hatziminaoglou}, {Herman}, {Hinde}, {Hristov}, {Huang}, {Imhof}, {Isaak},
  {Israelsson}, {Ivison}, {Jennings}, {Kiernan}, {King}, {Lange}, {Latter},
  {Laurent}, {Laurent}, {Leeks}, {Lellouch}, {Levenson}, {Li}, {Li},
  {Lilienthal}, {Lim}, {Liu}, {Lu}, {Madden}, {Mainetti}, {Marliani}, {McKay},
  {Mercier}, {Molinari}, {Morris}, {Moseley}, {Mulder}, {Mur}, {Naylor},
  {Nguyen}, {O'Halloran}, {Oliver}, {Olofsson}, {Olofsson}, {Orfei}, {Page},
  {Pain}, {Panuzzo}, {Papageorgiou}, {Parks}, {Parr-Burman}, {Pearce},
  {Pearson}, {P{\'e}rez-Fournon}, {Pinsard}, {Pisano}, {Podosek}, {Pohlen},
  {Polehampton}, {Pouliquen}, {Rigopoulou}, {Rizzo}, {Roseboom}, {Roussel},
  {Rowan-Robinson}, {Rownd}, {Saraceno}, {Sauvage}, {Savage}, {Savini},
  {Sawyer}, {Scharmberg}, {Schmitt}, {Schneider}, {Schulz}, {Schwartz},
  {Shafer}, {Shupe}, {Sibthorpe}, {Sidher}, {Smith}, {Smith}, {Smith},
  {Spencer}, {Stobie}, {Sudiwala}, {Sukhatme}, {Surace}, {Stevens}, {Swinyard},
  {Trichas}, {Tourette}, {Triou}, {Tseng}, {Tucker}, {Turner}, {Vaccari},
  {Valtchanov}, {Vigroux}, {Virique}, {Voellmer}, {Walker}, {Ward}, {Waskett},
  {Weilert}, {Wesson}, {White}, {Whitehouse}, {Wilson}, {Winter}, {Woodcraft},
  {Wright}, {Xu}, {Zavagno}, {Zemcov}, {Zhang}, \& {Zonca}}]{Griffin+10}
{Griffin}, M.~J., {Abergel}, A., {Abreu}, A., {et~al.} 2010, \aap, 518, L3

\bibitem[{{Groves} {et~al.}(2012){Groves}, {Krause}, {Sandstrom}, {Schmiedeke},
  {Leroy}, {Linz}, {Kapala}, {Rix}, {Schinnerer}, {Tabatabaei}, {Walter}, \&
  {da Cunha}}]{Groves+12}
{Groves}, B., {Krause}, O., {Sandstrom}, K., {et~al.} 2012, \mnras, 426, 892

\bibitem[{{Gruppioni} {et~al.}(1999){Gruppioni}, {Mignoli}, \&
  {Zamorani}}]{Gruppioni+99}
{Gruppioni}, C., {Mignoli}, M., \& {Zamorani}, G. 1999, \mnras, 304, 199

\bibitem[{{G{\"u}ltekin} {et~al.}(2009){G{\"u}ltekin}, {Cackett}, {Miller}, {Di
  Matteo}, {Markoff}, \& {Richstone}}]{Gultekin+09}
{G{\"u}ltekin}, K., {Cackett}, E.~M., {Miller}, J.~M., {et~al.} 2009, \apj,
  706, 404

\bibitem[{{Hardcastle} {et~al.}(2006){Hardcastle}, {Evans}, \&
  {Croston}}]{Hardcastle+06}
{Hardcastle}, M.~J., {Evans}, D.~A., \& {Croston}, J.~H. 2006, \mnras, 370,
  1893

\bibitem[{{Hardcastle} {et~al.}(2007){Hardcastle}, {Evans}, \&
  {Croston}}]{Hardcastle+07}
{Hardcastle}, M.~J., {Evans}, D.~A., \& {Croston}, J.~H. 2007, \mnras, 376,
  1849

\bibitem[{{Hasinger} {et~al.}(2005){Hasinger}, {Miyaji}, \&
  {Schmidt}}]{Hasinger+05}
{Hasinger}, G., {Miyaji}, T., \& {Schmidt}, M. 2005, \aap, 441, 417

\bibitem[{{Hayward} \& {Smith}(2015)}]{Hayward+15}
{Hayward}, C.~C. \& {Smith}, D.~J.~B. 2015, \mnras, 446, 1512

\bibitem[{{Heckman} \& {Best}(2014)}]{Heckman+14}
{Heckman}, T.~M. \& {Best}, P.~N. 2014, \araa, 52, 589

\bibitem[{{Herrera Ruiz} {et~al.}(2016){Herrera Ruiz}, {Middelberg}, {Norris},
  \& {Maini}}]{HerreraRuiz+16}
{Herrera Ruiz}, N., {Middelberg}, E., {Norris}, R.~P., \& {Maini}, A. 2016,
  \aap, 589, L2

\bibitem[{{Hickox} {et~al.}(2009){Hickox}, {Jones}, {Forman}, {Murray},
  {Kochanek}, {Eisenstein}, {Jannuzi}, {Dey}, {Brown}, {Stern}, {Eisenhardt},
  {Gorjian}, {Brodwin}, {Narayan}, {Cool}, {Kenter}, {Caldwell}, \&
  {Anderson}}]{Hickox+09}
{Hickox}, R.~C., {Jones}, C., {Forman}, W.~R., {et~al.} 2009, \apj, 696, 891

\bibitem[{{Hogan} {et~al.}(2015){Hogan}, {Edge}, {Hlavacek-Larrondo},
  {Grainge}, {Hamer}, {Mahony}, {Russell}, {Fabian}, {McNamara}, \&
  {Wilman}}]{Hogan+15}
{Hogan}, M.~T., {Edge}, A.~C., {Hlavacek-Larrondo}, J., {et~al.} 2015, \mnras,
  453, 1201

\bibitem[{{Hopkins} {et~al.}(2008){Hopkins}, {Hernquist}, {Cox}, \& {Kere{\v
  s}}}]{Hopkins+08}
{Hopkins}, P.~F., {Hernquist}, L., {Cox}, T.~J., \& {Kere{\v s}}, D. 2008,
  \apjs, 175, 356

\bibitem[{{Hopkins} \& {Quataert}(2010)}]{Hopkins+10}
{Hopkins}, P.~F. \& {Quataert}, E. 2010, \mnras, 407, 1529

\bibitem[{{Ilbert} {et~al.}(2006){Ilbert}, {Arnouts}, {McCracken},
  {Bolzonella}, {Bertin}, {Le F{\`e}vre}, {Mellier}, {Zamorani}, {Pell{\`o}},
  {Iovino}, {Tresse}, {Le Brun}, {Bottini}, {Garilli}, {Maccagni}, {Picat},
  {Scaramella}, {Scodeggio}, {Vettolani}, {Zanichelli}, {Adami}, {Bardelli},
  {Cappi}, {Charlot}, {Ciliegi}, {Contini}, {Cucciati}, {Foucaud}, {Franzetti},
  {Gavignaud}, {Guzzo}, {Marano}, {Marinoni}, {Mazure}, {Meneux}, {Merighi},
  {Paltani}, {Pollo}, {Pozzetti}, {Radovich}, {Zucca}, {Bondi}, {Bongiorno},
  {Busarello}, {de La Torre}, {Gregorini}, {Lamareille}, {Mathez}, {Merluzzi},
  {Ripepi}, {Rizzo}, \& {Vergani}}]{Ilbert+06}
{Ilbert}, O., {Arnouts}, S., {McCracken}, H.~J., {et~al.} 2006, \aap, 457, 841

\bibitem[{{Ilbert} {et~al.}(2009){Ilbert}, {Capak}, {Salvato}, {Aussel},
  {McCracken}, {Sanders}, {Scoville}, {Kartaltepe}, {Arnouts}, {Le Floc'h},
  {Mobasher}, {Taniguchi}, {Lamareille}, {Leauthaud}, {Sasaki}, {Thompson},
  {Zamojski}, {Zamorani}, {Bardelli}, {Bolzonella}, {Bongiorno}, {Brusa},
  {Caputi}, {Carollo}, {Contini}, {Cook}, {Coppa}, {Cucciati}, {de la Torre},
  {de Ravel}, {Franzetti}, {Garilli}, {Hasinger}, {Iovino}, {Kampczyk},
  {Kneib}, {Knobel}, {Kovac}, {Le Borgne}, {Le Brun}, {F{\`e}vre}, {Lilly},
  {Looper}, {Maier}, {Mainieri}, {Mellier}, {Mignoli}, {Murayama}, {Pell{\`o}},
  {Peng}, {P{\'e}rez-Montero}, {Renzini}, {Ricciardelli}, {Schiminovich},
  {Scodeggio}, {Shioya}, {Silverman}, {Surace}, {Tanaka}, {Tasca}, {Tresse},
  {Vergani}, \& {Zucca}}]{Ilbert+09}
{Ilbert}, O., {Capak}, P., {Salvato}, M., {et~al.} 2009, \apj, 690, 1236

\bibitem[{{Ilbert} {et~al.}(2013){Ilbert}, {McCracken}, {Le F{\`e}vre},
  {Capak}, {Dunlop}, {Karim}, {Renzini}, {Caputi}, {Boissier}, {Arnouts},
  {Aussel}, {Comparat}, {Guo}, {Hudelot}, {Kartaltepe}, {Kneib}, {Krogager},
  {Le Floc'h}, {Lilly}, {Mellier}, {Milvang-Jensen}, {Moutard}, {Onodera},
  {Richard}, {Salvato}, {Sanders}, {Scoville}, {Silverman}, {Taniguchi},
  {Tasca}, {Thomas}, {Toft}, {Tresse}, {Vergani}, {Wolk}, \&
  {Zirm}}]{Ilbert+13}
{Ilbert}, O., {McCracken}, H.~J., {Le F{\`e}vre}, O., {et~al.} 2013, \aap, 556,
  A55

\bibitem[{{Ilbert} {et~al.}(2010){Ilbert}, {Salvato}, {Le Floc'h}, {Aussel},
  {Capak}, {McCracken}, {Mobasher}, {Kartaltepe}, {Scoville}, {Sanders},
  {Arnouts}, {Bundy}, {Cassata}, {Kneib}, {Koekemoer}, {Le F{\`e}vre}, {Lilly},
  {Surace}, {Taniguchi}, {Tasca}, {Thompson}, {Tresse}, {Zamojski}, {Zamorani},
  \& {Zucca}}]{Ilbert+10}
{Ilbert}, O., {Salvato}, M., {Le Floc'h}, E., {et~al.} 2010, \apj, 709, 644

\bibitem[{{Jannuzi} \& {Dey}(1999)}]{Jannuzi+99}
{Jannuzi}, B.~T. \& {Dey}, A. 1999, in Astronomical Society of the Pacific
  Conference Series, Vol. 191, Photometric Redshifts and the Detection of High
  Redshift Galaxies, ed. R.~{Weymann}, L.~{Storrie-Lombardi}, M.~{Sawicki}, \&
  R.~{Brunner}, 111

\bibitem[{{Kellermann} \& {Pauliny-Toth}(1969)}]{kellermann+69}
{Kellermann}, K.~I. \& {Pauliny-Toth}, I.~I.~K. 1969, \apjl, 155, L71

\bibitem[{{Kennicutt}(1998)}]{Kennicutt98}
{Kennicutt}, Jr., R.~C. 1998, \apj, 498, 541

\bibitem[{{Kimball} {et~al.}(2011){Kimball}, {Kellermann}, {Condon},
  {Ivezi{\'c}}, \& {Perley}}]{Kimball+11}
{Kimball}, A.~E., {Kellermann}, K.~I., {Condon}, J.~J., {Ivezi{\'c}}, {\v Z}.,
  \& {Perley}, R.~A. 2011, \apjl, 739, L29

\bibitem[{{La Franca} {et~al.}(2010){La Franca}, {Melini}, \&
  {Fiore}}]{LaFranca+10}
{La Franca}, F., {Melini}, G., \& {Fiore}, F. 2010, \apj, 718, 368

\bibitem[{{Lagos} {et~al.}(2008){Lagos}, {Cora}, \& {Padilla}}]{Lagos+08}
{Lagos}, C.~D.~P., {Cora}, S.~A., \& {Padilla}, N.~D. 2008, \mnras, 388, 587

\bibitem[{{Laigle} {et~al.}(2016){Laigle}, {McCracken}, {Ilbert}, {Hsieh},
  {Davidzon}, {Capak}, {Hasinger}, {Silverman}, {Pichon}, {Coupon}, {Aussel},
  {Le Borgne}, {Caputi}, {Cassata}, {Chang}, {Civano}, {Dunlop}, {Fynbo},
  {Kartaltepe}, {Koekemoer}, {Le F{\`e}vre}, {Le Floc'h}, {Leauthaud}, {Lilly},
  {Lin}, {Marchesi}, {Milvang-Jensen}, {Salvato}, {Sanders}, {Scoville},
  {Smolcic}, {Stockmann}, {Taniguchi}, {Tasca}, {Toft}, {Vaccari}, \&
  {Zabl}}]{Laigle+16}
{Laigle}, C., {McCracken}, H.~J., {Ilbert}, O., {et~al.} 2016, \apjs, 224, 24

\bibitem[{{Lanzuisi} {et~al.}(2015){Lanzuisi}, {Perna}, {Delvecchio}, {Berta},
  {Brusa}, {Cappelluti}, {Comastri}, {Gilli}, {Gruppioni}, {Mignoli}, {Pozzi},
  {Vietri}, {Vignali}, \& {Zamorani}}]{Lanzuisi+15}
{Lanzuisi}, G., {Perna}, M., {Delvecchio}, I., {et~al.} 2015, \aap, 578, A120

\bibitem[{{Le F{\`e}vre} {et~al.}(2015){Le F{\`e}vre}, {Tasca}, {Cassata},
  {Garilli}, {Le Brun}, {Maccagni}, {Pentericci}, {Thomas}, {Vanzella},
  {Zamorani}, {Zucca}, {Amorin}, {Bardelli}, {Capak}, {Cassar{\`a}},
  {Castellano}, {Cimatti}, {Cuby}, {Cucciati}, {de la Torre}, {Durkalec},
  {Fontana}, {Giavalisco}, {Grazian}, {Hathi}, {Ilbert}, {Lemaux}, {Moreau},
  {Paltani}, {Ribeiro}, {Salvato}, {Schaerer}, {Scodeggio}, {Sommariva},
  {Talia}, {Taniguchi}, {Tresse}, {Vergani}, {Wang}, {Charlot}, {Contini},
  {Fotopoulou}, {L{\'o}pez-Sanjuan}, {Mellier}, \& {Scoville}}]{lefevre+15}
{Le F{\`e}vre}, O., {Tasca}, L.~A.~M., {Cassata}, P., {et~al.} 2015, \aap, 576,
  A79

\bibitem[{{Le Floc'h} {et~al.}(2009){Le Floc'h}, {Aussel}, {Ilbert},
  {Riguccini}, {Frayer}, {Salvato}, {Arnouts}, {Surace}, {Feruglio},
  {Rodighiero}, {Capak}, {Kartaltepe}, {Heinis}, {Sheth}, {Yan}, {McCracken},
  {Thompson}, {Sanders}, {Scoville}, \& {Koekemoer}}]{LeFloch+09}
{Le Floc'h}, E., {Aussel}, H., {Ilbert}, O., {et~al.} 2009, \apj, 703, 222

\bibitem[{{Lilly} {et~al.}(2009){Lilly}, {Le Brun}, {Maier}, {Mainieri},
  {Mignoli}, {Scodeggio}, {Zamorani}, {Carollo}, {Contini}, {Kneib}, {Le
  F{\`e}vre}, {Renzini}, {Bardelli}, {Bolzonella}, {Bongiorno}, {Caputi},
  {Coppa}, {Cucciati}, {de la Torre}, {de Ravel}, {Franzetti}, {Garilli},
  {Iovino}, {Kampczyk}, {Kovac}, {Knobel}, {Lamareille}, {Le Borgne}, {Pello},
  {Peng}, {P{\'e}rez-Montero}, {Ricciardelli}, {Silverman}, {Tanaka}, {Tasca},
  {Tresse}, {Vergani}, {Zucca}, {Ilbert}, {Salvato}, {Oesch}, {Abbas},
  {Bottini}, {Capak}, {Cappi}, {Cassata}, {Cimatti}, {Elvis}, {Fumana},
  {Guzzo}, {Hasinger}, {Koekemoer}, {Leauthaud}, {Maccagni}, {Marinoni},
  {McCracken}, {Memeo}, {Meneux}, {Porciani}, {Pozzetti}, {Sanders},
  {Scaramella}, {Scarlata}, {Scoville}, {Shopbell}, \& {Taniguchi}}]{Lilly+09}
{Lilly}, S.~J., {Le Brun}, V., {Maier}, C., {et~al.} 2009, \apjs, 184, 218

\bibitem[{{Lilly} {et~al.}(2007){Lilly}, {Le F{\`e}vre}, {Renzini}, {Zamorani},
  {Scodeggio}, {Contini}, {Carollo}, {Hasinger}, {Kneib}, {Iovino}, {Le Brun},
  {Maier}, {Mainieri}, {Mignoli}, {Silverman}, {Tasca}, {Bolzonella},
  {Bongiorno}, {Bottini}, {Capak}, {Caputi}, {Cimatti}, {Cucciati}, {Daddi},
  {Feldmann}, {Franzetti}, {Garilli}, {Guzzo}, {Ilbert}, {Kampczyk}, {Kovac},
  {Lamareille}, {Leauthaud}, {Borgne}, {McCracken}, {Marinoni}, {Pello},
  {Ricciardelli}, {Scarlata}, {Vergani}, {Sanders}, {Schinnerer}, {Scoville},
  {Taniguchi}, {Arnouts}, {Aussel}, {Bardelli}, {Brusa}, {Cappi}, {Ciliegi},
  {Finoguenov}, {Foucaud}, {Franceschini}, {Halliday}, {Impey}, {Knobel},
  {Koekemoer}, {Kurk}, {Maccagni}, {Maddox}, {Marano}, {Marconi}, {Meneux},
  {Mobasher}, {Moreau}, {Peacock}, {Porciani}, {Pozzetti}, {Scaramella},
  {Schiminovich}, {Shopbell}, {Smail}, {Thompson}, {Tresse}, {Vettolani},
  {Zanichelli}, \& {Zucca}}]{Lilly+07}
{Lilly}, S.~J., {Le F{\`e}vre}, O., {Renzini}, A., {et~al.} 2007, \apjs, 172,
  70

\bibitem[{{Lonsdale} {et~al.}(2015){Lonsdale}, {Lacy}, {Kimball}, {Blain},
  {Whittle}, {Wilkes}, {Stern}, {Condon}, {Kim}, {Assef}, {Tsai}, {Efstathiou},
  {Jones}, {Eisenhardt}, {Bridge}, {Wu}, {Lonsdale}, {Jones}, {Jarrett}, \&
  {Smith}}]{Lonsdale+15}
{Lonsdale}, C.~J., {Lacy}, M., {Kimball}, A.~E., {et~al.} 2015, \apj, 813, 45

\bibitem[{{Lusso} {et~al.}(2012){Lusso}, {Comastri}, {Simmons}, {Mignoli},
  {Zamorani}, {Vignali}, {Brusa}, {Shankar}, {Lutz}, {Trump}, {Maiolino},
  {Gilli}, {Bolzonella}, {Puccetti}, {Salvato}, {Impey}, {Civano}, {Elvis},
  {Mainieri}, {Silverman}, {Koekemoer}, {Bongiorno}, {Merloni}, {Berta}, {Le
  Floc'h}, {Magnelli}, {Pozzi}, \& {Riguccini}}]{Lusso+12}
{Lusso}, E., {Comastri}, A., {Simmons}, B.~D., {et~al.} 2012, \mnras, 425, 623

\bibitem[{{Lusso} {et~al.}(2010){Lusso}, {Comastri}, {Vignali}, {Zamorani},
  {Brusa}, {Gilli}, {Iwasawa}, {Salvato}, {Civano}, {Elvis}, {Merloni},
  {Bongiorno}, {Trump}, {Koekemoer}, {Schinnerer}, {Le Floc'h}, {Cappelluti},
  {Jahnke}, {Sargent}, {Silverman}, {Mainieri}, {Fiore}, {Bolzonella}, {Le
  F{\`e}vre}, {Garilli}, {Iovino}, {Kneib}, {Lamareille}, {Lilly}, {Mignoli},
  {Scodeggio}, \& {Vergani}}]{Lusso+10}
{Lusso}, E., {Comastri}, A., {Vignali}, C., {et~al.} 2010, \aap, 512, A34

\bibitem[{{Lutz} {et~al.}(2011){Lutz}, {Poglitsch}, {Altieri}, {Andreani},
  {Aussel}, {Berta}, {Bongiovanni}, {Brisbin}, {Cava}, {Cepa}, {Cimatti},
  {Daddi}, {Dominguez-Sanchez}, {Elbaz}, {F{\"o}rster Schreiber}, {Genzel},
  {Grazian}, {Gruppioni}, {Harwit}, {Le Floc'h}, {Magdis}, {Magnelli},
  {Maiolino}, {Nordon}, {P{\'e}rez Garc{\'{\i}}a}, {Popesso}, {Pozzi},
  {Riguccini}, {Rodighiero}, {Saintonge}, {Sanchez Portal}, {Santini}, {Shao},
  {Sturm}, {Tacconi}, {Valtchanov}, {Wetzstein}, \& {Wieprecht}}]{Lutz+11}
{Lutz}, D., {Poglitsch}, A., {Altieri}, B., {et~al.} 2011, \aap, 532, A90

\bibitem[{{Madau} \& {Dickinson}(2014)}]{Madau+14}
{Madau}, P. \& {Dickinson}, M. 2014, \araa, 52, 415

\bibitem[{{Magnelli} {et~al.}(2015){Magnelli}, {Ivison}, {Lutz}, {Valtchanov},
  {Farrah}, {Berta}, {Bertoldi}, {Bock}, {Cooray}, {Ibar}, {Karim}, {Le
  Floc'h}, {Nordon}, {Oliver}, {Page}, {Popesso}, {Pozzi}, {Rigopoulou},
  {Riguccini}, {Rodighiero}, {Rosario}, {Roseboom}, {Wang}, \&
  {Wuyts}}]{Magnelli+15}
{Magnelli}, B., {Ivison}, R.~J., {Lutz}, D., {et~al.} 2015, \aap, 573, A45

\bibitem[{{Magnelli} {et~al.}(2013){Magnelli}, {Popesso}, {Berta}, {Pozzi},
  {Elbaz}, {Lutz}, {Dickinson}, {Altieri}, {Andreani}, {Aussel},
  {B{\'e}thermin}, {Bongiovanni}, {Cepa}, {Charmandaris}, {Chary}, {Cimatti},
  {Daddi}, {F{\"o}rster Schreiber}, {Genzel}, {Gruppioni}, {Harwit}, {Hwang},
  {Ivison}, {Magdis}, {Maiolino}, {Murphy}, {Nordon}, {Pannella}, {P{\'e}rez
  Garc{\'{\i}}a}, {Poglitsch}, {Rosario}, {Sanchez-Portal}, {Santini}, {Scott},
  {Sturm}, {Tacconi}, \& {Valtchanov}}]{Magnelli+13}
{Magnelli}, B., {Popesso}, P., {Berta}, S., {et~al.} 2013, \aap, 553, A132

\bibitem[{{Magorrian} {et~al.}(1998){Magorrian}, {Tremaine}, {Richstone},
  {Bender}, {Bower}, {Dressler}, {Faber}, {Gebhardt}, {Green}, {Grillmair},
  {Kormendy}, \& {Lauer}}]{Magorrian+98}
{Magorrian}, J., {Tremaine}, S., {Richstone}, D., {et~al.} 1998, \aj, 115, 2285

\bibitem[{{Maini} {et~al.}(2016){Maini}, {Prandoni}, {Norris}, {Giovannini}, \&
  {Spitler}}]{Maini+16}
{Maini}, A., {Prandoni}, I., {Norris}, R.~P., {Giovannini}, G., \& {Spitler},
  L.~R. 2016, \aap, 589, L3

\bibitem[{{Marchesi} {et~al.}(2016){Marchesi}, {Civano}, {Elvis}, {Salvato},
  {Brusa}, {Comastri}, {Gilli}, {Hasinger}, {Lanzuisi}, {Miyaji}, {Treister},
  {Urry}, {Vignali}, {Zamorani}, {Allevato}, {Cappelluti}, {Cardamone},
  {Finoguenov}, {Griffiths}, {Karim}, {Laigle}, {LaMassa}, {Jahnke}, {Ranalli},
  {Schawinski}, {Schinnerer}, {Silverman}, {Smolcic}, {Suh}, \&
  {Trakhtenbrot}}]{Marchesi+16}
{Marchesi}, S., {Civano}, F., {Elvis}, M., {et~al.} 2016, \apj, 817, 34

\bibitem[{{Marconi} {et~al.}(2004){Marconi}, {Risaliti}, {Gilli}, {Hunt},
  {Maiolino}, \& {Salvati}}]{Marconi+04}
{Marconi}, A., {Risaliti}, G., {Gilli}, R., {et~al.} 2004, \mnras, 351, 169

\bibitem[{{Marulli} {et~al.}(2008){Marulli}, {Bonoli}, {Branchini},
  {Moscardini}, \& {Springel}}]{Marulli+08}
{Marulli}, F., {Bonoli}, S., {Branchini}, E., {Moscardini}, L., \& {Springel},
  V. 2008, \mnras, 385, 1846

\bibitem[{{McCracken} {et~al.}(2001){McCracken}, {Le F{\`e}vre}, {Brodwin},
  {Foucaud}, {Lilly}, {Crampton}, \& {Mellier}}]{McCracken+01}
{McCracken}, H.~J., {Le F{\`e}vre}, O., {Brodwin}, M., {et~al.} 2001, \aap,
  376, 756

\bibitem[{{McCracken} {et~al.}(2012){McCracken}, {Milvang-Jensen}, {Dunlop},
  {Franx}, {Fynbo}, {Le F{\`e}vre}, {Holt}, {Caputi}, {Goranova}, {Buitrago},
  {Emerson}, {Freudling}, {Hudelot}, {L{\'o}pez-Sanjuan}, {Magnard}, {Mellier},
  {M{\o}ller}, {Nilsson}, {Sutherland}, {Tasca}, \& {Zabl}}]{McCracken+12}
{McCracken}, H.~J., {Milvang-Jensen}, B., {Dunlop}, J., {et~al.} 2012, \aap,
  544, A156

\bibitem[{{Menci} {et~al.}(2008){Menci}, {Fiore}, {Puccetti}, \&
  {Cavaliere}}]{Menci+08}
{Menci}, N., {Fiore}, F., {Puccetti}, S., \& {Cavaliere}, A. 2008, \apj, 686,
  219

\bibitem[{{Merloni} \& {Heinz}(2007)}]{Merloni+07}
{Merloni}, A. \& {Heinz}, S. 2007, \mnras, 381, 589

\bibitem[{{Merloni} \& {Heinz}(2008)}]{Merloni+08}
{Merloni}, A. \& {Heinz}, S. 2008, \mnras, 388, 1011

\bibitem[{{Micha{\l}owski} {et~al.}(2014){Micha{\l}owski}, {Hayward}, {Dunlop},
  {Bruce}, {Cirasuolo}, {Cullen}, \& {Hernquist}}]{Michalowski+14}
{Micha{\l}owski}, M.~J., {Hayward}, C.~C., {Dunlop}, J.~S., {et~al.} 2014,
  \aap, 571, A75

\bibitem[{{Miettinen} {et~al.}(2017){Miettinen}, {Novak}, {Smol{\v c}i{\'c}},
  {Delvecchio}, {Aravena}, {Brisbin}, {Karim}, {Murphy}, {Schinnerer},
  {Albrecht}, {Aussel}, {Bertoldi}, {Capak}, {Casey}, {Civano}, {Hayward},
  {Herrera Ruiz}, {Ilbert}, {Jiang}, {Laigle}, {Le F{\`e}vre}, {Magnelli},
  {Marchesi}, {McCracken}, {Middelberg}, {Mu{\~n}oz Arancibia}, {Navarrete},
  {Padilla}, {Riechers}, {Salvato}, {Scott}, {Sheth}, {Tasca}, {Bondi}, \&
  {Zamorani}}]{miettinen+17}
{Miettinen}, O., {Novak}, M., {Smol{\v c}i{\'c}}, V., {et~al.} 2017, ArXiv
  e-prints [\eprint[arXiv]{1702.07527}]

\bibitem[{{Miettinen} {et~al.}(2015){Miettinen}, {Smol{\v c}i{\'c}}, {Novak},
  {Aravena}, {Karim}, {Masters}, {Riechers}, {Bussmann}, {McCracken}, {Ilbert},
  {Bertoldi}, {Capak}, {Feruglio}, {Halliday}, {Kartaltepe}, {Navarrete},
  {Salvato}, {Sanders}, {Schinnerer}, \& {Sheth}}]{Miettinen+15}
{Miettinen}, O., {Smol{\v c}i{\'c}}, V., {Novak}, M., {et~al.} 2015, \aap, 577,
  A29

\bibitem[{{Miller} {et~al.}(2013){Miller}, {Bonzini}, {Fomalont}, {Kellermann},
  {Mainieri}, {Padovani}, {Rosati}, {Tozzi}, \& {Vattakunnel}}]{Miller+13}
{Miller}, N.~A., {Bonzini}, M., {Fomalont}, E.~B., {et~al.} 2013, \apjs, 205,
  13

\bibitem[{{Miller} {et~al.}(1993){Miller}, {Rawlings}, \&
  {Saunders}}]{Miller+93}
{Miller}, P., {Rawlings}, S., \& {Saunders}, R. 1993, \mnras, 263, 425

\bibitem[{{Monaco} {et~al.}(2000){Monaco}, {Salucci}, \& {Danese}}]{Monaco+00}
{Monaco}, P., {Salucci}, P., \& {Danese}, L. 2000, \mnras, 311, 279

\bibitem[{{Morganti} {et~al.}(2013){Morganti}, {Frieswijk}, {Oonk},
  {Oosterloo}, \& {Tadhunter}}]{Morganti+13}
{Morganti}, R., {Frieswijk}, W., {Oonk}, R.~J.~B., {Oosterloo}, T., \&
  {Tadhunter}, C. 2013, \aap, 552, L4

\bibitem[{{Mori{\'c}} {et~al.}(2010){Mori{\'c}}, {Smol{\v c}i{\'c}}, {Kimball},
  {Riechers}, {Ivezi{\'c}}, \& {Scoville}}]{Moric+10}
{Mori{\'c}}, I., {Smol{\v c}i{\'c}}, V., {Kimball}, A., {et~al.} 2010, \apj,
  724, 779

\bibitem[{{Mullaney} {et~al.}(2011){Mullaney}, {Alexander}, {Goulding}, \&
  {Hickox}}]{Mullaney+11}
{Mullaney}, J.~R., {Alexander}, D.~M., {Goulding}, A.~D., \& {Hickox}, R.~C.
  2011, \mnras, 414, 1082

\bibitem[{{Noeske} {et~al.}(2007){Noeske}, {Weiner}, {Faber}, {Papovich},
  {Koo}, {Somerville}, {Bundy}, {Conselice}, {Newman}, {Schiminovich}, {Le
  Floc'h}, {Coil}, {Rieke}, {Lotz}, {Primack}, {Barmby}, {Cooper}, {Davis},
  {Ellis}, {Fazio}, {Guhathakurta}, {Huang}, {Kassin}, {Martin}, {Phillips},
  {Rich}, {Small}, {Willmer}, \& {Wilson}}]{Noeske+07}
{Noeske}, K.~G., {Weiner}, B.~J., {Faber}, S.~M., {et~al.} 2007, \apjl, 660,
  L43

\bibitem[{{Nordon} {et~al.}(2012){Nordon}, {Lutz}, {Genzel}, {Berta}, {Wuyts},
  {Magnelli}, {Altieri}, {Andreani}, {Aussel}, {Bongiovanni}, {Cepa},
  {Cimatti}, {Daddi}, {Fadda}, {F{\"o}rster Schreiber}, {Lagache}, {Maiolino},
  {P{\'e}rez Garc{\'{\i}}a}, {Poglitsch}, {Popesso}, {Pozzi}, {Rodighiero},
  {Rosario}, {Saintonge}, {Sanchez-Portal}, {Santini}, {Sturm}, {Tacconi},
  {Valtchanov}, \& {Yan}}]{Nordon+12}
{Nordon}, R., {Lutz}, D., {Genzel}, R., {et~al.} 2012, \apj, 745, 182

\bibitem[{{Oke}(1974)}]{Oke74}
{Oke}, J.~B. 1974, \apjs, 27, 21

\bibitem[{{Oliver} {et~al.}(2012){Oliver}, {Bock}, {Altieri}, {Amblard},
  {Arumugam}, {Aussel}, {Babbedge}, {Beelen}, {B{\'e}thermin}, {Blain},
  {Boselli}, {Bridge}, {Brisbin}, {Buat}, {Burgarella},
  {Castro-Rodr{\'{\i}}guez}, {Cava}, {Chanial}, {Cirasuolo}, {Clements},
  {Conley}, {Conversi}, {Cooray}, {Dowell}, {Dubois}, {Dwek}, {Dye}, {Eales},
  {Elbaz}, {Farrah}, {Feltre}, {Ferrero}, {Fiolet}, {Fox}, {Franceschini},
  {Gear}, {Giovannoli}, {Glenn}, {Gong}, {Gonz{\'a}lez Solares}, {Griffin},
  {Halpern}, {Harwit}, {Hatziminaoglou}, {Heinis}, {Hurley}, {Hwang}, {Hyde},
  {Ibar}, {Ilbert}, {Isaak}, {Ivison}, {Lagache}, {Le Floc'h}, {Levenson},
  {Faro}, {Lu}, {Madden}, {Maffei}, {Magdis}, {Mainetti}, {Marchetti},
  {Marsden}, {Marshall}, {Mortier}, {Nguyen}, {O'Halloran}, {Omont}, {Page},
  {Panuzzo}, {Papageorgiou}, {Patel}, {Pearson}, {P{\'e}rez-Fournon}, {Pohlen},
  {Rawlings}, {Raymond}, {Rigopoulou}, {Riguccini}, {Rizzo}, {Rodighiero},
  {Roseboom}, {Rowan-Robinson}, {S{\'a}nchez Portal}, {Schulz}, {Scott},
  {Seymour}, {Shupe}, {Smith}, {Stevens}, {Symeonidis}, {Trichas}, {Tugwell},
  {Vaccari}, {Valtchanov}, {Vieira}, {Viero}, {Vigroux}, {Wang}, {Ward},
  {Wardlow}, {Wright}, {Xu}, \& {Zemcov}}]{Oliver+12}
{Oliver}, S.~J., {Bock}, J., {Altieri}, B., {et~al.} 2012, \mnras, 424, 1614

\bibitem[{{Padovani} {et~al.}(2015){Padovani}, {Bonzini}, {Kellermann},
  {Miller}, {Mainieri}, \& {Tozzi}}]{Padovani+15}
{Padovani}, P., {Bonzini}, M., {Kellermann}, K.~I., {et~al.} 2015, \mnras, 452,
  1263

\bibitem[{{Padovani} {et~al.}(2011){Padovani}, {Miller}, {Kellermann},
  {Mainieri}, {Rosati}, \& {Tozzi}}]{Padovani+11}
{Padovani}, P., {Miller}, N., {Kellermann}, K.~I., {et~al.} 2011, \apj, 740, 20

\bibitem[{{Papovich} {et~al.}(2007){Papovich}, {Rudnick}, {Le Floc'h}, {van
  Dokkum}, {Rieke}, {Taylor}, {Armus}, {Gawiser}, {Huang}, {Marcillac}, \&
  {Franx}}]{Papovich+07}
{Papovich}, C., {Rudnick}, G., {Le Floc'h}, E., {et~al.} 2007, \apj, 668, 45

\bibitem[{{Perna} {et~al.}(2015){Perna}, {Brusa}, {Salvato}, {Cresci},
  {Lanzuisi}, {Berta}, {Delvecchio}, {Fiore}, {Lutz}, {Le Floc'h}, {Mainieri},
  \& {Riguccini}}]{Perna+15}
{Perna}, M., {Brusa}, M., {Salvato}, M., {et~al.} 2015, \aap, 583, A72

\bibitem[{{Poglitsch} {et~al.}(2010){Poglitsch}, {Waelkens}, {Geis},
  {Feuchtgruber}, {Vandenbussche}, {Rodriguez}, {Krause}, {Renotte}, {van
  Hoof}, {Saraceno}, {Cepa}, {Kerschbaum}, {Agn{\`e}se}, {Ali}, {Altieri},
  {Andreani}, {Augueres}, {Balog}, {Barl}, {Bauer}, {Belbachir}, {Benedettini},
  {Billot}, {Boulade}, {Bischof}, {Blommaert}, {Callut}, {Cara}, {Cerulli},
  {Cesarsky}, {Contursi}, {Creten}, {De Meester}, {Doublier}, {Doumayrou},
  {Duband}, {Exter}, {Genzel}, {Gillis}, {Gr{\"o}zinger}, {Henning},
  {Herreros}, {Huygen}, {Inguscio}, {Jakob}, {Jamar}, {Jean}, {de Jong},
  {Katterloher}, {Kiss}, {Klaas}, {Lemke}, {Lutz}, {Madden}, {Marquet},
  {Martignac}, {Mazy}, {Merken}, {Montfort}, {Morbidelli}, {M{\"u}ller},
  {Nielbock}, {Okumura}, {Orfei}, {Ottensamer}, {Pezzuto}, {Popesso},
  {Putzeys}, {Regibo}, {Reveret}, {Royer}, {Sauvage}, {Schreiber}, {Stegmaier},
  {Schmitt}, {Schubert}, {Sturm}, {Thiel}, {Tofani}, {Vavrek}, {Wetzstein},
  {Wieprecht}, \& {Wiezorrek}}]{Poglitsch+10}
{Poglitsch}, A., {Waelkens}, C., {Geis}, N., {et~al.} 2010, \aap, 518, L2

\bibitem[{{Polletta} {et~al.}(2007){Polletta}, {Tajer}, {Maraschi},
  {Trinchieri}, {Lonsdale}, {Chiappetti}, {Andreon}, {Pierre}, {Le F{\`e}vre},
  {Zamorani}, {Maccagni}, {Garcet}, {Surdej}, {Franceschini}, {Alloin},
  {Shupe}, {Surace}, {Fang}, {Rowan-Robinson}, {Smith}, \&
  {Tresse}}]{Polletta+07}
{Polletta}, M., {Tajer}, M., {Maraschi}, L., {et~al.} 2007, \apj, 663, 81

\bibitem[{{Pracy} {et~al.}(2016){Pracy}, {Ching}, {Sadler}, {Croom}, {Baldry},
  {Bland-Hawthorn}, {Brough}, {Brown}, {Couch}, {Davis}, {Drinkwater},
  {Hopkins}, {Jarvis}, {Jelliffe}, {Jurek}, {Loveday}, {Pimbblet}, {Prescott},
  {Wisnioski}, \& {Woods}}]{Pracy+16}
{Pracy}, M.~B., {Ching}, J.~H.~Y., {Sadler}, E.~M., {et~al.} 2016, \mnras, 460,
  2

\bibitem[{{Rosario} {et~al.}(2012){Rosario}, {Santini}, {Lutz}, {Shao},
  {Maiolino}, {Alexander}, {Altieri}, {Andreani}, {Aussel}, {Bauer}, {Berta},
  {Bongiovanni}, {Brandt}, {Brusa}, {Cepa}, {Cimatti}, {Cox}, {Daddi}, {Elbaz},
  {Fontana}, {F{\"o}rster Schreiber}, {Genzel}, {Grazian}, {Le Floch},
  {Magnelli}, {Mainieri}, {Netzer}, {Nordon}, {P{\'e}rez Garcia}, {Poglitsch},
  {Popesso}, {Pozzi}, {Riguccini}, {Rodighiero}, {Salvato}, {Sanchez-Portal},
  {Sturm}, {Tacconi}, {Valtchanov}, \& {Wuyts}}]{Rosario+12}
{Rosario}, D.~J., {Santini}, P., {Lutz}, D., {et~al.} 2012, \aap, 545, A45

\bibitem[{{Rowlands} {et~al.}(2014){Rowlands}, {Dunne}, {Dye},
  {Arag{\'o}n-Salamanca}, {Maddox}, {da Cunha}, {Smith}, {Bourne}, {Eales},
  {Gomez}, {Smail}, {Alpaslan}, {Clark}, {Driver}, {Ibar}, {Ivison},
  {Robotham}, {Smith}, \& {Valiante}}]{Rowlands+14}
{Rowlands}, K., {Dunne}, L., {Dye}, S., {et~al.} 2014, \mnras, 441, 1017

\bibitem[{{Saintonge} {et~al.}(2012){Saintonge}, {Tacconi}, {Fabello}, {Wang},
  {Catinella}, {Genzel}, {Graci{\'a}-Carpio}, {Kramer}, {Moran}, {Heckman},
  {Schiminovich}, {Schuster}, \& {Wuyts}}]{Saintonge+12}
{Saintonge}, A., {Tacconi}, L.~J., {Fabello}, S., {et~al.} 2012, \apj, 758, 73

\bibitem[{{Salvato} {et~al.}(2009){Salvato}, {Hasinger}, {Ilbert}, {Zamorani},
  {Brusa}, {Scoville}, {Rau}, {Capak}, {Arnouts}, {Aussel}, {Bolzonella},
  {Buongiorno}, {Cappelluti}, {Caputi}, {Civano}, {Cook}, {Elvis}, {Gilli},
  {Jahnke}, {Kartaltepe}, {Impey}, {Lamareille}, {Le Floc'h}, {Lilly},
  {Mainieri}, {McCarthy}, {McCracken}, {Mignoli}, {Mobasher}, {Murayama},
  {Sasaki}, {Sanders}, {Schiminovich}, {Shioya}, {Shopbell}, {Silverman},
  {Smol{\v c}i{\'c}}, {Surace}, {Taniguchi}, {Thompson}, {Trump}, {Urry}, \&
  {Zamojski}}]{Salvato+09}
{Salvato}, M., {Hasinger}, G., {Ilbert}, O., {et~al.} 2009, \apj, 690, 1250

\bibitem[{{Salvato} {et~al.}(2011){Salvato}, {Ilbert}, {Hasinger}, {Rau},
  {Civano}, {Zamorani}, {Brusa}, {Elvis}, {Vignali}, {Aussel}, {Comastri},
  {Fiore}, {Le Floc'h}, {Mainieri}, {Bardelli}, {Bolzonella}, {Bongiorno},
  {Capak}, {Caputi}, {Cappelluti}, {Carollo}, {Contini}, {Garilli}, {Iovino},
  {Fotopoulou}, {Fruscione}, {Gilli}, {Halliday}, {Kneib}, {Kakazu},
  {Kartaltepe}, {Koekemoer}, {Kovac}, {Ideue}, {Ikeda}, {Impey}, {Le Fevre},
  {Lamareille}, {Lanzuisi}, {Le Borgne}, {Le Brun}, {Lilly}, {Maier},
  {Manohar}, {Masters}, {McCracken}, {Messias}, {Mignoli}, {Mobasher}, {Nagao},
  {Pello}, {Puccetti}, {Perez-Montero}, {Renzini}, {Sargent}, {Sanders},
  {Scodeggio}, {Scoville}, {Shopbell}, {Silvermann}, {Taniguchi}, {Tasca},
  {Tresse}, {Trump}, \& {Zucca}}]{Salvato+11}
{Salvato}, M., {Ilbert}, O., {Hasinger}, G., {et~al.} 2011, \apj, 742, 61

\bibitem[{{Sanders} \& {Mirabel}(1996)}]{Sanders+96}
{Sanders}, D.~B. \& {Mirabel}, I.~F. 1996, \araa, 34, 749

\bibitem[{{Santini} {et~al.}(2012){Santini}, {Rosario}, {Shao}, {Lutz},
  {Maiolino}, {Alexander}, {Altieri}, {Andreani}, {Aussel}, {Bauer}, {Berta},
  {Bongiovanni}, {Brandt}, {Brusa}, {Cepa}, {Cimatti}, {Daddi}, {Elbaz},
  {Fontana}, {F{\"o}rster Schreiber}, {Genzel}, {Grazian}, {Le Floc'h},
  {Magnelli}, {Mainieri}, {Nordon}, {P{\'e}rez Garcia}, {Poglitsch}, {Popesso},
  {Pozzi}, {Riguccini}, {Rodighiero}, {Salvato}, {Sanchez-Portal}, {Sturm},
  {Tacconi}, {Valtchanov}, \& {Wuyts}}]{Santini+12}
{Santini}, P., {Rosario}, D.~J., {Shao}, L., {et~al.} 2012, \aap, 540, A109

\bibitem[{{Sargent} {et~al.}(2010){Sargent}, {Schinnerer}, {Murphy}, {Carilli},
  {Helou}, {Aussel}, {Le Floc'h}, {Frayer}, {Ilbert}, {Oesch}, {Salvato},
  {Smol{\v c}i{\'c}}, {Kartaltepe}, \& {Sanders}}]{Sargent+10}
{Sargent}, M.~T., {Schinnerer}, E., {Murphy}, E., {et~al.} 2010, \apjl, 714,
  L190

\bibitem[{{Schinnerer} {et~al.}(2010){Schinnerer}, {Sargent}, {Bondi}, {Smol{\v
  c}i{\'c}}, {Datta}, {Carilli}, {Bertoldi}, {Blain}, {Ciliegi}, {Koekemoer},
  \& {Scoville}}]{schinnerer+10}
{Schinnerer}, E., {Sargent}, M.~T., {Bondi}, M., {et~al.} 2010, \apjs, 188, 384

\bibitem[{{Schinnerer} {et~al.}(2007){Schinnerer}, {Smol{\v c}i{\'c}},
  {Carilli}, {Bondi}, {Ciliegi}, {Jahnke}, {Scoville}, {Aussel}, {Bertoldi},
  {Blain}, {Impey}, {Koekemoer}, {Le Fevre}, \& {Urry}}]{schinnerer+07}
{Schinnerer}, E., {Smol{\v c}i{\'c}}, V., {Carilli}, C.~L., {et~al.} 2007,
  \apjs, 172, 46

\bibitem[{{Schmidt}(1959)}]{Schmidt59}
{Schmidt}, M. 1959, \apj, 129, 243

\bibitem[{{Schreiber} {et~al.}(2015){Schreiber}, {Pannella}, {Elbaz},
  {B{\'e}thermin}, {Inami}, {Dickinson}, {Magnelli}, {Wang}, {Aussel}, {Daddi},
  {Juneau}, {Shu}, {Sargent}, {Buat}, {Faber}, {Ferguson}, {Giavalisco},
  {Koekemoer}, {Magdis}, {Morrison}, {Papovich}, {Santini}, \&
  {Scott}}]{Schreiber+15}
{Schreiber}, C., {Pannella}, M., {Elbaz}, D., {et~al.} 2015, \aap, 575, A74

\bibitem[{{Scott} {et~al.}(2008){Scott}, {Austermann}, {Perera}, {Wilson},
  {Aretxaga}, {Bock}, {Hughes}, {Kang}, {Kim}, {Mauskopf}, {Sanders},
  {Scoville}, \& {Yun}}]{Scott+08}
{Scott}, K.~S., {Austermann}, J.~E., {Perera}, T.~A., {et~al.} 2008, \mnras,
  385, 2225

\bibitem[{{Scoville} {et~al.}(2007){Scoville}, {Aussel}, {Brusa}, {Capak},
  {Carollo}, {Elvis}, {Giavalisco}, {Guzzo}, {Hasinger}, {Impey}, {Kneib},
  {LeFevre}, {Lilly}, {Mobasher}, {Renzini}, {Rich}, {Sanders}, {Schinnerer},
  {Schminovich}, {Shopbell}, {Taniguchi}, \& {Tyson}}]{Scoville+07}
{Scoville}, N., {Aussel}, H., {Brusa}, M., {et~al.} 2007, \apjs, 172, 1

\bibitem[{{Shao} {et~al.}(2010){Shao}, {Lutz}, {Nordon}, {Maiolino},
  {Alexander}, {Altieri}, {Andreani}, {Aussel}, {Bauer}, {Berta},
  {Bongiovanni}, {Brandt}, {Brusa}, {Cava}, {Cepa}, {Cimatti}, {Daddi},
  {Dominguez-Sanchez}, {Elbaz}, {F{\"o}rster Schreiber}, {Geis}, {Genzel},
  {Grazian}, {Gruppioni}, {Magdis}, {Magnelli}, {Mainieri}, {P{\'e}rez
  Garc{\'i}a}, {Poglitsch}, {Popesso}, {Pozzi}, {Riguccini}, {Rodighiero},
  {Rovilos}, {Saintonge}, {Salvato}, {Sanchez Portal}, {Santini}, {Sturm},
  {Tacconi}, {Valtchanov}, {Wetzstein}, \& {Wieprecht}}]{Shao+10}
{Shao}, L., {Lutz}, D., {Nordon}, R., {et~al.} 2010, \aap, 518, L26

\bibitem[{{Siebenmorgen} \& {Kr{\"u}gel}(2007)}]{Siebenmorgen+07}
{Siebenmorgen}, R. \& {Kr{\"u}gel}, E. 2007, \aap, 461, 445

\bibitem[{{Sijacki} {et~al.}(2007){Sijacki}, {Springel}, {Di Matteo}, \&
  {Hernquist}}]{Sijacki+07}
{Sijacki}, D., {Springel}, V., {Di Matteo}, T., \& {Hernquist}, L. 2007,
  \mnras, 380, 877

\bibitem[{{Silverman} {et~al.}(2008){Silverman}, {Green}, {Barkhouse}, {Kim},
  {Kim}, {Wilkes}, {Cameron}, {Hasinger}, {Jannuzi}, {Smith}, {Smith}, \&
  {Tananbaum}}]{Silverman+08}
{Silverman}, J.~D., {Green}, P.~J., {Barkhouse}, W.~A., {et~al.} 2008, \apj,
  679, 118

\bibitem[{{Silverman} {et~al.}(2009){Silverman}, {Kova{\v c}}, {Knobel},
  {Lilly}, {Bolzonella}, {Lamareille}, {Mainieri}, {Brusa}, {Cappelluti},
  {Peng}, {Hasinger}, {Zamorani}, {Scodeggio}, {Contini}, {Carollo}, {Jahnke},
  {Kneib}, {Le Fevre}, {Bardelli}, {Bongiorno}, {Brunner}, {Caputi}, {Civano},
  {Comastri}, {Coppa}, {Cucciati}, {de la Torre}, {de Ravel}, {Elvis},
  {Finoguenov}, {Fiore}, {Franzetti}, {Garilli}, {Gilli}, {Griffiths},
  {Iovino}, {Kampczyk}, {Koekemoer}, {Le Borgne}, {Le Brun}, {Maier},
  {Mignoli}, {Pello}, {Perez Montero}, {Ricciardelli}, {Tanaka}, {Tasca},
  {Tresse}, {Vergani}, {Vignali}, {Zucca}, {Bottini}, {Cappi}, {Cassata},
  {Marinoni}, {McCracken}, {Memeo}, {Meneux}, {Oesch}, {Porciani}, \&
  {Salvato}}]{Silverman+09}
{Silverman}, J.~D., {Kova{\v c}}, K., {Knobel}, C., {et~al.} 2009, \apj, 695,
  171

\bibitem[{{Smith} {et~al.}(2012){Smith}, {Dunne}, {da Cunha}, {Rowlands},
  {Maddox}, {Gomez}, {Bonfield}, {Charlot}, {Driver}, {Popescu}, {Tuffs},
  {Dunlop}, {Jarvis}, {Seymour}, {Symeonidis}, {Baes}, {Bourne}, {Clements},
  {Cooray}, {De Zotti}, {Dye}, {Eales}, {Scott}, {Verma}, {van der Werf},
  {Andrae}, {Auld}, {Buttiglione}, {Cava}, {Dariush}, {Fritz}, {Hopwood},
  {Ibar}, {Ivison}, {Kelvin}, {Madore}, {Pohlen}, {Rigby}, {Robotham},
  {Seibert}, \& {Temi}}]{Smith+12}
{Smith}, D.~J.~B., {Dunne}, L., {da Cunha}, E., {et~al.} 2012, \mnras, 427, 703

\bibitem[{{Smolcic} {et~al.}(2017{\natexlab{a}}){Smolcic}, {Delvecchio},
  {Zamorani}, {Baran}, {Novak}, {Delhaize}, {Schinnerer}, {Berta}, {Bondi},
  {Ciliegi}, {Capak}, {Civano}, {Karim}, {Le Fevre}, {Ilbert}, {Laigle},
  {Marchesi}, {McCracken}, {Tasca}, {Salvato}, \& {Vardoulaki}}]{smolcic+17b}
{Smolcic}, V., {Delvecchio}, I., {Zamorani}, G., {et~al.} 2017{\natexlab{a}},
  ArXiv e-prints [\eprint[arXiv]{1703.09719}]

\bibitem[{{Smolcic} {et~al.}(2017{\natexlab{b}}){Smolcic}, {Novak}, {Bondi},
  {Ciliegi}, {Mooley}, {Schinnerer}, {Zamorani}, {Navarrete}, {Bourke},
  {Karim}, {Vardoulaki}, {Leslie}, {Delhaize}, {Carilli}, {Myers}, {Baran},
  {Delvecchio}, {Miettinen}, {Banfield}, {Balokovic}, {Bertoldi}, {Capak},
  {Frail}, {Hallinan}, {Hao}, {Herrera Ruiz}, {Horesh}, {Ilbert}, {Intema},
  {Jelic}, {Kloeckner}, {Krpan}, {Kulkarni}, {McCracken}, {Laigle},
  {Middleberg}, {Murphy}, {Sargent}, {Scoville}, \& {Sheth}}]{smolcic+17a}
{Smolcic}, V., {Novak}, M., {Bondi}, M., {et~al.} 2017{\natexlab{b}}, ArXiv
  e-prints [\eprint[arXiv]{1703.09713}]

\bibitem[{{Smol{\v c}i{\'c}}(2009)}]{Smolcic+09b}
{Smol{\v c}i{\'c}}, V. 2009, \apjl, 699, L43

\bibitem[{{Smol{\v c}i{\'c}} {et~al.}(2012){Smol{\v c}i{\'c}}, {Aravena},
  {Navarrete}, {Schinnerer}, {Riechers}, {Bertoldi}, {Feruglio}, {Finoguenov},
  {Salvato}, {Sargent}, {McCracken}, {Albrecht}, {Karim}, {Capak}, {Carilli},
  {Cappelluti}, {Elvis}, {Ilbert}, {Kartaltepe}, {Lilly}, {Sanders}, {Sheth},
  {Scoville}, \& {Taniguchi}}]{Smolcic+12}
{Smol{\v c}i{\'c}}, V., {Aravena}, M., {Navarrete}, F., {et~al.} 2012, \aap,
  548, A4

\bibitem[{{Smol{\v c}i{\'c}} {et~al.}(2008){Smol{\v c}i{\'c}}, {Schinnerer},
  {Scodeggio}, {Franzetti}, {Aussel}, {Bondi}, {Brusa}, {Carilli}, {Capak},
  {Charlot}, {Ciliegi}, {Ilbert}, {Ivezi{\'c}}, {Jahnke}, {McCracken},
  {Obri{\'c}}, {Salvato}, {Sanders}, {Scoville}, {Trump}, {Tremonti}, {Tasca},
  {Walcher}, \& {Zamorani}}]{Smolcic+08}
{Smol{\v c}i{\'c}}, V., {Schinnerer}, E., {Scodeggio}, M., {et~al.} 2008,
  \apjs, 177, 14

\bibitem[{{Smol{\v c}i{\'c}} {et~al.}(2009){Smol{\v c}i{\'c}}, {Zamorani},
  {Schinnerer}, {Bardelli}, {Bondi}, {B{\^i}rzan}, {Carilli}, {Ciliegi},
  {Elvis}, {Impey}, {Koekemoer}, {Merloni}, {Paglione}, {Salvato}, {Scodeggio},
  {Scoville}, \& {Trump}}]{Smolcic+09a}
{Smol{\v c}i{\'c}}, V., {Zamorani}, G., {Schinnerer}, E., {et~al.} 2009, \apj,
  696, 24

\bibitem[{{Speagle} {et~al.}(2014){Speagle}, {Steinhardt}, {Capak}, \&
  {Silverman}}]{Speagle+14}
{Speagle}, J.~S., {Steinhardt}, C.~L., {Capak}, P.~L., \& {Silverman}, J.~D.
  2014, \apjs, 214, 15

\bibitem[{{Spergel} {et~al.}(2003){Spergel}, {Verde}, {Peiris}, {Komatsu},
  {Nolta}, {Bennett}, {Halpern}, {Hinshaw}, {Jarosik}, {Kogut}, {Limon},
  {Meyer}, {Page}, {Tucker}, {Weiland}, {Wollack}, \& {Wright}}]{Spergel+03}
{Spergel}, D.~N., {Verde}, L., {Peiris}, H.~V., {et~al.} 2003, \apjs, 148, 175

\bibitem[{{Steinhardt} {et~al.}(2014){Steinhardt}, {Speagle}, {Capak},
  {Silverman}, {Carollo}, {Dunlop}, {Hashimoto}, {Hsieh}, {Ilbert}, {Le Fevre},
  {Le Floc'h}, {Lee}, {Lin}, {Lin}, {Masters}, {McCracken}, {Nagao}, {Petric},
  {Salvato}, {Sanders}, {Scoville}, {Sheth}, {Strauss}, \&
  {Taniguchi}}]{Steinhardt+14}
{Steinhardt}, C.~L., {Speagle}, J.~S., {Capak}, P., {et~al.} 2014, \apjl, 791,
  L25

\bibitem[{{Strateva} {et~al.}(2001){Strateva}, {Ivezi{\'c}}, {Knapp},
  {Narayanan}, {Strauss}, {Gunn}, {Lupton}, {Schlegel}, {Bahcall}, {Brinkmann},
  {Brunner}, {Budav{\'a}ri}, {Csabai}, {Castander}, {Doi}, {Fukugita}, {Gy{\H
  o}ry}, {Hamabe}, {Hennessy}, {Ichikawa}, {Kunszt}, {Lamb}, {McKay},
  {Okamura}, {Racusin}, {Sekiguchi}, {Schneider}, {Shimasaku}, \&
  {York}}]{Strateva+01}
{Strateva}, I., {Ivezi{\'c}}, {\v Z}., {Knapp}, G.~R., {et~al.} 2001, \aj, 122,
  1861

\bibitem[{{Symeonidis} {et~al.}(2014){Symeonidis}, {Georgakakis}, {Page},
  {Bock}, {Bonzini}, {Buat}, {Farrah}, {Franceschini}, {Ibar}, {Lutz},
  {Magnelli}, {Magdis}, {Oliver}, {Pannella}, {Paolillo}, {Rosario},
  {Roseboom}, {Vaccari}, \& {Villforth}}]{Symeonidis+14}
{Symeonidis}, M., {Georgakakis}, A., {Page}, M.~J., {et~al.} 2014, \mnras, 443,
  3728

\bibitem[{{Szokoly} {et~al.}(2004){Szokoly}, {Bergeron}, {Hasinger}, {Lehmann},
  {Kewley}, {Mainieri}, {Nonino}, {Rosati}, {Giacconi}, {Gilli}, {Gilmozzi},
  {Norman}, {Romaniello}, {Schreier}, {Tozzi}, {Wang}, {Zheng}, \&
  {Zirm}}]{Szokoly+04}
{Szokoly}, G.~P., {Bergeron}, J., {Hasinger}, G., {et~al.} 2004, \apjs, 155,
  271

\bibitem[{{Tacconi} {et~al.}(2013){Tacconi}, {Neri}, {Genzel}, {Combes},
  {Bolatto}, {Cooper}, {Wuyts}, {Bournaud}, {Burkert}, {Comerford}, {Cox},
  {Davis}, {F{\"o}rster Schreiber}, {Garc{\'{\i}}a-Burillo}, {Gracia-Carpio},
  {Lutz}, {Naab}, {Newman}, {Omont}, {Saintonge}, {Shapiro Griffin}, {Shapley},
  {Sternberg}, \& {Weiner}}]{Tacconi+13}
{Tacconi}, L.~J., {Neri}, R., {Genzel}, R., {et~al.} 2013, \apj, 768, 74

\bibitem[{{Tasca} {et~al.}(2016){Tasca}, {Le Fevre}, {Ribeiro}, {Thomas},
  {Moreau}, {Cassata}, {Garilli}, {Le Brun}, {Lemaux}, {Maccagni},
  {Pentericci}, {Schaerer}, {Vanzella}, {Zamorani}, {Zucca}, {Amorin},
  {Bardelli}, {Cassara}, {Castellano}, {Cimatti}, {Cucciati}, {Durkalec},
  {Fontana}, {Giavalisco}, {Grazian}, {Hathi}, {Ilbert}, {Paltani}, {Pforr},
  {Scodeggio}, {Sommariva}, {Talia}, {Tresse}, {Vergani}, {Capak}, {Charlot},
  {Contini}, {de la Torre}, {Dunlop}, {Fotopoulou}, {Guaita}, {Koekemoer},
  {Lopez-Sanjuan}, {Mellier}, {Salvato}, {Scoville}, {Taniguchi}, \&
  {Wang}}]{Tasca+16}
{Tasca}, L.~A.~M., {Le Fevre}, O., {Ribeiro}, B., {et~al.} 2016, submitted
  [\eprint[arXiv]{1602.01842}]

\bibitem[{{Ueda} {et~al.}(2014){Ueda}, {Akiyama}, {Hasinger}, {Miyaji}, \&
  {Watson}}]{Ueda+14}
{Ueda}, Y., {Akiyama}, M., {Hasinger}, G., {Miyaji}, T., \& {Watson}, M.~G.
  2014, \apj, 786, 104

\bibitem[{{Veilleux} {et~al.}(2013){Veilleux}, {Mel{\'e}ndez}, {Sturm},
  {Gracia-Carpio}, {Fischer}, {Gonz{\'a}lez-Alfonso}, {Contursi}, {Lutz},
  {Poglitsch}, {Davies}, {Genzel}, {Tacconi}, {de Jong}, {Sternberg}, {Netzer},
  {Hailey-Dunsheath}, {Verma}, {Rupke}, {Maiolino}, {Teng}, \&
  {Polisensky}}]{Veilleux+13}
{Veilleux}, S., {Mel{\'e}ndez}, M., {Sturm}, E., {et~al.} 2013, \apj, 776, 27

\bibitem[{{Vito} {et~al.}(2014){Vito}, {Gilli}, {Vignali}, {Comastri}, {Brusa},
  {Cappelluti}, \& {Iwasawa}}]{Vito+14}
{Vito}, F., {Gilli}, R., {Vignali}, C., {et~al.} 2014, \mnras, 445, 3557

\bibitem[{{Whitaker} {et~al.}(2012){Whitaker}, {van Dokkum}, {Brammer}, \&
  {Franx}}]{Whitaker+12}
{Whitaker}, K.~E., {van Dokkum}, P.~G., {Brammer}, G., \& {Franx}, M. 2012,
  \apjl, 754, L29

\bibitem[{{White} {et~al.}(2015){White}, {Jarvis}, {H{\"a}u{\ss}ler}, \&
  {Maddox}}]{White+15}
{White}, S.~V., {Jarvis}, M.~J., {H{\"a}u{\ss}ler}, B., \& {Maddox}, N. 2015,
  \mnras, 448, 2665

\bibitem[{{Willott} {et~al.}(1999){Willott}, {Rawlings}, {Blundell}, \&
  {Lacy}}]{Willott+99}
{Willott}, C.~J., {Rawlings}, S., {Blundell}, K.~M., \& {Lacy}, M. 1999,
  \mnras, 309, 1017

\bibitem[{{Windhorst} {et~al.}(1985){Windhorst}, {Miley}, {Owen}, {Kron}, \&
  {Koo}}]{Windhorst+85}
{Windhorst}, R.~A., {Miley}, G.~K., {Owen}, F.~N., {Kron}, R.~G., \& {Koo},
  D.~C. 1985, \apj, 289, 494

\bibitem[{{Wuyts} {et~al.}(2011){Wuyts}, {F{\"o}rster Schreiber}, {Lutz},
  {Nordon}, {Berta}, {Altieri}, {Andreani}, {Aussel}, {Bongiovanni}, {Cepa},
  {Cimatti}, {Daddi}, {Elbaz}, {Genzel}, {Koekemoer}, {Magnelli}, {Maiolino},
  {McGrath}, {P{\'e}rez Garc{\'{\i}}a}, {Poglitsch}, {Popesso}, {Pozzi},
  {Sanchez-Portal}, {Sturm}, {Tacconi}, \& {Valtchanov}}]{Wuyts+11}
{Wuyts}, S., {F{\"o}rster Schreiber}, N.~M., {Lutz}, D., {et~al.} 2011, \apj,
  738, 106

\bibitem[{{Wuyts} {et~al.}(2008){Wuyts}, {Labb{\'e}}, {F{\"o}rster Schreiber},
  {Franx}, {Rudnick}, {Brammer}, \& {van Dokkum}}]{Wuyts+08}
{Wuyts}, S., {Labb{\'e}}, I., {F{\"o}rster Schreiber}, N.~M., {et~al.} 2008,
  \apj, 682, 985

\bibitem[{{Xue} {et~al.}(2010){Xue}, {Brandt}, {Luo}, {Rafferty}, {Alexander},
  {Bauer}, {Lehmer}, {Schneider}, \& {Silverman}}]{Xue+10}
{Xue}, Y.~Q., {Brandt}, W.~N., {Luo}, B., {et~al.} 2010, \apj, 720, 368

\bibitem[{{York} {et~al.}(2000){York}, {Adelman}, {Anderson}, {Anderson},
  {Annis}, {Bahcall}, {Bakken}, {Barkhouser}, {Bastian}, {Berman}, {Boroski},
  {Bracker}, {Briegel}, {Briggs}, {Brinkmann}, {Brunner}, {Burles}, {Carey},
  {Carr}, {Castander}, {Chen}, {Colestock}, {Connolly}, {Crocker}, {Csabai},
  {Czarapata}, {Davis}, {Doi}, {Dombeck}, {Eisenstein}, {Ellman}, {Elms},
  {Evans}, {Fan}, {Federwitz}, {Fiscelli}, {Friedman}, {Frieman}, {Fukugita},
  {Gillespie}, {Gunn}, {Gurbani}, {de Haas}, {Haldeman}, {Harris}, {Hayes},
  {Heckman}, {Hennessy}, {Hindsley}, {Holm}, {Holmgren}, {Huang}, {Hull},
  {Husby}, {Ichikawa}, {Ichikawa}, {Ivezi{\'c}}, {Kent}, {Kim}, {Kinney},
  {Klaene}, {Kleinman}, {Kleinman}, {Knapp}, {Korienek}, {Kron}, {Kunszt},
  {Lamb}, {Lee}, {Leger}, {Limmongkol}, {Lindenmeyer}, {Long}, {Loomis},
  {Loveday}, {Lucinio}, {Lupton}, {MacKinnon}, {Mannery}, {Mantsch}, {Margon},
  {McGehee}, {McKay}, {Meiksin}, {Merelli}, {Monet}, {Munn}, {Narayanan},
  {Nash}, {Neilsen}, {Neswold}, {Newberg}, {Nichol}, {Nicinski}, {Nonino},
  {Okada}, {Okamura}, {Ostriker}, {Owen}, {Pauls}, {Peoples}, {Peterson},
  {Petravick}, {Pier}, {Pope}, {Pordes}, {Prosapio}, {Rechenmacher}, {Quinn},
  {Richards}, {Richmond}, {Rivetta}, {Rockosi}, {Ruthmansdorfer}, {Sandford},
  {Schlegel}, {Schneider}, {Sekiguchi}, {Sergey}, {Shimasaku}, {Siegmund},
  {Smee}, {Smith}, {Snedden}, {Stone}, {Stoughton}, {Strauss}, {Stubbs},
  {SubbaRao}, {Szalay}, {Szapudi}, {Szokoly}, {Thakar}, {Tremonti}, {Tucker},
  {Uomoto}, {Vanden Berk}, {Vogeley}, {Waddell}, {Wang}, {Watanabe},
  {Weinberg}, {Yanny}, {Yasuda}, \& {SDSS Collaboration}}]{York+00}
{York}, D.~G., {Adelman}, J., {Anderson}, Jr., J.~E., {et~al.} 2000, \aj, 120,
  1579

\bibitem[{{Yun} {et~al.}(2001){Yun}, {Reddy}, \& {Condon}}]{Yun+01}
{Yun}, M.~S., {Reddy}, N.~A., \& {Condon}, J.~J. 2001, \apj, 554, 803

\bibitem[{{Zamorani} {et~al.}(1981){Zamorani}, {Henry}, {Maccacaro},
  {Tananbaum}, {Soltan}, {Avni}, {Liebert}, {Stocke}, {Strittmatter},
  {Weymann}, {Smith}, \& {Condon}}]{Zamorani+81}
{Zamorani}, G., {Henry}, J.~P., {Maccacaro}, T., {et~al.} 1981, \apj, 245, 357

\end{thebibliography}

\begin{appendix}

\onecolumn
\section{Results from \textit{Herschel} stacking}

In Table \ref{table:stack}, we list the fluxes and corresponding 1$\sigma$ uncertainties obtained by means of \textit{Herschel} stacking. The results reported below refer to the sample of {2\,203} radio sources without ($>$3$\sigma$) detection in any \textit{Herschel} band, and not classified as HLAGN. See Sect. \ref{ir_stacking} for a more detailed description.

\begin{table}[!h]
   \centering
   \caption{\small Median stacked fluxes derived in each redshift bin through a bootstrapping procedure. Lower and upper errors (in mJy) correspond to the 16$^{\rm th}$ and 84$^{\rm th}$ percentiles of the cumulative flux distribution. In case the error is larger than the median flux, we report the 1$\sigma$ upper flux. The number of stacked sources $N_{\rm stack}$ is reported for each redshift bin.}
\begin{tabular}{l cccccc }
\hline
\hline
 redshift bin         & $N_{\rm stack}$  &    PACS 100   &      PACS 160       &      SPIRE 250          &        SPIRE 350        &       SPIRE 500          \\
                      &                  &   [mJy]       &       [mJy]         &       [mJy]             &         [mJy]           &      [mJy]               \\
\hline
\bigskip
0.01 $\leq z <$ 0.30  &  37  &  $<$0.42  &  
1.90$^{+0.44}_{-0.69}$  &  2.20$^{+1.92}_{-0.66}$  &  
2.24$^{+1.09}_{-0.47}$  &  $<$3.02  \\ 
  \bigskip 
0.30 $\leq z <$ 0.70  &  366  &  1.09$^{+0.15}_{-0.11}$  &  
2.22$^{+0.30}_{-0.27}$  &  1.64$^{+0.20}_{-0.37}$  &  
0.89$^{+0.68}_{-0.62}$  &  $<$1.09  \\ 
  \bigskip 
0.70 $\leq z <$ 1.20  &  642  &  1.02$^{+0.08}_{-0.09}$  &  
1.91$^{+0.20}_{-0.14}$  &  1.98$^{+0.28}_{-0.27}$  &  
1.47$^{+0.44}_{-0.37}$  &  0.70$^{+0.30}_{-0.43}$  \\ 
  \bigskip 
1.20 $\leq z <$ 1.80  &  620  &  1.44$^{+0.10}_{-0.08}$  &  
3.03$^{+0.20}_{-0.24}$  &  4.21$^{+0.47}_{-0.49}$  &  
4.25$^{+0.52}_{-0.50}$  &  2.75$^{+0.20}_{-0.37}$  \\ 
  \bigskip 
1.80 $\leq z <$ 2.50  &  306  &  0.95$^{+0.13}_{-0.11}$  &  
2.49$^{+0.24}_{-0.30}$  &  3.33$^{+0.25}_{-0.38}$  &  
3.77$^{+0.45}_{-0.77}$  &  3.10$^{+0.50}_{-0.58}$  \\ 
  \bigskip 
2.50 $\leq z <$ 3.50  &  175  &  1.00$^{+0.10}_{-0.13}$  &  
2.25$^{+0.34}_{-0.30}$  &  5.44$^{+0.99}_{-1.36}$  &  
5.73$^{+0.88}_{-0.54}$  &  4.55$^{+1.13}_{-1.13}$  \\ 
  \bigskip 
3.50 $\leq z <$ 5.70  &  57  &  0.79$^{+0.22}_{-0.37}$  &  
2.15$^{+0.88}_{-0.40}$  &  5.19$^{+1.19}_{-0.98}$  &  
7.68$^{+0.94}_{-1.35}$  &  6.80$^{+0.60}_{-1.35}$  \\ 

\hline

\end{tabular}

\label{table:stack}
\end{table}


\section{Value-added 3~GHz radio catalogue} \label{catalog_table}

For guidance, we show 20 lines of the value-added catalogue used in this work in Table \ref{table:catalogue}, following the same format introduced in Sect. \ref{catalog}.

\begin{landscape}
\begin{table}[!h]
   \centering
   \small
   \caption{\small Table listing the properties and classification for the first 20 radio sources (sorted by ID~3~GHz) used in this work. The full Table will be made available through the IPAC/IRSA database. }
\begin{tabular}{l ccccccccccccccc }
\hline
\hline
\small 

(1) & (2) & (3) & (4) & (5) & (6) & (7) & (8) & (9) & (10) & (11) & (12) & (13) & (14-17) & (18) \\
ID & RA 3~GHz & Dec 3~GHz & Redshift  &  Redshift  &   $S_{\rm 3~GHz}$ & $L_{\rm 3~GHz}$ & $L_{\rm 1.4\, GHz}$ & $L_{\rm IR, SF}$  &  $\geq$3$\sigma$ \textit{Herschel}   &    $M_{\star}$  &  SFR$_{\rm IR}$ &  {\sc [NUV-$r$]}   & AGN  &  Class \\
 3~GHz &  [deg]  & [deg]  &   &   type   & [$\mu$Jy]   &  log[W~Hz$^{-1}$] &  log[W~Hz$^{-1}$] & log[$L_{\odot}$]  & detection   & log[$M_{\odot}$] & $M_{\odot}$~yr$^{-1}$ &   AB mag  &  criteria  & \\
  \hline \\
1  &  149.64771  &  2.09546  &  1.546  &  phot  &  16147.0  &  25.84  &  
25.72  &  11.66  &  true  &  11.22  &  46.0  &  2.42  &         0       0
       01  &   MLAGN \\
3  &  150.33360  &  2.57880  &  1.555  &  phot  &  11369.2  &  25.80  &  
25.77  &  12.12  &  true  &  10.21  &  131.4  &  -0.29  &         1       1
       11  &   HLAGN \\
5  &  150.72035  &  1.93047  &  2.446  &  phot  &  7746.0  &  26.57  &  
26.90  &  12.01  &  false  &  11.45  &  102.5  &  1.84  &         0       0
       01  &   MLAGN \\
6  &  150.47405  &  2.83167  &  1.259  &  spec  &  9212.9  &  25.90  &  
26.20  &  11.26  &  false  &  11.23  &  18.3  &  3.14  &         1       0
       01  &   HLAGN \\
8  &  150.00256  &  2.25863  &  2.450  &  spec  &  6749.5  &  25.79  &  
25.68  &  11.48  &  true  &  11.08  &  30.3  &  2.07  &         1       1
       11  &   HLAGN \\
16  &  149.51338  &  2.23267  &  2.238  &  phot  &  13523.3  &  26.62  &  
26.88  &  11.42  &  false  &  11.03  &  26.2  &  1.34  &         1       0
       01  &   HLAGN \\
17  &  150.72133  &  1.58238  &  1.051  &  phot  &  6278.1  &  25.04  &  
24.80  &  11.41  &  false  &  11.22  &  25.8  &  2.49  &         1       0
       11  &   HLAGN \\
18  &  149.95944  &  1.80146  &  0.684  &  spec  &  3235.1  &  24.35  &  
23.99  &  10.49  &  false  &  10.92  &  3.1  &  2.85  &         1       0
       01  &   HLAGN \\
19  &  149.42624  &  2.07387  &  1.081  &  phot  &  8626.8  &  25.75  &  
26.10  &  11.91  &  true  &  10.76  &  81.3  &  0.63  &         0       0
       11  &   HLAGN \\
23  &  150.56024  &  2.58613  &  2.099  &  phot  &  3151.0  &  25.85  &  
26.07  &  11.40  &  false  &  11.24  &  25.3  &  2.26  &         0       0
       01  &   MLAGN \\
25  &  150.06906  &  2.44399  &  2.436  &  phot  &  3178.2  &  26.05  &  
26.30  &  12.70  &  true  &  10.61  &  501.8  &  -0.18  &         0       1
       11  &   HLAGN \\
29  &  150.44730  &  2.05394  &  0.323  &  spec  &  4133.0  &  24.12  &  
24.36  &  9.98  &  false  &  11.45  &  0.9  &  4.90  &         1       0       0
1  &   HLAGN \\
30  &  149.96638  &  2.09516  &  1.356  &  phot  &  2277.0  &  24.84  &  
24.67  &  12.27  &  true  &  11.52  &  187.5  &  1.17  &         0       0
       01  &   MLAGN \\
31  &  150.61989  &  2.28940  &  2.625  &  spec  &  2372.8  &  25.88  &  
26.07  &  11.82  &  false  &  11.23  &  65.5  &  1.17  &         1       0
       01  &   HLAGN \\
33  &  150.37977  &  2.49021  &  0.349  &  spec  &  2969.6  &  24.06  &  
24.31  &  10.27  &  false  &  11.30  &  1.9  &  4.42  &         1       0
       01  &   HLAGN \\
37  &  149.78699  &  1.60178  &  1.680  &  phot  &  2405.4  &  25.75  &  
26.15  &  11.64  &  false  &  10.74  &  43.8  &  1.22  &         0       0
       01  &   MLAGN \\
39  &  149.61919  &  1.91632  &  0.913  &  spec  &  1868.1  &  24.87  &  
25.17  &  10.87  &  false  &  10.67  &  7.4  &  2.42  &         0       0
       01  &   MLAGN \\
40  &  149.57238  &  2.26267  &  0.706  &  spec  &  1825.5  &  24.52  &  
24.71  &  11.49  &  true  &  11.31  &  30.9  &  1.97  &         1       0
       11  &   HLAGN \\
41  &  149.55937  &  1.63104  &  0.901  &  spec  &  2102.8  &  24.71  &  
24.77  &  11.10  &  false  &  11.32  &  12.6  &  2.32  &         1       0
       01  &   HLAGN \\
43  &  150.67227  &  1.71565  &  0.609  &  phot  &  1510.3  &  23.92  &  
23.54  &  11.35  &  true  &  11.53  &  22.4  &  2.33  &         0       0
       01  &   MLAGN \\

\end{tabular}

\label{table:catalogue}
\end{table}

\end{landscape}

\end{appendix}

\end{document}